\documentclass[twocolumn,showpacs,nofootinbib,preprintnumbers,amsmath,amssymb,
showkeys,superscriptaddress]{revtex4-1}

\usepackage[pdftex]{graphicx}  
\usepackage{color}
\usepackage{slashed}
\usepackage[normalem]{ulem}
\usepackage{soul}
\usepackage{dsfont}

\newcommand \beq{\begin{eqnarray}}
\newcommand \eeq{\end{eqnarray}}

\begin{document}

\title{Two-loop corrections to the QCD propagators\\ within the Curci-Ferrari model}

\author{Nahuel Barrios\vspace{.4cm}}%
\affiliation{%
Instituto de F\'{\i}sica, Facultad de Ingenier\'{\i}a, Universidad de
la Rep\'ublica, J. H. y Reissig 565, 11000 Montevideo, Uruguay.
\vspace{.1cm}}%
\affiliation{%
Centre de Physique Th\'eorique (CPHT), CNRS, Ecole Polytechnique,\\ Institut Polytechnique de Paris,  Route de Saclay, F-91128 Palaiseau, France.
\vspace{.1cm}}%

\author{John A. Gracey}
\affiliation{Theoretical Physics Division,
Department of Mathematical Sciences,
University of Liverpool,
P.O. Box 147,
Liverpool,
L69 3BX,
United Kingdom}%

\author{Marcela Pel\'aez\vspace{.4cm}}%
\affiliation{%
Instituto de F\'{\i}sica, Facultad de Ingenier\'{\i}a, Universidad de
la Rep\'ublica, J. H. y Reissig 565, 11000 Montevideo, Uruguay.
\vspace{.1cm}}%

\author{Urko Reinosa}%
\affiliation{%
Centre de Physique Th\'eorique (CPHT), CNRS, Ecole Polytechnique,\\ Institut Polytechnique de Paris,  Route de Saclay, F-91128 Palaiseau, France.
\vspace{.1cm}}%

\date{\today}

\begin{abstract}
We evaluate all two-point correlation functions of the Curci-Ferrari (CF) model in four dimensions and in the presence of mass-degenerate fundamental quark flavors, as a natural extension of an earlier investigation in the quenched approximation. In principle, the proper account of chiral symmetry breaking ($\chi$SB) and the corresponding dynamical generation of a quark mass function within the CF model requires one to go beyond perturbation theory \cite{Pelaez:2020ups}. However, it is interesting to assess whether a perturbative description applies to correlation functions that are not directly sensitive to $\chi$SB, such as the gluon, ghost and quark dressing functions. We compare our two-loop results for these form factors to QCD lattice data in the two flavor case for two different values of the pion mass, one that is relatively far from the chiral limit, and one that is closer to the physical value. Our results confirm that the QCD gluon and ghost dressing functions are well described by a perturbative approach within the CF model, as already observed at one-loop order in Ref.~\cite{Pelaez:2014mxa}. Our new main result is that the quark dressing function is also well captured by the perturbative approach, but only starting at two-loop order, as also anticipated in Ref.~\cite{Pelaez:2014mxa}. The quark mass function predicted by the CF model at two-loop order is in good agreement with the data if the quarks are not too light but shows some clear tension with respect to the two-loop CF dressing functions in the close to physical case, as expected. Interestingly, however, we find that there is much less tension between the non-perturbative quark mass function, as it can be obtained from lattice simulations or from \cite{Pelaez:2020ups}, and the two-loop CF dressing functions, which confirms the perturbative nature of the latter.
\end{abstract}

\preprint{LTH 1255}

\pacs{}
\keywords{}
\maketitle

\section{Introduction}
The success of the Standard Model of particle physics in describing three out of the four fundamental interactions is not in any doubt. Nonetheless, while the properties
of the electroweak sector are very well understood over a large
range of energies, that part describing the strong sector is not. At high
energy, the quarks and gluons, the fundamental fields of the $SU(3)$
gauge theory of the strong sector known as Quantum Chromodynamics (QCD),
behave asymptotically as free entities, \cite{grosswilczek,politzer}. This is 
only a high energy property, however, as in reality such quark and gluon states are never 
realized in Nature as observable particles. Instead, they are confined within 
nucleons and from lattice gauge studies of their propagators, it has become clear that they do not share the same fundamental behaviour 
as the electrons and photons of Quantum Electrodynamics. A distinctive feature is that, as a 
function of the momentum $p^2$, the propagators do not have a simple real pole. See, 
for instance,
\cite{glmq12,glmq13,glmq14,glmq15,glmq16,glmq17,glmq18,glmq19,glmq20}. 

Consequently there have been numerous theoretical attempts to explain the 
behaviour of the gluon propagator analytically. The most common approaches rely on non-perturbative methods such 
as the Dyson-Schwinger equations \cite{Huber:2020keu} or the functional renormalization group \cite{Cyrol:2016tym}. Alongside these non-perturbative studies, it has also been advocated that valuable information could be obtained from perturbative methods  \cite{Tissier:2010ts,Reinosa:2017qtf,Pelaez:2021tpq}. All these approaches centre  around a common theme of there being a non-zero mass scale of some sort in the pure gauge sector of QCD -- also referred to in what follows as the Yang-Mills (YM) sector -- that is active primarily at low energies.

Ideally, one aims at generating this scale from first principles, as for instance in the original work of Gribov, 
\cite{gribov78}, where it arose out of endeavouring to globally fix the 
Landau gauge uniquely.  A more phenomenological approach relies on the inclusion of a non-zero gluon mass term in the 
Landau gauge-fixed YM Lagrangian as a way to model the effect of the non-perturbative gauge-fixing. This modified Lagrangian corresponds in fact to one particular case of the Curci-Ferrari (CF) model \cite{Curci:1976bt}. That approach from nearly half a century ago fell out of 
favour despite leading to renormalizable actions. Indeed, it transpired 
that the BRST charge is not nilpotent in the presence of the explicit mass. 
Consequently the standard definition of the physical state space contained 
states with negative norm
\cite{deBoer:1995dh,Curci:1976kh,Ojima:1981fs}. Since then,
however, lattice simulations have identified positivity 
violations in the gluon propagator \cite{Cucchieri:2004mf,Bowman:2007du}. This empirical observation, 
together with the decoupling behaviour  of the gluon propagator  observed for dimensions strictly greater than two \cite{Cucchieri:2007rg,Cucchieri:2008fc}, has made of the CF model one new avenue for exploring the infrared behaviour of the gluon 
and Faddeev-Popov ghost propagators. 

In fact, over the past years, the model has been extensively used to examine the infrared behaviour of YM correlation functions in the vacuum 
\cite{Tissier:2010ts,Tissier:2011ey,Pelaez:2013cpa} as well as the corresponding phase structure at nonzero 
temperature \cite{Reinosa:2014ooa,Reinosa:2020mnx,VanEgmond:2021mlj}, both using rather simple one-loop calculations. One of the reasons why the model may be regarded as a
credible candidate for describing infrared gluon dynamics is that it was argued that the mass parameter can
be interpreted as a necessary second gauge parameter \cite{Tissier:2017fqf,Serreau:2012cg} . Its origin derives from 
taking into account the presence and effect of Gribov copies; see also \cite{Nous:2020vdq} for more recent developments. In the 
ultraviolet, such a mass is unnecessary and absent as it runs to zero consistent
with the fact that the Landau gauge is uniquely fixed in that region. On the other hand, the success of one-loop CF calculations rests on the fact that the pure gauge coupling of the model remains perturbative\footnote{More precisely, this is the Taylor coupling that can easily be mapped to the coupling in the IR-safe scheme used in Refs.~\cite{Tissier:2010ts,Tissier:2011ey}, see also Ref.~\cite{Reinosa:2017qtf}.} in the whole energy range and even decreases to zero at low energies, in agreement with what is observed in lattice simulations \cite{Bogolubsky:2009dc,Duarte:2016iko}.

While the one loop studies of the gluon and ghost propagators using the CF 
model \cite{Pelaez:2014mxa,Pelaez:2015tba} were very encouraging and gave good 
coverage of lattice data to all energies, the natural question that arose 
concerned whether this could be improved with the inclusion of higher loop corrections. This question was examined at two loop order in Ref.~\cite{glmq8} for the case of YM two-point correlation functions where a much closer 
agreement with lattice data over all momenta emerged. Similar observations were made in studies at finite temperature \cite{Reinosa:2014zta,Reinosa:2015gxn}. While this does not imply
that a gluon mass term should be included in Landau gauge-fixed YM theory, it did at least 
demonstrate that perturbative computations could be used to quantitatively 
probe the deeper infrared regions of pure YM theory that at first might not 
seem possible. More recently, a similar investigation was pursued for the case of the ghost-antighost-gluon vertex in one particular momentum configuration \cite{Barrios:2020ubx}, with the added difficulty that all relevant parameters had been fixed in Ref.~\cite{glmq8}, thus representing a stringent test of the method.

Having demonstrated that a gluon mass term gives a window into the 
infrared, the next natural extension of this core idea is to include massive 
quarks on top of the YM gluon mass term of the CF model and thereby 
endeavour to access QCD in the infrared. This is certainly a challenge in particular within the CF model 
as one needs to consider the quark wave (or dressing)  and the mass functions as extra form factors on top of the gluon and 
ghost dressing functions. Moreover, all these form factors depend a priori on two mass scales. 

More importantly, including a quark mass one aims at probing chiral symmetry breaking ($\chi$SB), another central aspect of the infrared that is not fully understood. As is well known, $\chi$SB lies out of the reach of any perturbative approach. Thus, even though the perturbative CF model still remains competitive for studies where the quark masses are artificially large \cite{Reinosa:2015oua,Maelger:2017amh,Maelger:2018vow}, it is doomed to fail in (and close to) the chiral limit, in particular regarding the dynamical generation of a non-zero quark mass function.

We stress that this does not necessarily point to a limitation of the model itself, but rather to a limitation of the considered method. As a matter of fact, the CF model has been investigated beyond perturbation theory using a double expansion scheme, dubbed the Rainbow-Improved (RI) expansion scheme, that exploits the perturbative nature of the pure gauge sector of the CF model together with an expansion in the inverse number of colors \cite{Pelaez:2017bhh,Pelaez:2020ups}. At leading order, this approach essentially boils down to the Rainbow-Ladder approximation (see e.g. \cite{Alkofer,Roberts:2007jh}) with a definite choice for the gluon propagator and the quark-gluon vertex and a consistent inclusion of the running of the parameters.\footnote{One distinctive feature of this approach with respect to the ever growing accurate description of QCD correlation functions based on other nonperturbative approaches such as Dyson-Schwinger or functional Renormalization Group equations, (see for instance Refs.~\cite{Fischer:2003rp,Fischer:2004wf,Aguilar:2014lha, Williams:2015cvx,Cyrol:2017ewj,Aguilar:2018epe,Gao:2021wun}) is that it is based on a perturbative expansion of the pure gauge vertices. This dictates, at each order, the form of the gluon propagator and quark-vertex to be considered in the quark equation. In particular, at the level of approximation considered in \cite{Pelaez:2017bhh,Pelaez:2020ups} they both need to be taken at tree level.} It has been shown to capture $\chi$SB while providing an accurate account of the quark mass function, even close to the chiral limit.

It should not be deduced from the previous considerations, however, that the perturbative CF approach is to be completely abandoned in the case of QCD. It is true that the quark mass function close to the chiral limit cannot be satisfactorily reproduced within this approach because it is directly sensitive to $\chi$SB breaking. However, many other form factors, including the gluon, ghost and quark dressing functions, are certainly less sensitive to these symmetry considerations, and, therefore, potentially within the reach of perturbative CF calculations.

 With this line of thought, a one-loop investigation of the CF model in the presence of massive quarks was carried out in Ref.~\cite{Pelaez:2014mxa} where the various form factors were evaluated using the IR-safe renormalization scheme for an arbitrary number of colors ($N$), degenerate flavors ($N_f$) and dimensions ($d$), and compared with lattice data for $\smash{d=4}$, $\smash{N=3}$, and  $\smash{N_f=2}$, $2+1$ or $2+1+1$ flavors, \cite{Sternbeck:2012qs,Bowman:2004jm,Bowman:2005vx,Ayala:2012pb}. It was shown in particular that the gluon and ghost propagators are correctly accounted for by this perturbative approach. Unexpectedly, however, the quark dressing function was not properly reproduced and even featured the wrong monotonicity as a function of the momentum. At first sight, this seems to go against the above expectations and to signal again a limitation of the perturbative approach within the CF model.
 
However, as was pointed out in Ref.~\cite{Pelaez:2014mxa}, there is a way to reconcile and potentially cure these results within the perturbative CF paradigm. The key observation is that the one-loop correction to the quark dressing function is finite and even vanishes identically in the limit of a massless gluon in the Landau gauge. In effect, this means that this leading order perturbative correction is abnormally small and cannot be commensurate 
with the other form factors. Therefore, to extract results that are meaningful at the same level of precision as \cite{glmq8} for instance, a full two-loop study is absolutely necessary. In fact, an estimate of the two-loop corrections to this quantity indicates that they could greatly contribute to resolve the tension with the lattice data for this function \cite{Pelaez:2014mxa}. The main goal of the present paper is to show that this is indeed what happens and therefore that, just as the gluon and ghost correlators, the quark wave function admits an accurate description within the perturbative CF paradigm, not only far from the chiral limit but also close to the physical case. 

This analysis is subtle because the quark mass function coincides with the running of the quark mass parameter in the renormalization scheme that we consider. Therefore, it is inevitably coupled to the dressing functions. Since the perturbative CF approach fails in reproducing the quark mass function close to the physical case (as we also illustrate for completeness) and even though our first estimation of the dressing functions will feature a two-loop running quark mass, we will have to investigate how the quality of these perturbative estimates is impacted by the use, instead, of a non-perturbative running, as obtained from lattice simulations or as dynamically generated within the CF model in Ref.~\cite{Pelaez:2020ups}. This impact will turn out to be marginal, confirming the perturbative nature of the dressing functions.

The paper is organized as follows. We provide the necessary background details
for the Curci-Ferrari model in Section II. This includes the definition 
of the form factors that are computed to two loops as well as a general review 
of the finer points of the renormalization scheme that allows us to probe the 
infrared. A summary of the one loop work of Ref.~\cite{Pelaez:2014mxa} is also 
provided together with the definition of the Infrared Safe renormalization scheme to be used throughout this work. Section III describes the technical aspects of calculating 
the necessary two loop Feynman graphs contributing to each of the two-point 
functions when there are two independent mass scales.\footnote{We stress that going to two-loop order in the present set-up is not a straightforward task. In Ref.~\cite{glmq8} the focus was on pure YM where there was only one mass scale. Here we will have two distinct masses when the dynamical quarks are included. Therefore we have to evaluate all possible  two loop massive Feynman integrals contributing to the gluon, ghost and quark two-point functions in the Landau gauge. Indeed aside from the one loop correction to the quark two-point function, it is not until two loops that graphs with both mass scales are present in individual diagrams. It is only at this point that we truly have a tool to fully explore the interrelationship  between the mass parameters behind color confinement and chiral symmetry breaking.} A substantial part of the discussion 
is devoted to internal checks in various limits that ensure the results are 
reliable prior to constructing plots. The implementation of the Infrared Safe renormalization
scheme at two-loop order is discussed in Section IV which completes the analytic aspect of
the computation. Our results are presented in Section \ref{sec:results}. In particular, Section \ref{sec:VA} focuses on the main goal of this work, namely the two-loop evaluation of the gluon, quark and dressing functions and their comparison to several lattice data sets corresponding to various pion masses. We find that the two-loop perturbative expressions for these functions in the CF model provide a good account of the data both far from the chiral limit and close to the physical case. Section \ref{sec:VB} illustrates the failure of the perturbative CF approach with regard to the quark mass function as one approaches the physical case, while Section \ref{sec:VA3} investigates the impact of a non-perturbative running for the quark mass parameter on the quality of the perturbative determination of the dressing functions.
After concluding remarks in Section VI there are four Appendices. The first 
illustrates all the graphs we have computed while the next discusses finer 
aspects of the two loop renormalization group flow. These ideas are illustrated in a third appendix using the simple case of the minimal subtraction scheme which we used as benchmark before implementing the Infrared Safe renormalization scheme and which could also serve as a pedagogical introduction to two-loop running. The final appendix gathers next-to-leading order UV and IR asymptotic expansions of the various anomalous dimensions used in the present work.

\section{The Curci-Ferrari model}
We turn to more specific aspects of our study and discuss the necessary
background to the Curci-Ferrari model. In Ref.~\cite{Curci:1976bt} the model was considered for an arbitrary covariant gauge parameter which featured a mass for the
Faddeev-Popov ghosts as well as one for the gluons. However as the former
depends linearly on the gauge parameter, the ghost mass vanishes in the Landau gauge limit on which we focus in this work. This is not unconnected with the massless longitudinal mode of the
gluon.

\subsection{Generalities}
In the Landau gauge limit, the Euclidean CF Lagrangian density in the presence of $N_f$ mass-degenerate quark flavors (in the fundamental representation of the color group) reads
\begin{align}
\label{eq_action}
      {\cal L} = & \,\frac{1}{4} F_{\mu\nu}^aF_{\mu\nu}^a+ih^a\partial_\mu A_\mu^a +\partial_\mu\bar c^a(D_\mu c)^a\nonumber\\
& +\frac 12 m^2 (A_\mu^a)^2 + \sum_{i=1}^{N_f}\bar\psi_i({\cal D}\!\!\!\!\slash + M)\psi_i\,,
\end{align}
where $\smash{F_{\mu\nu}^a\equiv \partial_\mu A_\nu^a -\partial_\nu A_\mu^a+ g f^{abc}A_\mu^b A_\nu ^c}$ is the field-strength tensor, $h^a$ a Nakanashi-Lautrup field, $(c^a,\bar c^a)$ a pair of ghost and antighost fields, and $(\psi_i,\bar\psi_i)$ a pair of quark and antiquark fields for each flavor $i$. The covariant derivatives in the adjoint
($\phi$) and fundamental ($\psi$) representations read respectively
\begin{align}
(D_{\mu}\phi)^a&\equiv\partial_{\mu}\phi^a+g f^{abc}A_{\mu}^b \phi^c,\\
{\cal D}_{\mu}\psi&\equiv\partial_{\mu}\psi-ig A_{\mu}^a t^a \psi\,,
\end{align}
with $f^{abc}$ the structure constants of the SU($N$) gauge group and $t^a$
the generators of the corresponding Lie algebra,
normalized such that ${\rm tr}(t^at^b)=\delta^{ab}/2$. The parameters 
$g$, $m$ and $M$ denote respectively the bare coupling
constant, bare gluon mass and bare quark mass. 

In what follows, we choose a Euclidean convention for the Dirac matrices, such that $\smash{\{\gamma_\mu,\gamma_\nu\}=2\delta_{\mu\nu}\mathds{1}}$, with $\mathds{1}$ the identity matrix in spinor space. The Feynman slash notation $\smash{{\cal D}\!\!\!\!\slash\,\equiv\gamma_\mu {\cal D}_\mu}$ is defined in terms of those Euclidean matrices. The formulas to be derived below are valid for an arbitrary number of colors and an arbitrary number of degenerate quark flavors (in the Landau gauge), but we shall restrict the comparison to the lattice data to the case of three colors and two degenerate flavors.\\

The model is regularized by working in $\smash{d=4-2\epsilon}$ dimensions. This allows us to take full advantage of the symmetries of the model, in particular the BRST symmetry mentioned above. These symmetries, together with the fact that the tree-level gluon propagator is transverse and decreases with two powers of the momentum, ensure the renormalizability of the model. Renormalization proceeds along the usual lines. One first rescales the bare fields and bare parameters in terms of their renormalized counterparts. Denoting the bare quantities that appear in the action (\ref{eq_action}) with a subscript $B$, this step writes
\beq
 & & A_B^{a\,\mu}=\sqrt{Z_A}\,A^{a\,\mu}, \,\,\, c_B^{a}= \sqrt{Z_c}\,c^{a}, \,\,\, \bar c_B^{a}= \sqrt{Z_c}\,\bar c^{a}\,,\nonumber\\
 & & \hspace{1.5cm} \psi_B= \sqrt{Z_\psi}\,\psi, \,\,\, \bar\psi_B= \sqrt{Z_\psi}\,\bar\psi\,, 
 \eeq
 and
 \beq\label{eq:rescale}
g_B = Z_g\,g\,, \quad m_B^2= Z_{m^2}\,m^2\,, \quad M_B= Z_M M\,.
\eeq
Then, the divergences present in the $n$-point functions are absorbed into the various renormalization factors $Z_X$ with $X\in\{A,c,\bar c,\psi,\bar\psi,g,m^2,M\}$ and  the finite parts of these factors are fixed via a choice of renormalization scheme. In this work, we consider the  Infrared-safe renormalization scheme whose definition in terms renormalization conditions is reviewed below together with its main properties.

Let us recall here that, in dimensional regularization, the bare coupling acquires the mass dimension $\epsilon$, which it is usually convenient to make explicit by introducing a scale. In this article, we denote this scale as $\Lambda$ in such a way that the bare and renormalized couplings in (\ref{eq:rescale}) are rescaled as $g_B\to\Lambda^\epsilon g_B$ and $g\to \Lambda^\epsilon g$ respectively. The reason for this unusual choice is that this scale has a priori nothing to do with the renormalization scale $\mu$ that is introduced via the renormalization conditions. The scale $\Lambda$ is in fact a scale associated with the regularization procedure, and, as such, the renormalized quantities do not depend on its choice in the continuum limit (corresponding to $\epsilon\to 0$) while they depend in general on the renormalization scale $\mu$. We shall illustrate this below when evaluating the anomalous dimensions and the beta functions in the IR-safe scheme.  We will also see that, in intermediate computational steps, that is prior to taking the continuum limit, it is convenient to keep the two scales $\Lambda$ and $\mu$ independent of each other.\footnote{Of course, it is also possible to make the standard choice $\Lambda=\mu$. This hides, however, some of the simplifying features, while obscuring the true source of $\mu$-dependence of the renormalized quantities. A well known scheme where this happens is the minimal subtraction scheme: in this case, there are no renormalization conditions that introduce a $\mu$-dependence and the only source of $\mu$-dependence seems to originate from the regulating scale $\Lambda$ which is taken equal to $\mu$ in this scheme. We shall revisit the minimal subtraction scheme in App.~\ref{sec:MS}, show how the paradox is solved and how this peculiar scheme fits the general picture.}

\subsection{Two-point functions}
Our focus in this article is on the two-point functions of the model. These are obtained by inverting the second field derivative of the effective action $\Gamma[A,ih,c,\bar c,\psi,\bar\psi]$. In the ghost sector, this second derivative will be written as
\beq
\Gamma^{(2)}_{c^a\bar c^b}(k)\equiv\delta^{ab}\Gamma(k)\,.
\eeq
Similarly, in the gluon and quark sectors, we shall use the notation
\beq
\Gamma^{(2)}_{A^a_\mu A^b_\nu}(k)\equiv \delta^{ab}\Big(P_{\mu\nu}^\perp(k)\Gamma^\perp(k)+P_{\mu\nu}^\parallel(k)\Gamma^\parallel(k)\Big)\label{eq:G2}
\eeq
and
\beq
\Gamma^{(2)}_{\psi\bar\psi}(k)\equiv-ik\!\!\!\slash \,\Gamma^{\gamma}(k)+\mathds{1}\,\Gamma^{\mathds{1}}(k)\,,\label{eq:G3}
\eeq
where
\beq
P^\perp_{\mu\nu}(k)\equiv\delta_{\mu\nu}-\frac{k_\mu k_\nu}{k^2} \quad \mbox{and} \quad P^\parallel_{\mu\nu}(k)\equiv\frac{k_\mu k_\nu}{k^2}
\eeq
are the transverse and longitudinal projectors.\\

The ghost propagator is obtained as $G_{\rm gh}(k)\equiv1/\Gamma(k)$. From the derivative nature of the ghost-antighost-gluon tree-level vertex and the transverse nature of the tree-level gluon propagator, it is easily argued that $\Gamma(k)$ vanishes at least as $k^2$ in the limit $k\to 0$.\footnote{This is because each loop contribution to $\Gamma(k)$ involves a factor $k$ from the vertex attached to the external antighost leg, and another factor $(k+q)_\mu P^\perp_{\mu\nu}(q)=k_\mu P^\perp_{\mu\nu}(q)$ from the vertex attached to the external ghost leg, with $q$ the momentum associated with the internal gluon propagator attached to this vertex.} It is then convenient to define the ghost dressing function 
\beq
F(k)\equiv k^2G_{\rm gh}(k)=k^2/\Gamma(k)\,.
\eeq 

As for the gluon propagator, it is obtained by first inverting the second derivative of the effective action in the $A/ih$-sector  and then restricting the so-obtained inverse to the $A$-sector. The $ih$-sector cannot be disregarded because it couples to the $A$-sector. However, since the $ih$-dependent part of the action is not renormalized \cite{Tissier:2011ey}, one is led to the inversion of the following matrix
\beq
\left(\begin{array}{cc}
P_{\mu\nu}^\perp(k)\Gamma^\perp(k)+P_{\mu\nu}^\parallel(k)\Gamma^\parallel(k) & ik_\mu\\
-ik_\nu & 0
\end{array}\right).
\eeq
The inverse is easily found to be
\beq
\left(\begin{array}{cc}
P_{\mu\nu}^\perp(k)/\Gamma^\perp(k) & -ik_\mu/k^2\\
ik_\nu/k^2 & \Gamma^\parallel(k)/k^2
\end{array}\right),
\eeq
from which it follows that the gluon propagator is transverse, $P^\perp_{\mu\nu}(k)G(k)$, with $\smash{G(k)=1/\Gamma^\perp(k)}$. By analogy with the ghost sector, and despite the fact that $\Gamma^\perp(k)$ does not vanish as $k\to 0$, it is customary to introduce a gluon dressing function 
\beq
D(k)\equiv k^2 G(k)=k^2/\Gamma^\perp(k)\,.
\eeq 
\vglue1mm

Finally, the quark propagator is obtained by inverting $\Gamma^{(2)}_{\psi\bar\psi}(k)$. Multiplying Eq.~(\ref{eq:G3}) by $ik\!\!\!\slash \,\Gamma^{\gamma}(k)+\mathds{1}\,\Gamma^{\mathds{1}}(k)$ and owing to the property $k\!\!\!\slash^2=k^2$, one finds the propagator
\beq
S(k)=\frac{ik\!\!\!\slash \,\Gamma^{\gamma}(k)+\mathds{1}\,\Gamma^{\mathds{1}}(k)}{k^2(\Gamma^\gamma(k))^2+(\Gamma^{\mathds{1}}(k))^2}\,.
\eeq
It is customary to rewrite this as
\beq
S(k)=Z(k)\frac{ik\!\!\!\slash+\mathds{1}M(k)}{k^2+M^2(k)}\,,\label{eq:param}
\eeq
with 
\beq
Z(k)\equiv 1/\Gamma^\gamma(k) \quad \mbox{and} \quad  M(k)\equiv \Gamma^{\mathds{1}}(k)/\Gamma^\gamma(k)\,.
\eeq 
The benefit of this rewriting is that $M(k)$ appears as the ratio of two tensor components of the same two-point function and, as such, is a finite, renormalization group invariant quantity, known as the quark mass function. As for the function $Z(k)$, we shall refer to it as the quark dressing function.\\

Although we shall not be dealing directly with three-point vertices in this work, let us mention here that a similar argument to the one used for the ghost propagator leads to the conclusion that loop corrections to the ghost-antighost-gluon vertex vanish in the limit of vanishing ghost momentum $k\to 0$:
\beq
\Gamma^{(3)}_{c^aA^b_\mu\bar c^c }(0,l,h)=-if^{abc}g_B\Lambda^\epsilon h_\mu\,.\label{eq:nrt1}
\eeq
This is Taylor's non-renormalization theorem in the CF model \cite{taylor71, Doria,Gracey:2002yt,Dudal:2002pq,Wschebor:2007vh,Tissier:2011ey}. Another such theorem holds for the combination $\Gamma^\parallel(k)F^{-1}(k)$ which is related to the bare gluon mass via the Slavnov-Taylor identity \cite{Tissier:2011ey}:
\beq
\Gamma^\parallel(k) F^{-1}(k)=m^2_B\,.\label{eq:nrt2}
\eeq
Upon renormalization, the two identities (\ref{eq:nrt1}) and (\ref{eq:nrt2})  constrain the combinations $Z_g\sqrt{Z_A}Z_c$ and $Z_{m^2}Z_AZ_c$ of renormalization factors to remain finite. These constraints are fully exploited within the Infrared-safe renormalization scheme which we now review.

\subsection{Infrared safe renormalization scheme}\label{sec:IS}
The Infrared safe (or IR-safe in short) renormalization scheme is defined by extending the relations between the divergent parts of the renormalization factors $Z_g$, $Z_{m^2}$, $Z_A$ and $Z_c$ discussed in the previous section so as to include their finite parts. One then requires that
\beq
Z_g\sqrt{Z_A}Z_c=1\,, \quad Z_{m^2}Z_AZ_c=1\,.\label{eq:ZZZ}
\eeq
The benefit of these conditions is that they give access to $Z_g$ and $Z_{m^2}$ solely in terms of $Z_A$ and $Z_c$. The latter are fixed by requiring that the renormalized ghost and gluon two-point functions (which depend both on the external momentum $k$ and on the renormalization scale $\mu$) satisfy the conditions
\beq
\Gamma(k=\mu;\mu)=1\,, \quad \Gamma^\perp(k=\mu;\mu)=\mu^2+m^2(\mu)\,.\label{eq:ren_glue}
\eeq
As for the quark renormalization factors $Z_\psi$ and $Z_M$ they are fixed by imposing the conditions
\beq
\Gamma^\gamma(k=\mu;\mu)=1\,, \quad \Gamma^{\mathds{1}}(k=\mu;\mu)= M(\mu)\,.\label{eq:ren_quark}
\eeq
Here, we are deliberately using the same notation for the renormalized mass and for the quark mass function defined in the previous section. In a generic renormalization scheme, these two functions do not need to coincide. In the present scheme however, they do coincide because the bare components $\Gamma^\gamma(k)$ and $\Gamma^{\mathds{1}}(k)$ renormalize identically, so that one has
\beq
\frac{\Gamma^{\mathds{1}}(k)}{\Gamma^{\gamma}(k)}=\frac{\Gamma^{\mathds{1}}(k;\mu)}{\Gamma^{\gamma}(k;\mu)}=\frac{\Gamma^{\mathds{1}}(k;k)}{\Gamma^{\gamma}(k;k)}\,,
\eeq
with the left-hand side corresponding to the quark-mass function and the right-hand side corresponding to the renormalized mass in the present scheme and at scale $\mu=k$. 

Once all the renormalization factors are known from (\ref{eq:ZZZ})-(\ref{eq:ren_quark}), one can determine the various anomalous dimensions and beta functions. These are necessary in order to obtain a controlled perturbative description of the various propagators, in those cases where large logarithms (associated with large separations of scales) would invalidate the use of a na\"\i ve perturbative expansion. For the moment we skip all details concerning the practical implementation of the renormalization group (RG) as they will be recalled in full detail when considering the RG flow at two-loop order in Sec.~\ref{sec:RG}.

One of the main benefits of the IR-safe renormalization scheme is that it features renormalization group trajectories that are free of any Landau singularity and along which the running coupling remains relatively small, allowing for a perturbative investigation of the CF model over all scales.\footnote{Other useful features of the IR-safe renormalization scheme will be reviewed in Sec.~\ref{sec:RG}.} As already stated in the introduction, in the case of QCD, such a perturbative investigation makes sense a priori for those correlation functions that are not directly sensitive to $\chi$SB. In what follows, we shall thus concentrate primarily on the gluon, ghost and quark dressing functions. We shall also evaluate the quark mass function to two-loop order. This will allow us both to illustrate the limitations of the perturbative CF approach and to estimate the impact on the perturbative dressing functions of the use of either a two-loop or a non-perturbative running quark mass.

In the next section, we provide details on the evaluation of the two-loop corrections to all the two-point functions of the CF model. The implementation of the renormalization group at two-loop order in the IR-safe scheme will be dealt with in Sec.~\ref{sec:RG}. Our results are finally discussed in Sec.~\ref{sec:results}.

\section{Unquenched two-point functions at two-loop order}
We devote this section to the details of how the two loop corrections to the Landau gauge gluon, ghost and quark two-point functions are evaluated in the presence of non-zero gluon and quark masses. Once the two-point functions are determined as functions of the bare parameters, we
carry out the renormalization at two-loop order. Aside from being necessary for our 
ultimate goal, it provides an intermediate check on our original set-up. 
Additional cross-checks will also be discussed. Some of these entail checking
that previous results, such as the case when quarks are massless, correctly
emerge in the limit $M$~$\to$~$0$ for example.

\subsection{Notation}
Since we shall often refer simultaneously to the various two-point functions $\Gamma$, $\Gamma^\perp$, $\Gamma^\parallel$, $\Gamma^\gamma$, $\Gamma^{\mathds{1}}$ introduced in the previous section, it will be convenient to denote them generically as $\Gamma^C$ with $C\in\{\emptyset,\perp,\parallel,\gamma,\mathds{1}\}$ and where the empty set $\emptyset$ is used to refer to the ghost component $\Gamma(k)$. Moreover, we write
\beq\label{eq:twop}
\Gamma^C(k) & = & \Gamma^C_0(k^2,m^2_B,M_B)\nonumber\\
& + & \lambda_B\,\Gamma^C_1(k^2,m^2_B,M_B)\nonumber\\
& + & \lambda_B^2\,\Gamma^C_2(k^2,m^2_B,M_B)\,,
\eeq
where $\Gamma^C_n(k^2,m_B^2,M_B)$ represents the sum of $n$-loop Feynman diagrams contributing to $\Gamma^C(k)$. For convenience, we have factored out $\lambda_B^n$, with 
\beq
\lambda_B\equiv \frac{g^2_BN}{16\pi^2}\,,
\eeq
in front of $\Gamma^C_n(k^2,m_B^2,M_B)$. In practice, this means that, in computing Feynman diagrams contributing to $\Gamma^C_n(k^2,m_B^2,M_B)$, the $d$-dimensional momentum integrals are replaced by
\beq
\int \frac{d^dp}{(2\pi)^d}\to\int_p\equiv 16\pi^2\Lambda^{2\epsilon}\int \frac{d^dp}{(2\pi)^d}\,,
\eeq 
and the color factors are all systematically divided by $N^n$. 

The tree-level contributions $\Gamma^C_0(k^2,m_B^2,M_B)$ are linear in any of their arguments. More precisely, we have 
\beq
\Gamma_0=k^2\,,\,\Gamma^\perp_0=k^2+m^2_B\,,\,\Gamma^\parallel_0=m^2_B\,,\,\Gamma^\gamma_0=1\,,\,\Gamma^{\mathds{1}}_0=M_B\,.\nonumber
\eeq 
The one-loop contributions $\Gamma^C_1(k^2,m_B^2,M_B)$ have been systematically evaluated in Ref.~\cite{Pelaez:2014mxa} and expressed in terms of the two one-loop master integrals
\beq
A_{m_a} & \!\!\equiv\!\! & \int_p G_{m_a}(p)\,,\label{eq:A}\\
 B_{m_am_b}(k^2) & \!\!\equiv\!\! & \int_pG_{m_a}(p)G_{m_b}(p+k)\,.
\eeq
As for the two-loop contributions $\Gamma^C_2(k^2,m_B^2,M_B)$, as we now explain, they can be systematically reduced to the evaluation of the two-loop master integrals
\beq
& & S_{m_am_bm_c}(k^2)\nonumber\\
& & \hspace{0.6cm}\equiv\,\int_pG_{m_a}(p)B_{m_bm_c}((p+k)^2)\,,\\
& & U_{m_am_bm_cm_d}(k^2)\nonumber\\
& & \hspace{0.6cm}\equiv\,\int_pG_{m_b}(p)G_{m_a}(p+k)B_{m_cm_d}(p^2)\,,\\
& & M_{m_am_bm_cm_dm_e}(k^2)\nonumber\\
& & \hspace{0.6cm}\equiv \int_pG_{m_a}(p)G_{m_c}(p+k)\nonumber\\
&  & \hspace{0.8cm}\times\,\int_qG_{m_b}(q)G_{m_d}(q+k)G_{m_e}(q-p)\,,\label{eq:M}
\eeq
which can then be evaluated numerically using  the {\sc Tsil} package \cite{glmq9}.\\

\subsection{Reduction to master integrals}
One starts with the generation of the two-loop Feynman graphs 
contributing to each of the two-point functions. This is achieved using the 
{\sc Fortran} based {\sc Qgraf} package, \cite{glmq1}. There are $23$, $7$ and $7$ graphs at two loops for the gluon, ghost and quark two-point functions respectively compared with $4$, $1$ and $1$ graphs respectively at one loop. These are illustrated in App. A. In generating the graphs we have included snail topologies. Ordinarily, 
such graphs are excluded when there is no gluon mass since they would vanish 
in dimensional regularization in this particular case.

Once the graphs have been generated for each two-point function, the next stage,
after appending colour and Lorentz indices, is to write each Green's function
in terms of scalar integrals. The reason for this resides in the techniques
we use to evaluate the large set of integrals. The path to the scalar integrals
proceeds in several steps. First, for the gluon and quark two-point functions 
we have to project out the transverse and longitudinal components in the former
case, and in the latter case we have to isolate the contributions proportional 
to $k\!\!\!\slash$ and $\mathds{1}$, see Eqs.~(\ref{eq:G2}) and (\ref{eq:G3}). Of course, no projection is necessary 
for the ghost two-point function. 

While this converts tensor Feynman integrals with no free Lorentz or spinor
indices into scalar ones, the resultant integrals still contain scalar products 
of loop and external momenta. To two-loop order, all such scalar products can be 
rewritten in terms of the squared length of the propagator momenta, using 
for instance
\begin{equation}
k\cdot p =\frac{1}{2}\left[ k^2 +p^2 - (k-p)^2 \right]
\end{equation}
for the massless case, where $p$ is a loop momentum. When a non-zero mass $m$ is
present one merely makes the extra replacement
\begin{equation}
(k-p)^2 = [ (k-p)^2 + m^2 ] - m^2
\end{equation}
\vglue1mm
\noindent{for the appropriate propagator. This produces integrals with no scalar products
but rational polynomials of the propagators. It is in this representation that 
each Feynman integral of the large set of integrals appearing in the two-point functions 
has to be written in order to implement the standard integration technique now 
widely used in multi-loop computations. This is the Laporta algorithm, 
\cite{glmq2}, which is based on a systematic use of integration by parts. In particular, we 
used the {\sc Reduze} implementation, \cite{glmq3,glmq4}, written in C$++$ with
{\sc GiNaC}, \cite{glmq5}, as the core algebra foundation component. To 
organize the tedious algebra associated with writing the integrals contributing
to a Green's function, we have employed the symbolic manipulation language 
{\sc Form}, \cite{glmq6,glmq7}.}

\begin{widetext}
\begin{center}
{\begin{figure}[h]
\begin{center}
\includegraphics[width=10.0cm,height=2.6cm]{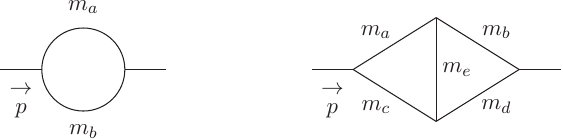}
\end{center}
\caption{Graphical representations of $I_{1ab}(n_1,n_2)$ and
$I_{abcde}(n_1,n_2,n_3,n_4,n_5)$ defined in (\ref{int1def}) and 
(\ref{int2def}).}
\label{glmqfig1}
\end{figure}}
\end{center}
\end{widetext}

The consequence is that the two-loop integrals  can all be written
in terms of two basic integrals which are a one-loop one and a two-loop one. The one loop one is
\begin{equation}
I_{1 ab}(n_1,n_2) ~=~ \int_{p} \frac{1}{[p^2+m_a^2]^{n_1} 
[(p-k)^2+m_b^2]^{n_2}}\,,
\label{int1def}
\end{equation}
where $n_i$ are integers both positive and negative. We use $m_a$ and $m_b$ as
generic masses which can both take values from the set $\{0,m,M\}$ of
the three possible masses that will concern us here. The two-loop core 
integral is
\begin{widetext}
\begin{equation}
I_{abcde}(n_1,n_2,n_3,n_4,n_5) ~=~
\int_{pq} \frac{1}{[p^2+m_a^2]^{n_1} [q^2+m_b^2]^{n_2} [(p-k)^2+m_c^2]^{n_3}
[(q-k)^2+m_d^2]^{n_4} [(p-q)^2+m_e^2]^{n_5}}
\label{int2def}
\end{equation}
\end{widetext}
in the same notation as (\ref{int1def}) which extends that used in Ref.~\cite{glmq8}.
Moreover this syntax is the one we used for defining the integral families of the Laporta algorithm. The two integrals have the graphical representations given in Fig. \ref{glmqfig1}. While (\ref{int2def}) is the most general massive two-loop self-energy structure, we will only be concerned with two non-zero masses. To understand the types of integrals that can actually appear in the evaluation of the two-point functions, we provide two examples for each of the gluon and quark two-point functions in Fig. \ref{glmqfig2}. The respective labels shown underneath each graph indicate one of the set of integrals of (\ref{int2def}) that can arise. However for lines involving gluons some of the propagators that emerge will be massless. So in addition to $I_{ababb}$ the structures $I_{0babb}$, $I_{ab0bb}$ and $I_{0b0bb}$ will be present. When $0$ appears in the label it indicates a massless propagator with the convention that $m_0$~$\equiv$~$0$. So for the other graphs $I_{bbbb0}$ will be present in the other gluon self-energy diagram. For the two quark self-energy graphs $I_{0bbab}$, $I_{abb0b}$ and $I_{0bb0b}$ will also occur in addition to $I_{abbab}$. For that labelled $I_{aabba}$ there are seven other contributions which are $I_{aabb0}$, $I_{a0bba}$, $I_{0abba}$, $I_{00bba}$, $I_{0abb0}$, 
$I_{a0bb0}$ and $I_{00bb0}$. 

\begin{widetext}
\begin{center}
{\begin{figure}[h]
\begin{center}
\includegraphics[width=10.0cm,height=5.5cm]{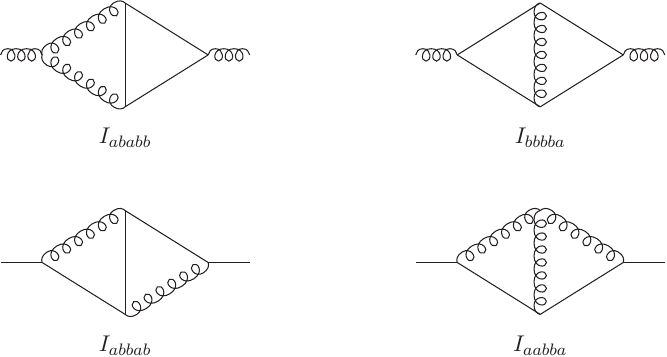}
\end{center}
\caption{Graphs in gluon and quark $2$-point functions containing the labelled
integrals as examples. Gluon propagators are represented by curly lines while 
quarks are denoted by straight ones.}
\label{glmqfig2}
\end{figure}}
\end{center}
\end{widetext}

While the actual non-zero masses in our computations are $m$ and $M$ we use
$m_a$ and $m_b$ for the integral definitions since in the process to write each 
Green's function in terms of scalar integrals, other topologies are present at 
two loops which are illustrated in Fig. \ref{glmqfig3}. For example, each 
graph is contained in $I_{abcde}$ through $I_{abcde}(0,n_2,0,n_4,n_5)$ for the 
sunset diagram and $I_{abcde}(n_1,n_2,0,n_4,n_5)$ for the graph with four 
propagators. The sunset integral $I_{aaaab}(0,n_2,0,n_4,n_5)$ that arises in 
the graph labelled $I_{abbab}$ in Fig. \ref{glmqfig2} is equivalent to 
$I_{ababa}(0,n_2,0,n_4,n_5)$ which occurs in that labelled by $I_{ababb}$ of
the same figure. One can see this by noting that if the propagator is absent 
then the argument of the function corresponding to it is zero. So the
respective mass label can be anything or $0$, $a$ or $b$ in this case. In 
addition, the sunset topology has a sixfold permutation symmetry that ensures 
the equality. The outcome is that the labels $a$ and $b$ on the general two 
loop integral in the two mass case could correspond to either quark or gluon 
mass or vice versa depending on the Green's function and ultimate topology. 
Therefore in applying the Laporta algorithm we have built the system of 
integration by parts equations for a generic set of mass configurations based 
on arbitrary masses $m_a$ and $m_b$. Therefore in (\ref{int2def}) the elements 
of the two loop integral family is given by allowing each of the labels in the 
integral to be one of the set $\{0,m_a,m_b\}$. While this would produce $3^5$ 
core integrals the actual number is fewer due to using rotational symmetry such
as
\begin{eqnarray}
I_{0aaba}(n_1,n_2,n_3,n_4,n_5) &=& I_{a0baa}(n_2,n_1,n_4,n_3,n_5) ~~~~~
\nonumber \\
&=& I_{baa0a}(n_4,n_3,n_2,n_1,n_5) 
\label{rotex}
\end{eqnarray}
and similar relations for others with related label patterns. Equally if the
exponent of a massive propagator is zero then we relabel the corresponding
index on $I_{abcde}(n_1,n_2,n_3,n_4,n_5)$ as $0$ by default.

\begin{center}
{\begin{figure}[ht]
\begin{center}
\includegraphics[width=8.0cm,height=1.5cm]{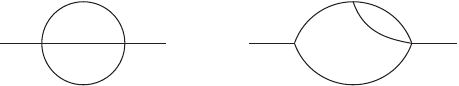}
\end{center}
\caption{Additional two loop topologies that arise in each $2$-point function.}
\label{glmqfig3}
\end{figure}}
\end{center}

Dwelling on this notational aspect of the calculation is important since it is 
in a language that can be coded for the {\sc Reduze} version of the Laporta
algorithm. For instance, we have used the labels in (\ref{int1def}) and
(\ref{int2def}) to define the integral families for the application of the
{\sc Reduze} package. There are not $3^5$ cases in total since we reduce the 
number by using the separate left-right and up-down symmetries of the two loop 
graph of Fig. \ref{glmqfig1}. This substantially lowers the number of cases. 
An example of this was given in (\ref{rotex}). The result of 
applying the Laporta algorithm, \cite{glmq2}, is to reduce the evaluation of 
all the graphs and integrals in the two-point functions to a set of master 
integrals which is significantly smaller than the original input set. However 
the coefficients of each master are functions of the two masses, the external 
momentum and the spacetime dimension $d$. The presence of two non-zero masses 
means the set of master integrals is larger than that for the single scale 
problem of Ref.~\cite{glmq8}. Given this, we follow the same approach in the sense we
choose a basis for the masters that tallies with the integrals of the 
{\sc Tsil} package \cite{glmq9} which we use extensively. It evaluates two loop
self-energy integrals with non-zero masses numerically and allows us to 
determine the behaviour of the Green's function over all momenta. More 
specifically the mapping to the master integrals given above (which are the ones defined in the {\sc Tsil} package) is
\begin{equation}
I_{1a0}(1,0) ~=~ A_{m_a} ~~,~~
I_{1ab}(1,1) ~=~ B_{m_am_b}
\end{equation}
at one loop. At two loops we have 
\begin{eqnarray}
I_{ab00c}(1,1,0,0,1) &=& I_{m_am_bm_c} \nonumber \\
I_{0ab0c}(0,1,1,0,1) &=& S_{m_am_bm_c} \nonumber \\
I_{abc0d}(1,1,1,0,1) &=& U_{m_cm_am_bm_d} \nonumber \\
I_{abcde}(1,1,1,1,1) &=& M_{m_am_bm_cm_dm_e}\,,
\end{eqnarray}
where the first mapping corresponds to the two loop vacuum bubble $I_{m_am_bm_c}=S_{m_am_bm_c}(k^2=0)$.
We also encounter
\begin{eqnarray}
I_{0ab0c}(0,2,1,0,1) &=& T_{m_am_bm_c} \nonumber \\
I_{abc0d}(2,1,1,0,1) &=& V_{m_cm_am_bm_d}\,,
\end{eqnarray}
with $T_{m_am_bm_c}=-\partial S_{m_am_bm_c}/\partial m_a^2$ and $V_{m_cm_am_bm_d}=-\partial U_{m_cm_am_bm_d}/\partial m_a^2$ but we note that these mass derivatives can be expressed in terms of the other master integrals. The 
remaining masters are the product of one loop masters since
\begin{eqnarray}
I_{ab000}(1,1,0,0,0) &=& I_{1a0}(1,0) I_{1b0}(1,0) \nonumber \\
I_{ab000}(1,1,1,0,0) &=& I_{1a0}(1,1) I_{1b0}(1,0) \nonumber \\
I_{ab0c0}(1,1,0,1,0) &=& I_{1a0}(1,0) I_{1bc}(1,1) \nonumber \\
I_{abcd0}(1,1,1,1,0) &=& I_{1ac}(1,1) I_{1bd}(1,1) ~.
\end{eqnarray}
We note that the electronic version of each of our two-point functions can be
found in Ref.~\cite{glmq11}.

\vspace{0.3cm}

\subsection{Renormalization}
Once written in terms of the master integrals, it is fairly easy to isolate the UV divergences in each two-point function (\ref{eq:twop}), the renormalization of which proceeds along the usual lines. First, one rescales the corresponding function by the appropriate factor, $\Gamma^C\to Z_C\Gamma^ C$, with
\beq
Z_\emptyset=Z_c\,, \quad Z_\perp=Z_\parallel=Z_A\,, \quad Z_\gamma=Z_{\mathds{1}}=Z_\psi\,.
\eeq 
Next, one expresses the bare parameters in terms of renormalized ones, $m^2_B=Z_{m^2}m^2$ and $\lambda_B=Z_\lambda\lambda$. Finally, one writes each renormalization factor $Z$ as $Z=1+\delta Z$ with $\delta Z$ a formal series in powers of the renormalized coupling $\lambda$, which one expands to the relevant order. At one-loop order for instance, the renormalized two-point functions read
\beq
\Gamma^C(k) & = & \Gamma^C_0(k^2,m^2,M)+\lambda\,\Gamma^C_1(k^2,m^2,M)\nonumber\\
& + & {\cal R}^{{\rm 1l}}\,\Gamma^C_0(k^2,m^2,M)\,,\label{eq:ren1l}
\eeq
where ${\cal R}$ is the operator
\beq
{\cal R}\equiv\delta Z_C+\delta Z_{m^2}m^2\frac{\partial}{\partial m^2}+\delta Z_{M}M\frac{\partial}{\partial M}\,,\eeq
and ${\cal R}^{{\rm nl}}$ refers to its $n$-loop truncation, obtained by truncating the counterterms accordingly.
It should be mentioned that, because the tree-level contribution $\Gamma^C_0(k^2,m^2,M)=u^C k^2+v^C m^2+w^C M$ is linear with respect to any of its arguments (with $u^C$, $v^C$, $w^C$ equal to $0$ or $1$),  the action of the operator ${\cal R}$ on $\Gamma^C_0(k^2,m^2,M)$ writes, at any order,
\beq
u^C\delta Z_C k^2+v^C(\delta Z_C+\delta Z_{m^2})m^2+w^C(\delta Z_C+\delta Z_{M})M\,.\nonumber\\
\label{eq:lin}
\eeq
This applies in particular to the term in the second line of Eq.~(\ref{eq:ren1l}). Therefore, each counterterm appearing in this term allows one to absorb the one-loop divergences that are present in the first line of (\ref{eq:ren1l}) and that are proportional to $k^2$, $m^2$ and $M$. More precisely, writing the one-loop counterterms $\delta Z_X^{{\rm 1l}}$ with $X\in\{C,m^2,M\}$ as
\beq
\delta Z_X^{\rm 1l}=\lambda\,\frac{z_{X,1}}{\epsilon}\,,
\eeq
with $z_{X,1}=z_{X,11}+\epsilon z_{X,10}$, the elimination of divergences amounts to the proper adjustment of the factors $z_{X,11}$. We mention that these factors  are universal numbers that do not depend on the considered renormalization scheme. We checked that the values we obtained agree with the well known results, see for instance \cite{Tarasov:1976ef,Egorian:1978zx}. In particular, we find that $z_{\psi,11}=0$, in line with the fact that the one-loop corrections to $Z(k)$ vanish in the limit of a massless gluon, and, therefore, that they are UV finite for a non-zero $m$. On the other hand, the factors $z_{X,10}$ (which produce finite contributions to the one-loop counterterms) have to do with the scheme specification. They can depend on the scales $\Lambda$ and $\mu$ as well as on the various masses present in the problem and will enter directly the anomalous dimensions and beta functions which we discuss in Sec.~\ref{sec:RG}.

Similarly, at two-loop order, one finds
\begin{widetext}
\beq
\Gamma^C(k) & = & \Gamma^C_0(k^2,m^2,M)+\lambda\Gamma^C_1(k^2,m^2,M)+\lambda^2\Gamma^C_2(k^2,m^2,M)\nonumber\\
& + & \lambda \left(\delta Z^{{\rm 1l}}_{\lambda}+{\cal R}^{{\rm 1l}}\right)\Gamma^C_1(k^2,m^2,M)+{\cal R}^{{\rm 2l}}\,\Gamma^C_0(k^2,m^2,M)\,.\label{eq:ren2l}
\eeq
\end{widetext}
The role of the first term in the second line of (\ref{eq:ren2l}) is to absorb the subdivergences hidden in the first line. We mention that all the divergent parts of the counterterms appearing in this term have been determined in the previous step, with the exception of the one in $\delta Z_\lambda^{{\rm 1l}}$. However, the latter can be easily determined from the fact that, after this divergent part is fixed, there should only remain divergences that are proportional to $k^2$, $m^2$ and $M$, so that they can be absorbed in the second term of (\ref{eq:ren2l}) which has again the form (\ref{eq:lin}). The two-loop counterterms in this term can be written as
\beq
\delta Z_X^{{\rm 2l}}=\lambda\frac{z_{X,1}}{\epsilon}+\lambda^2\frac{z_{X,2}}{\epsilon^2}\,,\label{eq:zzz1}
\eeq
where $z_{X,1}$ has already been determined at one-loop order and $z_{X,2}=z_{X,22}+\epsilon z_{X,21}+\epsilon^2 z_{X,20}$. Again, the factors $z_{X,22}$ are pure constants that do not depend on the renormalization scheme, and we have checked that the values we obtained match known results \cite{Tarasov:1976ef,Egorian:1978zx}. The factors $z_{X,21}$, even though they have also to do with divergences, are not universal and are impacted by the choice of scheme at one-loop order. Obviously, the factor $z_{X,20}$ is also impacted by the choice of scheme. It will enter the anomalous dimensions and beta functions at two-loop order, as we show in Sec.~\ref{sec:RG}.

Before closing this section, let us make an important remark. Of course, the main purpose of eliminating the divergences is to obtain finite expressions for the two-point functions in the continuum limit $\epsilon\to 0$. In this respect, one should not forget certain terms that survive in this limit from cancellations of the form $\epsilon\times 1/\epsilon$. One important such contributions arises from $\epsilon^2$ corrections to the factor $z_{X,1}$ in (\ref{eq:zzz1}). In principle, when implementing a given renormalization scheme at one-loop order, the factor $z_{X,1}$ receives such a contribution and in fact any power of $\epsilon$. Of course, when it comes to evaluating the one-loop order two-point functions in the continuum limit, these higher powers of $\epsilon$ are irrelevant. However, the $\epsilon^2$ contribution to $z_{X,1}$ is not irrelevant in the first term of the second line of (\ref{eq:ren2l}) because it produces a term of order $\epsilon^0$ that persists in the continuum limit. In this term, one should take instead $z_{X,1}=z_{X,11}+\epsilon z_{X,10}+\epsilon^2z_{X,1(-1)}$ where $z_{X,1(-1)}$ is determined by implementing the renormalization scheme at one-loop order and for a finite value of $\epsilon$. For similar reasons, the factors $z_{X,1(-1)}$ also enter the anomalous dimension and beta functions at two-loop order, as we show in Sec.~\ref{sec:RG}.

\subsection{Cross-checks}
As a result of the reduction of the two-loop two-point functions, one obtains expressions in terms of master integrals multiplied by rational functions of $k^2$, $m^2$ and $M^2$.  Since these expressions are rather lengthy, it is preferable to test them as much as possible before any serious practical application. In this section, we review the various tests that we used in order to cross-check our expressions. 

We mention that all these tests can be performed prior to renormalization. On the other hand, the renormalization of the two-loop expressions represents a test in itself since the cancellation of subdivergences by the counterterms determined at one-loop occurs only if the diagrams are computed and combined correctly in order to generate the correct subdivergences, as we described in the previous section. Another test related to renormalization that we considered was to retrieve the correct renormalization factors in the minimal subtraction scheme. Although this is not the scheme we use eventually for our comparison to lattice data, it is useful in order to understand certain features in a simpler setting and we provide a self-contained discussion in App.~\ref{sec:MS}. Let us just mention here that, in this scheme, one has $z_{X,10}=z_{X,1(-1)}=z_{X,20}=0$ by definition. We checked that the values obtained for $z_{X,11}$, $z_{X,22}$ and $z_{X,21}$ correspond to the well known results of Ref.~\cite{Tarasov:1976ef,Egorian:1978zx}.

We now describe our other tests in detail.

\subsubsection{Quenched limit}
In Ref.~\cite{glmq8}, the ghost and gluon two-point functions were studied in the quenched limit. We have checked that our unquenched expressions for these functions lead to the expressions of that reference in the limit $N_f \to 0$.

\subsubsection{Ultraviolet behaviour}
Based on the superficial degree of divergence of the diagrams contributing to each of the two-point functions, we expect the following large momentum behaviour to hold true from Weinberg's theorem, \cite{Weinberg:1959nj},
\beq
& & \lim_{k\to\infty} \frac{\Gamma(k)}{\lvert k \rvert^3}=0\,, \quad \lim_{k\to\infty}\frac{\Gamma^\perp(k)}{\lvert k \rvert^3}=0\,,\label{eq:FandGuvlim}\\
& & \lim_{k\to\infty} \frac{\Gamma^{\gamma}(k)}{\lvert k \rvert}=0\,, \quad \mbox{and} \quad \lim_{k\to\infty}\frac{\Gamma^{\mathds{1}}(k)}{\lvert k \rvert}=0\,.\label{eq:quarkuvlim}
\eeq
One difficulty in checking this behavior is that they are not obeyed by all the terms that make the reduced expression of each two-point function. Rather, they emerge after certain cancellations occur between these terms. Since it is in general difficult to check these cancellations numerically, we resorted to an analytical check using UV asymptotic expansions of the various master integrals, which were derived through our own implementation of the algorithm described in Ref.~\cite{Davydychev:1993pg}. An earlier version of this algorithm was already used in Ref.~\cite{glmq8}. For the present investigation, we had to extend it to the case where two mass scales are present in the master integrals. At leading order, we obtain the expressions:
\beq\label{eq:largepexp3}
& & \frac{\Gamma(k)}{k^2}=1-\lambda\Bigg[1+\frac{3}{4}\ln\left(\frac{\mu^2}{k^2}\right)\Bigg]\nonumber\\
& & \hspace{0.4cm}-\,\lambda^2\Bigg[\frac{1751}{192}-\frac{15}{16}\zeta(3)-\frac{95}{48}\frac{N_f}{N}\nonumber\\
& & \hspace{2.0cm}+\,\left(\frac{235}{48}-\frac{13}{12}\frac{N_f}{N}\right)\ln\left(\frac{\mu^2}{k^2}\right)\nonumber\\
& & \hspace{3.0cm}+\,\left(\frac{35}{32}-\frac{1}{4}\frac{N_f}{N}\right)\ln\left(\frac{\mu^2}{k^2}\right)^2\Bigg]\nonumber\\
& & +\,\mathcal{O}\left(\frac{m^2}{k^2},\frac{M^2}{k^2}\right),
\eeq
\beq\label{eq:largepexp4}
& & \frac{\Gamma^{\perp}(k)}{k^2}=1-\lambda \Bigg[\frac{97}{36}-\frac{10}{9}\frac{N_f}{N}\nonumber\\
& & \hspace{2.5cm}+\,\left(\frac{13}{6}-\frac{2}{3}\frac{N_f}{N}\right)\ln \left(\frac{\mu^2}{k^2}\right)\Bigg]\nonumber\\
& & \hspace{0.4cm}-\,\lambda^2 \Bigg[\frac{2381}{96}-\frac{59}{8}\frac{N_f}{N}-\frac{55}{6}\frac{C_F}{N}\frac{N_f}{N}\nonumber\\
& & \hspace{1.4cm}-\,\zeta(3)\left(3+4\frac{N_f}{N}-8\frac{C_F}{N} \frac{N_f}{N}\right)\nonumber\\
& & \hspace{1.9cm}+\,\left(\frac{137}{12}-\frac{25}{6}\frac{N_f}{N}-2\frac{C_F}{N} \frac{N_f}{N}\right)\ln\left(\frac{\mu^2}{k^2}\right)\nonumber\\
& & \hspace{4.0cm}+\,\left(\frac{13}{8}-\frac{1}{2}\frac{N_f}{N}\right)\ln\left(\frac{\mu^2}{k^2}\right)^2\Bigg]\nonumber\\
& & +\,\mathcal{O}\left(\frac{m^2}{k^2},\frac{M^2}{k^2}\right),
\eeq
\beq\label{eq:largepexp1}
& & \Gamma^{\gamma}(k)=1+\lambda^2 \frac{C_F}{N}\Bigg[\frac{41}{4}-3 \zeta(3)-\frac{5}{8}\frac{C_F}{N}-\frac{7}{4}\frac{N_f}{N}\nonumber\\
& & \hspace{0.4cm}+\,\left(\frac{25}{4}-\frac{3}{2}\frac{C_F}{N}-\frac{N_f}{N} \right)\ln\left(\frac{\mu^2}{k^2}\right)
\Bigg]\nonumber\\
& & +\,\mathcal{O}\left(\frac{m^2}{k^2},\frac{M^2}{k^2}\right),
\eeq
and
\beq\label{eq:largepexp2}
& & \frac{\Gamma^{\mathds{1}}(k)}{M}=1+\lambda \frac{C_F}{N} \Bigg[4+3\ln\left(\frac{\mu^2}{k^2}\right)
\Bigg]\nonumber\\
& & \hspace{0.4cm}+\,\lambda^2 \frac{C_F}{N}\Bigg[\frac{1531}{24}+13\frac{C_F}{N}-\frac{26}{3}\frac{ N_f}{N}\nonumber\\
& & \hspace{1.9cm} -\,\zeta(3)\left(21-12\frac{C_F}{N}\right)\nonumber\\
& & \hspace{2.3cm}+\,\left(\frac{445}{12}+12\frac{C_F}{N}-\frac{16}{3}\frac{N_f}{N}\bigg)\ln\bigg(\frac{\mu^2}{k^2}\right)\nonumber\\
& & \hspace{3.0cm}+\,\left(\frac{11}{2}+\frac{9}{2}\frac{C_F}{N}-\frac{N_f}{N}\right)\ln\bigg(\frac{\mu^2}{k^2}\bigg)^2\Bigg]\nonumber\\
& & +\,\mathcal{O}\left(\frac{m^2}{k^2},\frac{M^2}{k^2}\right),
\eeq
which indeed verify \eqref{eq:FandGuvlim} and \eqref{eq:quarkuvlim}. In the equations above, $\smash{C_F=(N^2-1)/(2N)}$ denotes the fundamental SU($N$) Casimir. We have also used that the adjoint Casimir is $\smash{C_A=N}$. For the sake of simplicity, we provide the asymptotic behaviors as obtained in the minimal subtraction scheme. We could easily derive them in a generic renormalization scheme, in which case the corresponding expressions depend explicitly on $z_{X,10}$, $z_{X,1(-1)}$ and $z_{X,20}$. Let us also mention that the absence of terms of order $\lambda$ in the leading order contribution of $\Gamma^\gamma(k)$ relates again to the fact that these terms cancel in the limit of a vanishing gluon mass.

We mention that Weinberg's theorem \cite{Weinberg:1959nj} implies also that $\lim_{k\to\infty}\Gamma^\parallel(k)/|k|^3=0$, but in fact, from the Slavnov-Taylor identity (\ref{eq:nrt2}), we have a stronger constraint, namely $\lim_{k\to\infty}\Gamma^\parallel(k)/|k|=0$. By plugging (\ref{eq:twop}) into (\ref{eq:nrt2}) and expanding up to the relevant order, we find
\beq
m^2_B\Gamma_1+k^2\Gamma^\parallel_1=0
\eeq
and
\beq
m^2_B\Gamma_2+k^2\Gamma^\parallel_2+\Gamma_1\Gamma^\parallel_1=0\,.
\eeq
We have checked that these identities hold true, thus confirming the Slavnov-Taylor identity (\ref{eq:nrt2}) at two-loop order and the corresponding UV suppression of $\Gamma^\parallel(k)$ with respect to the na\"\i ve counting.

\subsubsection{Infrared behaviour}\label{sec:IR}
In the opposite momentum range, we expect the two-point functions $\Gamma^C(k)$ to be regular. This is because, these functions are built out of Euclidean Feynman integrals and there are always enough massive propagators to regularize the $k\to 0$ limit. As we have already discussed, in the case of the ghost two-point function, $\Gamma(k)$ is not only regular in this limit, but vanishes at least as $k^2$. 

Again, these expectations might be difficult to check numerically because they typically emerge as the result of cancellations between various terms in the reduced expressions for $\Gamma^C(k)$, which themselves do not behave accordingly. We then resorted to an analytical check that requires the expansion of the various master integrals in powers of $k^2$. 

We first checked that the various master integrals that produce the wrongly behaving terms always involve enough massive propagators in such a way that the routing of $k$ inside the integral can always be chosen to avoid massless propagators. In this situation, we can employ the strategy of Ref.~\cite{Davydychev:1992mt} that leads to a regular expansion in powers of $k^2$ with coefficients given by the momentum independent master integrals $A_{m_a}$ and $\smash{I_{m_am_bm_c}\equiv S_{m_am_bm_c}(k^2=0)}$ and their mass derivatives. The latter mass derivatives can always be conveniently re-expressed in terms of $A_{m_a}$ and $I_{m_am_bm_c}$ using
\beq
\frac{\partial}{\partial m_a^2}A_{m_a} & = & (d/2-1)\frac{A_{m_a}}{m^2_a}
\eeq
that follows from dimensional analysis and
\beq
& & \Delta_{m_am_bm_c}\frac{\partial }{\partial m_c^2}I_{m_am_bm_c}\nonumber\\
& & \hspace{0.5cm}=\,(d-3)(m^2_a+m^2_b-m^2_c)I_{m_am_bm_c}\nonumber\\
& & \hspace{0.5cm}+\,(d-2)\left[A_{m_a}A_{m_b}+\frac{m^2_a-m^2_b-m^2_c}{2m^2_c}A_{m_a}A_{m_c}\right.\nonumber\\
& & \hspace{2.5cm}\left.+\,\frac{m^2_b-m^2_a-m^2_c}{2m^2_c}A_{m_b}A_{m_c}\right]\,,\label{eq:derI}
\eeq
with
\beq
\Delta_{m_am_bm_c} & = & m^4_a+m^4_b+m^4_c\nonumber\\
& - & 2\,(m^2_am^2_b+m^2_bm^2_c+m^2_cm^2_a)\,,
\eeq
that follows from integration by parts techniques \cite{Davydychev:1992mt,Caffo:1998du}. In this way, the coefficients of the Taylor expansion at small $k$ are functions of these two master integrals. Using these expansions, we could check that the various two-point functions behave as expected.

Let us also mention that the regularity of $\Gamma^\parallel(k)$ in the limit $k\to 0$ can alternatively be seen as a consequence of the Slavnov-Taylor identity (\ref{eq:nrt2}) and the fact that $\Gamma(k)$ vanishes at least as $k^2$, or, equivalently, the fact that $\Gamma(k)$ vanishes at least as $k^2$ can be seen as a consequence of the Slavnov-Taylor identity and the regularity of $\Gamma^\parallel(k)$.

\subsubsection{Spurious singularities}
The limit $k\to 0$ is not the only one where individual terms in the reduced expression for $\Gamma^C(k)$ behave in a singular manner. In the case of the ghost and gluon two-point functions, we find that certain terms are singular as $k^2$ approaches $2m^2$ or $2M^2$. Of course, the two-point function in these limits should be regular, thus providing a test for the reduced expressions. We have checked that this is indeed the case since the residue of $1/(k^2-2x^2)$ with $x=m$ or $x=M$ vanishes thanks to the following identity between master integrals ($k_x\equiv\sqrt{2}x$)
\beq\label{eqeq}
& & 2(d-3)x^2\Big[6(d-4)x^4M_{xxxx0}(k^2_x)-(3d-8)S_{xx0}(k^2_x)\Big]\nonumber\\
& & =\!\Big[(d-2)A_x\!-\!2(d-3)x^2B_{xx}(k^2_x)\Big]^2\!\!\!\!-\!8(d-3)^2x^4B^2_{xx}(k^2_x)\,,\nonumber\\
\eeq
that one can derive using the Laporta algorithm. When expanded in $\epsilon$, it is easily shown that this combination is finite and reproduces the combination of finite master integrals in Eq.~(23) of Ref.~\cite{glmq8} that was also found to vanish using the results in Ref.~\cite{glmq9}.

In addition, all two-point functions contain terms that are singular as $m$ approaches $2M$. Since the Euclidean two-point functions have no reason to be singular in this limit, some cancellations need to occur among these terms, providing a further check on the reduced expressions. For instance, in the case of $\Gamma^{\mathds{1}}(k)$, we found a potentially singular term at $m=2M$, the residue of $1/(m-2M)$ being proportional to
\beq
& & (d-2)(2A_M-A_{2M})B_{M(2M)}(k)\nonumber\\
& & \hspace{0.4cm}+\,4M^2\Big(T_{M(2M)(2M)}(k)
-\,2T_{(2M)(2M)M}(k)\nonumber\\
& & \hspace{2.2cm}-\,(d-3)U_{(2M)M(2M)M}(k)\Big)\,.
\eeq
Using the Laporta algorithm, we verified that this combination of master integrals indeed vanishes. In particular we built a different {\sc Reduze} database to the two mass scale one described earlier. Instead a single mass scale database was constructed where we set $m_a$~$=$~$M$ and $m_b$~$=$~$2M$ at the outset. These cancellations played a role in other two-point functions. In the case of the gluon two-point function, we needed several other vanishing combinations of master integrals. These are
\beq
& & (d-2)(2A_M-A_{2M})B_{MM}(k)\nonumber\\
& & \hspace{0.4cm}-\,4M^2\Big(T_{M(2M)M}(k)\nonumber\\
& & \hspace{2.2cm}-\,2T_{(2M)MM}(k)\nonumber\\
& & \hspace{3.0cm}-\,(d-3)U_{MM(2M)M}(k)\Big)\,,
\eeq
\beq
& & (d-2)A_M B_{(2M)(2M)}\nonumber\\
& & \hspace{0.6cm}+\,2M^2\Big(T_{M(2M)M}+(d-3)U_{(2M)(2M)MM}\Big)\,,
\eeq
and finally
\begin{widetext}
\beq
& & (d-3)\Big[(d-4)k^2x^2(k^2+4x^2)^2M_{xxxx0}-2(3d-8)(k^2+4x^2)S_{xx0}\Big]\nonumber\\
& & \hspace{1.0cm}=\,\Big[(d-2)^2A_x^2-2(d-2)(d-3)k^2A_xB_{xx}-2(d-3)^2k^2x^2 B_{xx}^2\Big](k^2+4x^2)\nonumber\\
& & \hspace{1.2cm}-\,4(d-3)\Big[3p^2T_{xx0}+4(d-3)(k^2+x^2)U_{0(2x)xx}-2(d-2)(k^2+x^2)A_xB_{(2x)x}\Big](k^2-2x^2)
\eeq
that we also substantiated using the Laporta algorithm.
This later cancellation boils down to (\ref{eqeq}) when $k^2=2x^2$.
\end{widetext}

\subsubsection{Zero mass limit} 
This is more an internal cross-check since we computed independently the propagators in the case of vanishing gluon and quark mass ($M=m=0$) with the goal of recovering them from the zero mass limit of the massive propagators. In order to compute this limit it is useful to bare in mind that any of the master integrals presented above can be written as
\begin{equation}
\begin{split}
(\mu^{2\epsilon})^{L}{\cal F}(k^2,m^2,M^2)=&(\mu^{2\epsilon})^{L}(k^2)^{D}\\
&\times {\cal F}(1,m^2/k^2,M^2/k^2)\,,
\end{split}
\end{equation}
where $L$ is the number of loops and $D$ the mass dimension of the integral (leaving aside the powers of $\mu$ that multiply it). As a result of this simple dimensional analysis it is clear that the low mass expansion ($m\ll k$ and $M\ll k$ simultaneously) is equivalent to the large momentum expansion. Consequently, the zero quark and gluon mass limits for $\Gamma(k)$, $\Gamma^\perp(k)$, $\Gamma^{\gamma}(k)$ are nothing but the leading terms in the expansions \eqref{eq:largepexp3}-\eqref{eq:largepexp1}. We have checked that these expressions coincide with the results of a direct calculation with massless fields (in the minimal subtraction scheme). Of course, in this limit, $\Gamma^\parallel(k)$ and $\Gamma^{\mathds{1}}(k)$ are just zero.

\section{Renormalization Group}\label{sec:RG}
In principle, in order to compare the renormalized two-point functions computed within a given approach to those obtained within lattice simulations, it is enough to evaluate the renormalized two-point functions at a given renormalization scale, that is $\Gamma^C(k;\mu_0)$. Indeed, the momentum dependence of renormalized two-point functions as computed within different approaches should differ only to within an overall constant which is easily adjusted.

In practice, however, a direct perturbative evaluation of $\Gamma^C(k;\mu_0)$, such as the one described in the previous section, is not accurate in the case of a large separation of scales between $k$ and $\mu_0$. Indeed, in this case, large logarithms $\ln k/\mu_0$ effectively modify the expansion parameter $\lambda$ into $\lambda\ln k/\mu_0$ which has no reason to be small, even when $\lambda$ is small. As is well known, the way to cope with these large logarithms is to use the renormalization group (RG).

The renormalized functions $\Gamma^C(k;\mu)$ for different values of $\mu$ are trivially related to each other as they arise from the same bare function $\Gamma^C(k)$. The differential equation governing the evolution of $\Gamma^C(k;\mu)$ with $\mu$ is the Callan-Szymanzik equation. In its integrated form it can be written as
\beq\label{eq:RG2}
& & \Gamma^C(k;m^2_0,M_0,\lambda_0,\mu_0)\nonumber\\
& & \hspace{0.4cm}=\,z_C^{-1}(\mu,\mu_0)\,\Gamma^C(k;m^2(\mu),M(\mu),\lambda(\mu),\mu)
\eeq
and relates a given $n$-point function at the fixed scale $\mu_0$ to the same $n$-point function at the running scale $\mu$. The benefit of Eq.~(\ref{eq:RG2}) is that it allows one to evaluate $\Gamma^C(k;\mu_0)$ while maintaining perturbative control at any scale. This is achieved by evaluating the right-hand side of Eq.~(\ref{eq:RG2}) with the choice $\smash{\mu=k}$ that prevents the appearance of large logarithms of the form $\ln k/\mu$. This requires in turn the evaluation of the rescaling factor $z_C(\mu,\mu_0)$ as well as the running $m^2(\mu)$, $M(\mu)$ and $\lambda(\mu)$ of the various parameters. 

The rescaling factor is given by
\beq
z_C(\mu,\mu_0)=\,\exp\left(\int_{\mu_0}^\mu d\nu\,\gamma_C(\nu)\right),\label{eq:rescaling}
\eeq
where $\gamma_C$ is the anomalous dimension related to the corresponding renormalization factor $Z_C$ as
\beq
\gamma_C\equiv \frac{d\ln Z_C}{d\ln\mu}\,.\label{eq:gamma22}
\eeq
The running of the parameters is given by the beta functions
\beq
\beta_{m^2}\equiv \frac{d m^2}{d\ln\mu}\,, \quad \beta_M\equiv \frac{d M}{d\ln\mu}\,, \quad \beta_{g^2}\equiv \frac{dg^2}{d\ln\mu}\,.\label{eq:beta22}
\eeq
It should be noted that the derivatives $d/d\ln\mu$ in Eqs.~(\ref{eq:gamma22})-(\ref{eq:beta22}) are to be taken for fixed bare masses and dimensionful bare coupling $\Lambda^{2\epsilon}Z_{g^2}g^2$. These constraints imply\footnote{These equations are easily obtained by requiring that the logarithms of $Z_{m^2}m^2$, $Z_M M$ and $\Lambda^{2\epsilon}Z_{g^2}g^2$ do not depend on $\ln\mu$. Here we are taking a $\mu$-independent $\Lambda$. In App.~\ref{app:2l_RG}, we discuss the case of a $\mu$-dependent $\Lambda$, including the conventional choice of $\Lambda=\mu$.}
\beq\label{eq:beta_gamma}
0=\gamma_{m^2}+\frac{\beta_{m^2}}{m^2}=\gamma_{M}+\frac{\beta_{M}}{M}=\gamma_{g^2}+\frac{\beta_{g^2}}{g^2}\,,
\eeq
where
\beq
\gamma_{m^2}\equiv \frac{d\ln Z_{m^2}}{d\ln\mu}\,,\,\,  \gamma_{M}\equiv \frac{d\ln Z_{M}}{d\ln\mu} \,\, \mbox{and} \,\, \gamma_{g^2}\equiv\frac{d\ln Z_{g^2}}{d\ln\mu}\nonumber\\
\eeq
are the anomalous dimension associated with the parameters.

Thus, in a sense, the renormalization group allows us to evaluate $\Gamma^C(k;\mu_0)$ by using perturbation theory indirectly: rather than using perturbation theory to evaluate $\Gamma^C(k;\mu_0)$, one uses perturbation theory to determine all the anomalous dimensions
\beq
\gamma_X\equiv \frac{d\ln Z_X}{d\ln\mu}\,,\label{eq:def}
\eeq
(with $X\in\{C,m^2,M,g^2\}$) from which one can reconstruct the running of the parameters and the rescaling factors that enter (\ref{eq:RG2}). 
In the next section, we explain how the two-loop anomalous dimensions are evaluated.

Finally, we stress that the choice $\smash{\mu=k}$ prevents the appearance of large logarithms of the form $\ln k/\mu$ which are the only ones present in the ultraviolet. In a theory with massless degrees of freedom such as the CF model, one may find other logarithms of the form $\ln m/\mu$ in the infrared. As we discuss in \ref{app:IR_asymptotic}, we show that despite the presence of these logarithms in the anomalous dimensions, the two-loop contributions remain under perturbative control in the infrared.\\

\subsection{Two-loop anomalous dimensions and beta functions in the IR-safe scheme}
In a generic renormalization scheme, the renormalization conditions allow us to access the various renormalization factors $Z_X$ with $X\in\left\{A,c,\psi,m^2,M,\lambda\right\}$ from which one can evaluate the corresponding anomalous dimensions $\gamma_X$. More precisely, from 
\beq
Z_X=1+\lambda \frac{z_{X,1}}{\epsilon}+\lambda^2\frac{z_{X,2}}{\epsilon}\,,
\eeq
with $z_{X,1}=z_{X,11}+z_{X,10}\epsilon+z_{X,1(-1)}$ and $z_{X,2}=z_{X,22}+z_{X,21}\epsilon+z_{X,20}\epsilon^2$ where the $z_{X,ab}$ are functions of $\Lambda$, $\mu$, $m^2$ and $M$, it is possible to derive the following generic expression for the anomalous dimension:
\begin{widetext}
\beq\label{eq:nice2}
\gamma_X=g^2\frac{\partial z_{X,10}}{\partial\ln\mu}+g^4\!\left(\frac{\partial z_{X,20}}{\partial\ln\mu}\!-\!\left(\frac{\partial z_{X,10}}{\partial\ln\mu}+\frac{\partial z_{g^2,10}}{\partial\ln\mu}\right)\!z_{X,10}\!-\!\left(\frac{\partial z_{X,1(-1)}}{\partial\ln\mu}+\frac{\partial z_{g^2,1(-1)}}{\partial\ln\mu}\right)\!z_{X,11}-\!\sum_i\frac{\partial z_{m^2_i,10}}{\partial\ln \mu}\frac{\partial z_{X,10}}{\partial\ln m^2_i}\right),\nonumber\\
\eeq
\end{widetext}
where $\sum_i$ sums over all possible masses in the problem, here $m_i=m$ and $m_i=M$, see App.~\ref{app:2l_RG} for details. Moreover, the finiteness of the anomalous dimensions requires the following constraints to hold true
\beq
0 & = & \frac{\partial z_{X,11}}{\partial\ln\mu}=\frac{\partial z_{X,22}}{\partial\ln\mu}\nonumber\\
& = & \frac{\partial}{\partial\ln\mu}\Big(z_{X,21}-(z_{X,10}+z_{g^2,10})z_{X,11}\Big)\,.
\eeq
The first two are trivial since, as we have already seen $z_{X,11}$ and $z_{X,22}$ are pure constants. The last constraint is less trivial and we have checked that it holds true in the particular renormalization scheme considered here. It should of course hold true in any other renormalization scheme. In particular, we show in App.~\ref{sec:MS} that this constraint is nothing but a generalization of the constraint $z_{X,22}=z_{X,11}(z_{X,11}+z_{g^2,11})/2$ that arises as a consequence of the finiteness of the anomalous dimensions within the minimal subtraction renormalization scheme. 

We mention also that, in the case where $\smash{z_{X,11}\neq 0}$, the above constraints can be used to simplify the formula (\ref{eq:nice2}) by rewriting the second term within the round bracket as $-\frac{z_{X,10}}{z_{X,11}}\frac{\partial z_{X,21}}{\partial\ln\mu}$. When $\smash{z_{X,11}=0}$, this replacement cannot be made but the formula simplifies as well because the third term within the bracket vanishes. In the present model, this occurs for the quark anomalous dimension since $z_{\psi,11}=0$.\\

As we have already mentioned above, in this work we consider the IR-safe renormalization scheme defined by the conditions (\ref{eq:ZZZ})-(\ref{eq:ren_quark}). In addition to the benefits of this choice which were already reviewed in Sec.~\ref{sec:IS}, we note that the use of (\ref{eq:ZZZ}) allows us to bypass the calculation of the anomalous dimensions for $m^2$ and $\lambda$ since they are directly given in terms of the anomalous dimensions for $A$ and $c$ via
\beq
\gamma_\lambda=-(\gamma_A+2\gamma_c)\,, \quad \gamma_{m^2}=-(\gamma_A+\gamma_c)\,,
\eeq
leading to the beta functions
\beq
\frac{\beta_\lambda}{\lambda}=\gamma_A+2\gamma_c\,, \quad \frac{\beta_{m^2}}{m^2}=\gamma_A+\gamma_c\,.
\eeq
Moreover, one can formally solve this system for $\gamma_A$ and $\gamma_c$ in terms of linear combinations of $\beta_{m^2}/m^2$ and $\beta_\lambda/\lambda$ giving
\beq
\gamma_A=\frac{\beta_\lambda}{\lambda}-2\frac{\beta_{m^2}}{m^2}\,, \quad \gamma_c=\frac{\beta_{m^2}}{m^2}-\frac{\beta_\lambda}{\lambda}\,.
\eeq
Then, one can explicitly integrate the rescaling factors $z_A(\mu,\mu_0)$ and $z_c(\mu,\mu_0)$ in terms of the running parameters $m^2(\mu)$ and $\lambda(\mu)$
\beq
z_A(\mu,\mu_0)=\frac{m^4_0}{\lambda_0}\frac{\lambda(\mu)}{m^4(\mu)}\,, \quad z_c(\mu,\mu_0)=\frac{\lambda_0}{m^2_0}\frac{m^2(\mu)}{\lambda(\mu)}\,. ~~~
\eeq
This, combined with the renormalization conditions (\ref{eq:ren_glue}), provides explicit expressions for the gluon and ghost dressing functions in terms of the running parameters \cite{Tissier:2011ey}
\beq
D(p;\mu_0) & = & \frac{\lambda_0}{m_0^4}\frac{m^4(p)}{\lambda(p)}\frac{p^2}{p^2+m^2(p)}\,,\\
F(p;\mu_0) & = & \frac{m^2_0}{\lambda_0}\frac{\lambda(p)}{m^2(p)}\,.
\eeq
For the quark propagator, we need to determine the quark mass anomalous dimension in order to extract the corresponding beta function, as well as the quark anomalous dimension in order to obtain the corresponding rescaling factor. However, with the parametrization (\ref{eq:param}), the rescaling factor applies only to $Z(k)$, and because of the renormalization condition, we have
\beq
Z(k;\mu_0)=\exp\left(-\int_{\mu_0}^k d\nu\,\gamma_\psi(\nu)\right).
\eeq
As already mentioned, the quark mass function $M(k)$ identifies with the running mass in the chosen scheme.

\subsection{Asymptotic behaviors}\label{sec:asymptotic}
In Apps.~\ref{app:UV_asymptotic} and \ref{app:IR_asymptotic}, the interested reader can find the UV and IR asymptotic expansion of the various two-loop anomalous dimensions at next-to-leading order, which we used in order to control the RG flow in these regimes.

With the RG flow at our disposal, we can now evaluate the various two-point functions and compare to available lattice data.

\section{Results}\label{sec:results}
In this section, we investigate to which extent the lattice data for the QCD two-point correlation functions can be described within the perturbative CF model at two-loop order. We consider SU($3$) data sets for two mass-degenerate quark flavors and for two values of the pion mass (used as a label to the various data sets), one relatively far from the chiral limit ($\smash{M_\pi=422}$~MeV) and one close to the physical value ($\smash{M_\pi=150}$~MeV). 

Our main focus are the ghost, gluon and quark dressing functions which we analyze in Sec.~\ref{sec:VA}. As already explained, these functions are not directly impacted by the spontaneous breaking of chiral symmetry and it is reasonable to expect that they can be captured by perturbation theory. Our results in Secs.~\ref{sec:VA1} and \ref{sec:VA2} support these expectations. 

The quark mass function is discussed in Sec.~\ref{sec:VB} for completeness and also as an illustration of the limitations of the perturbative CF approach. We stress that these limitations do not necessarily imply a failure of the CF model. In fact, the spontaneous breaking of chiral symmetry can be captured within the CF model using resummations that allow to dynamically generate a quark mass function in pretty good agreement with lattice data, with corrections controlled by two small parameters \cite{Pelaez:2020ups}. However, this means that it is mandatory to assess how the quality of the perturbative description of the dressing functions discussed in Sec.~\ref{sec:VA} depends on the use of either the two-loop perturbative quark mass function or the fully non-perturbative quark mass function such as the one generated on the lattice or in Ref.~\cite{Pelaez:2020ups}. Indeed, in the considered renormalization scheme, the quark mass function coincides with the quark mass parameter and is thus inevitably coupled to the dressing functions. This analysis is provided in Sec.~\ref{sec:VA3}.

\subsection{Dressing functions}\label{sec:VA}
In this section, we fit the one- and two-loop expressions for the dressing functions to the lattice data. The perturbative expressions depend on three parameters defined at the initial scale $\mu_0$ of the RG flow: the renormalized coupling $\smash{\lambda_0\equiv\lambda(\mu_0)}$, the renormalized gluon mass $\smash{m_0\equiv m(\mu_0)}$ and the renormalized quark mass $\smash{M_0\equiv M(\mu_0)}$. In addition to these three parameters, we have adjustable normalization factors ${\cal N}_X$, with $\smash{X\in\{D,F,Z\}}$. In order to find the best fit to the lattice data, the parameters and the normalizations need to be chosen so as to minimize a joint error function $\chi_{DFZ}$ combining the individual errors $\chi_X$, with $\smash{X\in\left\{D,F,Z\right\}}$:
\begin{equation} \label{joint_error}
\chi^2_{DFZ}\equiv\frac{1}{3}\Big[\chi_D^2+\chi_F^2+\chi_Z^2\Big]\,.
\end{equation}
The individual error for $\smash{X\in\{D,F,Z\}}$ is taken to be
\beq\label{eq:chi_X}
\chi_{X}^2=\frac{1}{N}\sum_i \left({\cal N}_X\frac{ X_{\text{th.}}(k_i)}{X_{\text{lt.}}(k_i)}-1\right)^2\!,
\eeq
which simply averages, over the available data points, the relative error of the appropriately rescaled theoretical values $X_{\text{th.}}(k_i)$ to the data $X_{\text{lt.}}(k_i)$. Because the error function (\ref{joint_error}) depends quadratically on the normalizations ${\cal N}_X$, the latter can be determined explicitly in terms of the lattice and theoretical data. One finds
\beq
{\cal N}_X=\frac{\sum_i X_{\text{th.}}(k_i)/X_{\text{lt.}}(k_i)}{\sum_i X^2_{\text{th.}}(k_i)/X^2_{\text{lt.}}(k_i)}\,.
\eeq
The fitting problem reduces then to the minimization of $\chi^2$ with respect to the three remaining parameters, $\lambda_0$, $m_0$ and $M_0$. Unless otherwise stated, the parameters will be defined at the scale $\mu_0=1$\,GeV.

To test the quality of the perturbative approach, we proceed in two ways. We first consider a global fit of the three dressing functions and quantify both the total and individual errors as one changes from one-loop to two-loop accuracy. We also consider partial fits of two of the dressing functions supplemented by a ``prediction'' of the third dressing function.

\subsubsection{Global fit}\label{sec:VA1}
\begin{figure}[t]
\hglue-4mm \includegraphics[width=0.45\textwidth]{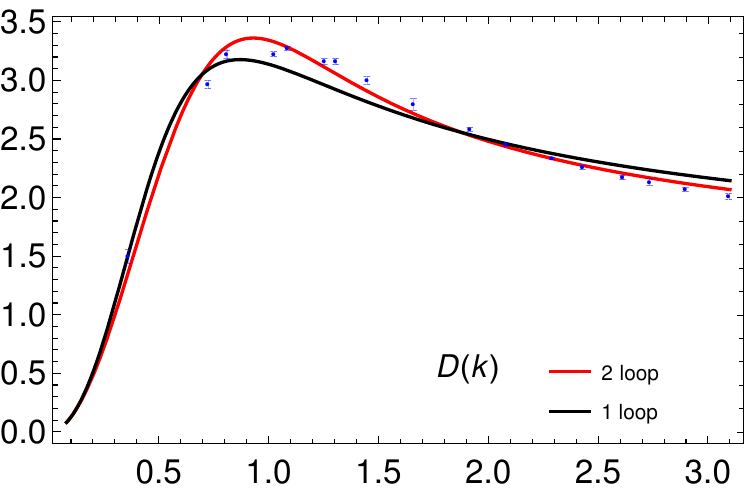}
\hglue-4mm \includegraphics[width=0.45\textwidth]{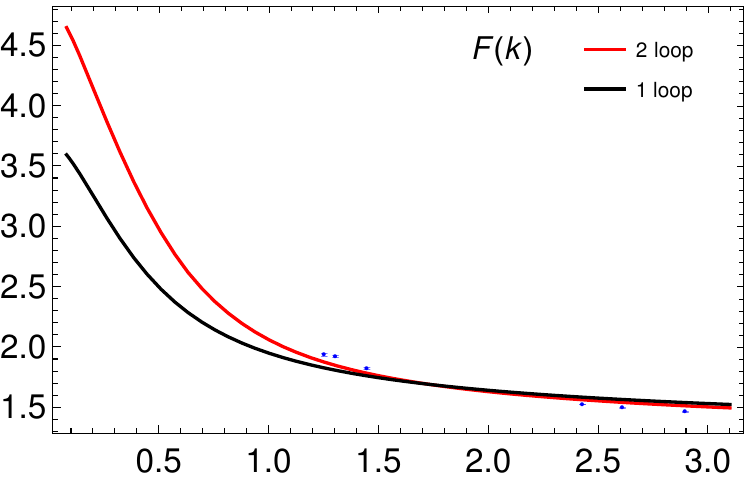}
 \hglue-4mm \includegraphics[width=0.45\textwidth]{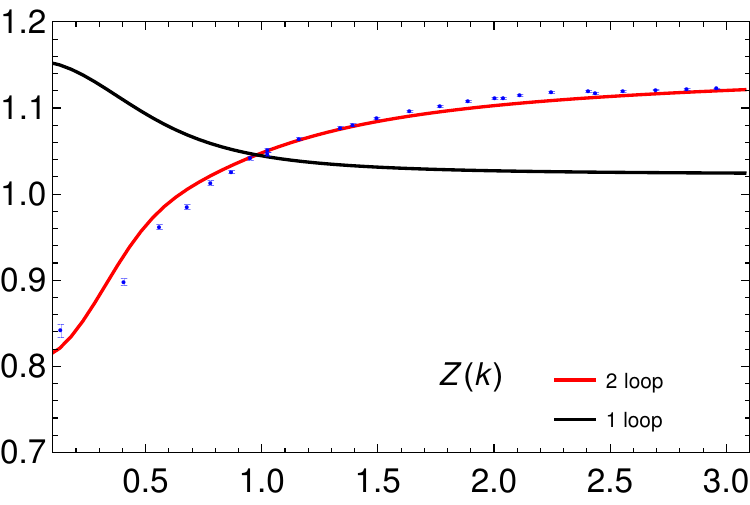}
 \caption{\label{fig:M422all} Comparison of the one- and two-loop CF results for the gluon (top), ghost (middle) and quark (bottom) dressing functions to the lattice data of Ref.~\cite{Sternbeck:2012qs,Oliveira:2018lln} using $\smash{M_\pi=422}$ MeV. The parameters (determined from a global fit using the three functions $D$, $F$, $Z$) are found to be $\lambda_0=0.28$, $m_0=390$~MeV, $M_0=300$~MeV in the one-loop case, and $\lambda_0=0.32$, $m_0=350$~MeV, $M_0=100$~MeV in the two-loop case.}
\end{figure}

We first compare our one- and two-loop results with lattice data far from the chiral limit, simulated using a pion mass $\smash{M_\pi=422}$ MeV, see Refs.~\cite{Sternbeck:2012qs,Oliveira:2018lln}. The global and individual errors at one- and two-loop order are gathered in Tab.~\ref{table1} while Fig.~\ref{fig:M422all} shows the corresponding plots and gives the relevant parameters. From now on, the horizontal axis of all the plots refers to momenta in GeV, whereas the unit used on each vertical axis is in GeV elevated to the mass dimension of the plotted quantity.

\begin{table}[h]
\begin{tabular}{|| c|c|c|c|c ||}
\hline
order & $\chi_{DFZ}(\%)$ & $\chi_D(\%)$ & $\chi_F(\%)$ & $\chi_Z(\%)$ \\
\hline \hline
\;1-loop\; & \;\;7.3\;\; & \;\;4.6\;\; &\;\;4.8\;\;  & \;\;10.8\;\;  \\
\hline
\;2-loop\; & \;\;2.7\;\;  & \;\;3.2\;\; &\;\;3.1\;\; & \;\;1.2\;\;\\
\hline
\end{tabular}
\caption{\label{table1} Global and individual errors as obtained from the global fit of $D$, $F$ and $Z$ in the case $M_\pi=422$~MeV.}
\end{table}

We observe that the global agreement with lattice data greatly improves once two-loop corrections are included. The two-loop contributions appear to be small in the ghost-gluon sector, as expected \cite{Pelaez:2017bhh}. This suggests that perturbation theory is well controlled in the gauge sector of the CF model. On the other hand, the improvement on the quark dressing function is quite remarkable, given the inconsistent results obtained at one-loop for this quantity \cite{Pelaez:2014mxa}. As already mentioned earlier, this is an indication that the quark dressing function is well described by perturbation theory within the CF model, the mismatch of the one-loop results just meaning that one needs to go at least up to two-loop order to start having a good account of the function. In fact the error $\chi_Z$ is comparable to $\chi_F$ and $\chi_D$. This confirms earlier expectations based on estimates of the two-loop corrections \cite{Pelaez:2014mxa}.\footnote{In a certain sense, the leading order perturbative contribution to the quark dressing function is the two-loop contribution. Based on this remark, it would be even more consistent to fit the lattice propagators using the two-loop expressions for $F$, $D$ and the three-loop expression for $Z$. A complete three-loop evaluation of $Z$ is a difficult task but one could imagine doing a rough estimate similar to the estimate made in \cite{Pelaez:2014mxa} for the two-loop corrections.}

\begin{figure}[t!]
\hglue-4mm \includegraphics[width=0.45\textwidth]{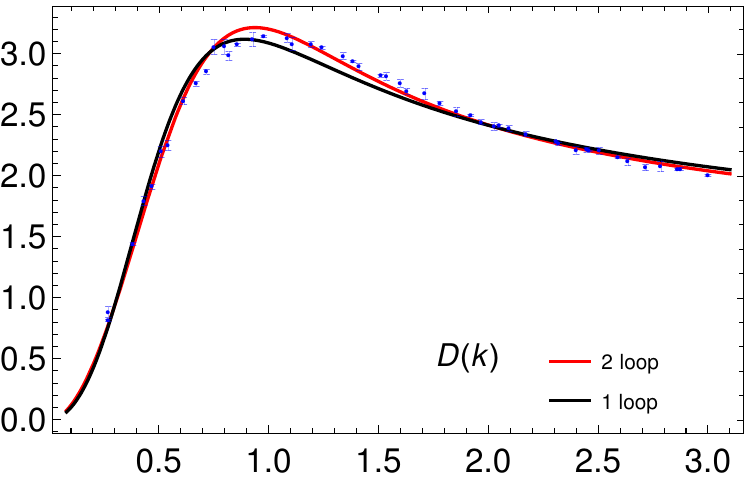}
\hglue-2mm \includegraphics[width=0.44\textwidth]{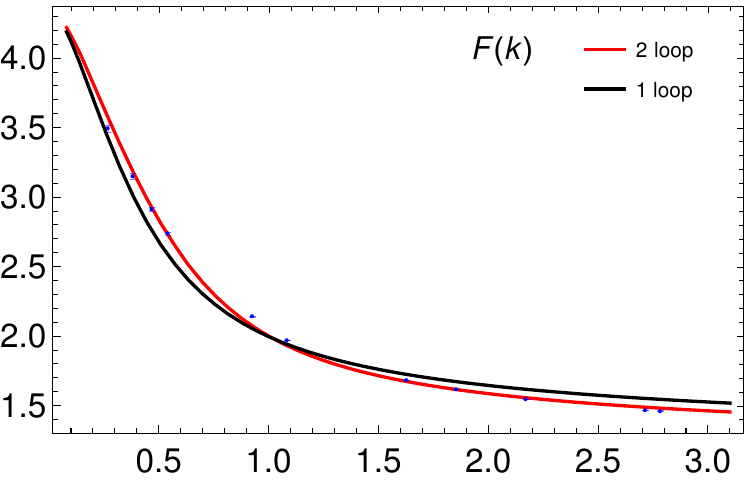}
\hglue-4mm \includegraphics[width=0.45\textwidth]{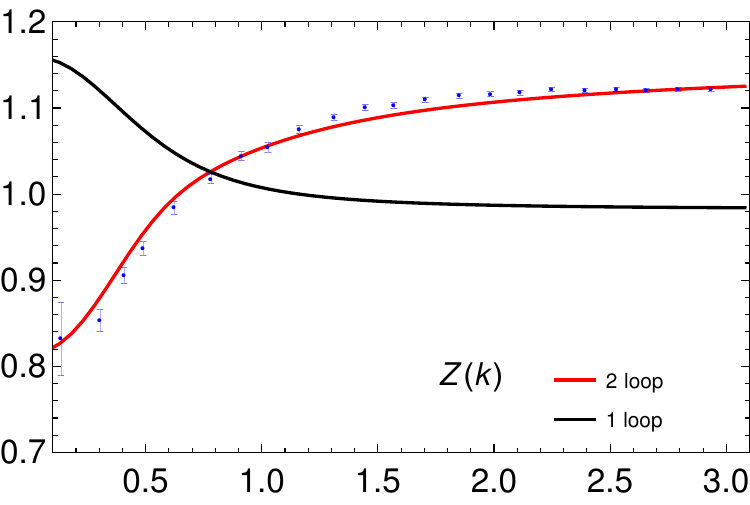}
 \caption{\label{fig:M150all} Comparison of the one- and two-loop CF results for the gluon (top), ghost (middle) and quark (bottom) dressing functions to the lattice data of Ref.~\cite{Sternbeck:2012qs,Oliveira:2018lln} using $M_\pi=150$ MeV. The parameters (determined from a global fit using the three functions $D$, $F$, $Z$) are found to be $\lambda_0=0.33$, $m_0=410$~MeV, $M_0=250$~MeV in the one-loop case, and $\lambda_0=0.32$, $m_0=370$~MeV, $M_0=160$~MeV in the two-loop case.}
\end{figure}

\begin{figure*}[t]
\includegraphics[width=0.3\textwidth]{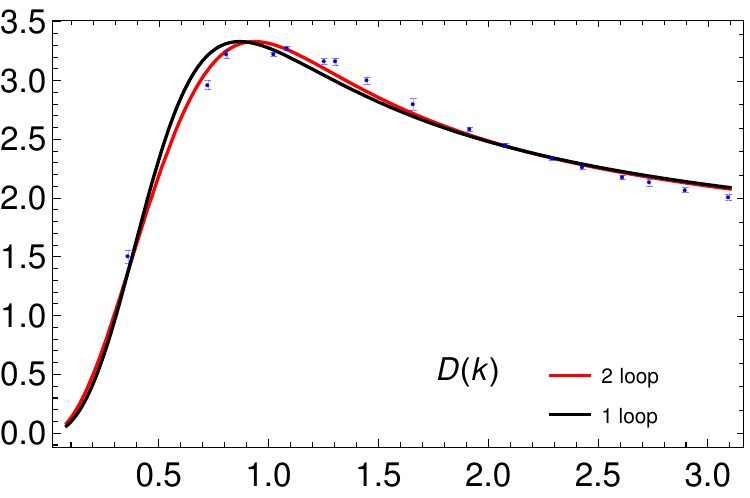}
\includegraphics[width=0.3\textwidth]{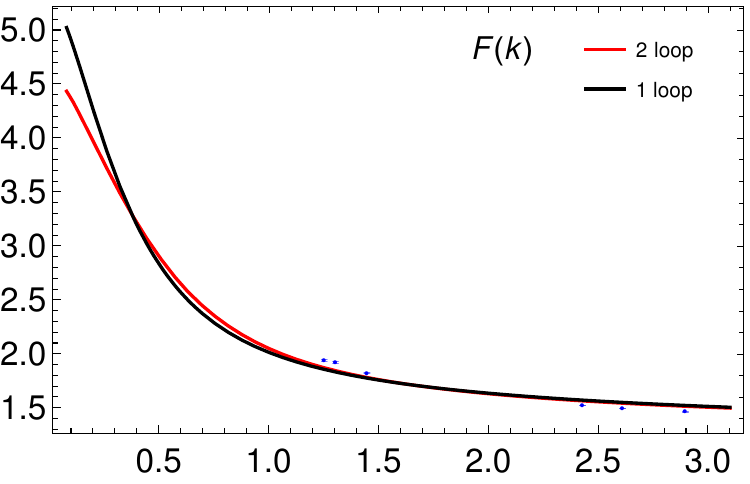}
\includegraphics[width=0.3\textwidth]{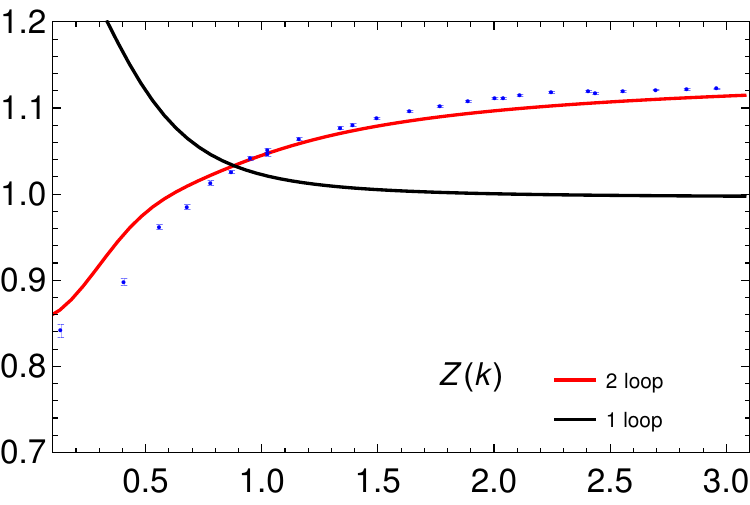}\\
\includegraphics[width=0.3\textwidth]{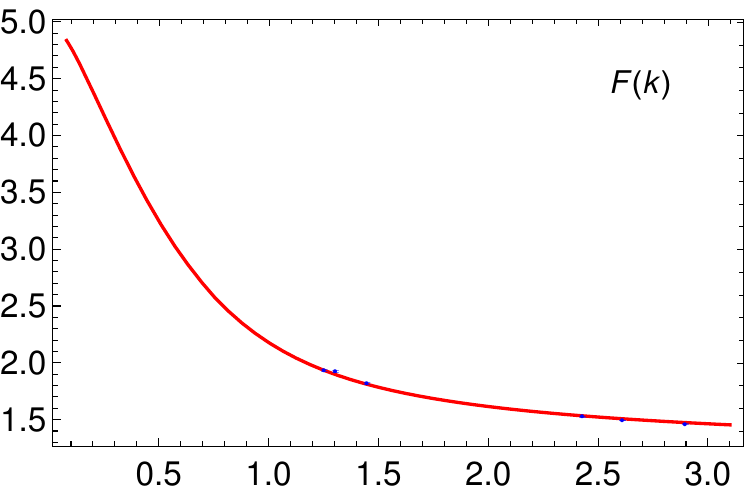}
\includegraphics[width=0.3\textwidth]{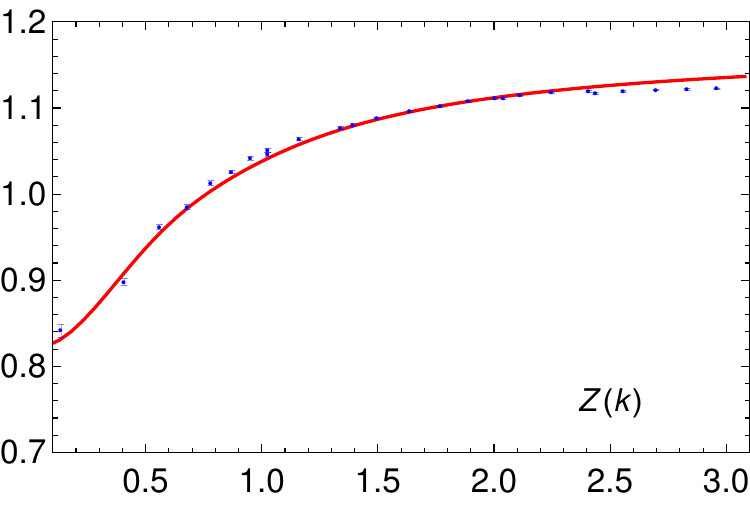}
\includegraphics[width=0.3\textwidth]{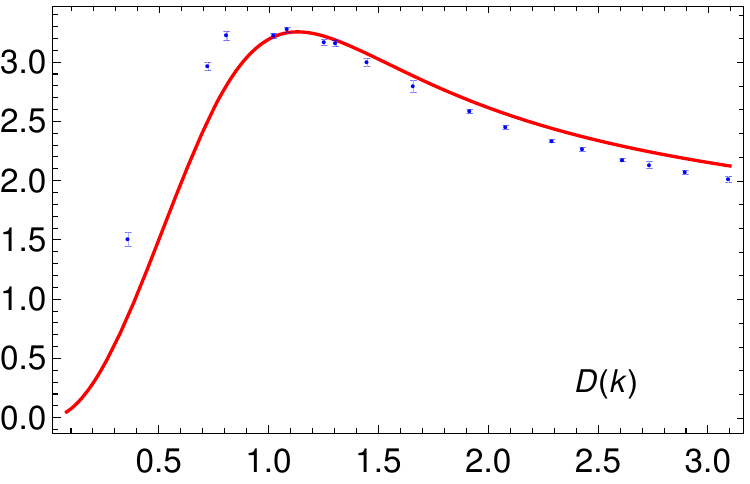}\\
\includegraphics[width=0.3\textwidth]{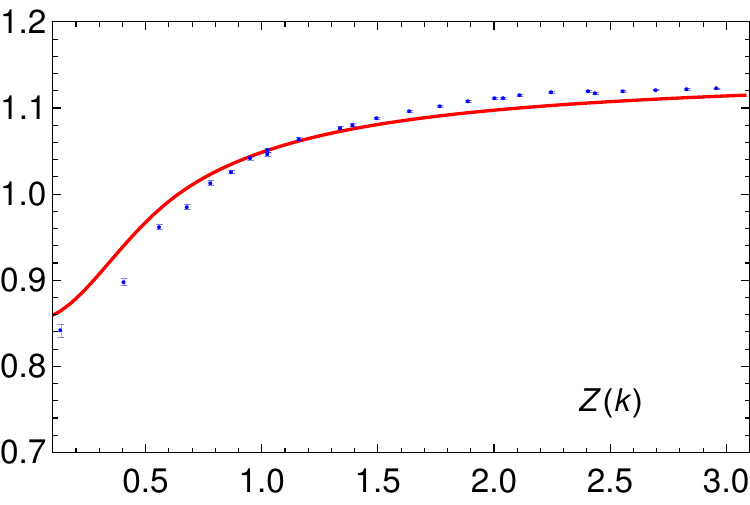}
\includegraphics[width=0.3\textwidth]{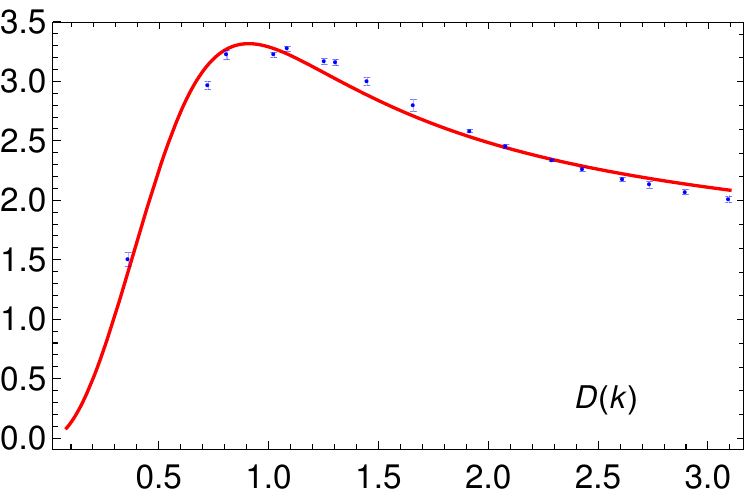}
\includegraphics[width=0.3\textwidth]{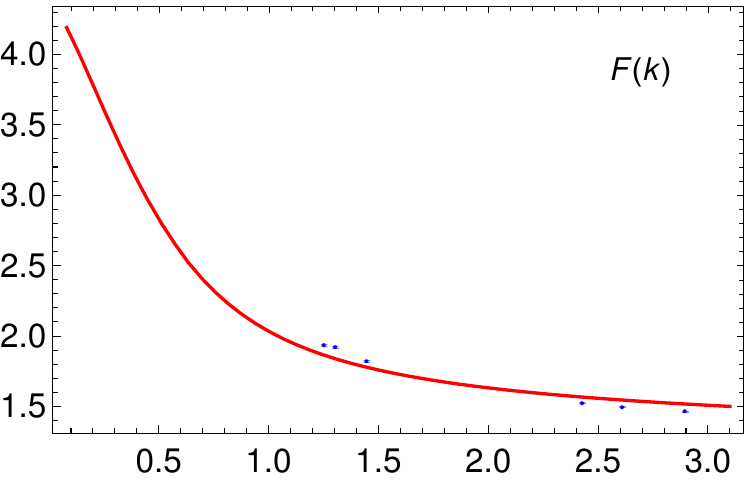}
 \caption{\label{fig:M426_2gg_ghostgluon} Fit of the two-loop CF results for the dressing functions $X$ (left) and $Y$ (middle) to the lattice data of Ref.~\cite{Sternbeck:2012qs,Oliveira:2018lln} using $\smash{M_\pi=422}$ MeV and the corresponding prediction of the third dressing function compared to data from the same references. The parameters (determined from a fit minimizing the reduced joint error $\chi_{XY}$) are found to be $\lambda_0=0.31$, $m_0=350$~MeV, $M_0=90$~MeV in the case $XY=DF$, $\lambda_0=0.43$, $m_0=490$~MeV, $M_0=200$~MeV in the case $XY=FZ$, and $\lambda_0=0.31$, $m_0=350$~MeV, $M_0=120$~MeV in the case $XY=ZD$.}
\end{figure*}

We can proceed to the same analysis with lattice data close to the physical case, simulated using a pion mass $\smash{M_\pi=150}$ MeV, see Refs.~\cite{Sternbeck:2012qs,Oliveira:2018lln}. The global and individual errors at one- and two-loop order are gathered in Tab.~\ref{table2} while Fig.~\ref{fig:M150all} shows the corresponding plots and gives the relevant parameters.

\begin{table}[h]
\begin{tabular}{|| c|c|c|c|c|c ||}
\hline
order & $\chi_{DFZ}(\%)$ & $\chi_D(\%)$ & $\chi_F(\%)$ & $\chi_Z(\%)$ \\
\hline \hline
\;1-loop\; & \;\;9.2\;\; & \;\;3.6\;\; & \;\;4.4\;\; & \;\;14.9\;\;  \\
\hline
\;2-loop\; & \;\;1.8\;\; & \;\;2.6\;\; & \;\;1.5\;\; & \;\;1.1\;\; \\
\hline
\end{tabular}
\caption{\label{table2} Global and individual errors as obtained from the global fit of $D$, $F$, $Z$ in the case $M_\pi=150$~MeV.}
\end{table}

The three dressing functions are very well reproduced at two-loop order and the quality of the fit is comparable (and even slightly better than in the previous case) confirming that these three quantities admit a good perturbative description within the CF model, irrespectively of the considerations on chiral symmetry breaking.

\pagebreak

\subsubsection{Partial fits}\label{sec:VA2}
To further test the quality of the perturbative evaluation of the dressing functions within the CF model, we also perform partial fits of two of these quantities, leaving the third one as a pure prediction of the model. That is, for two different quantities $X$ and $Y$, with $X$, $Y$ $\in \{D,F,Z\}$, we choose the parameters $m_0$, $\lambda_0$ and $M_0$ in such a way that they minimize the joint error $\chi_{XY}$, defined as:
\begin{equation}\label{joint_error_2}
\chi_{XY}^2=\frac{1}{2}\Big[\chi_X^2+\chi_Y^2\Big].
\end{equation}

\begin{figure*}[t]
\includegraphics[width=0.3\textwidth]{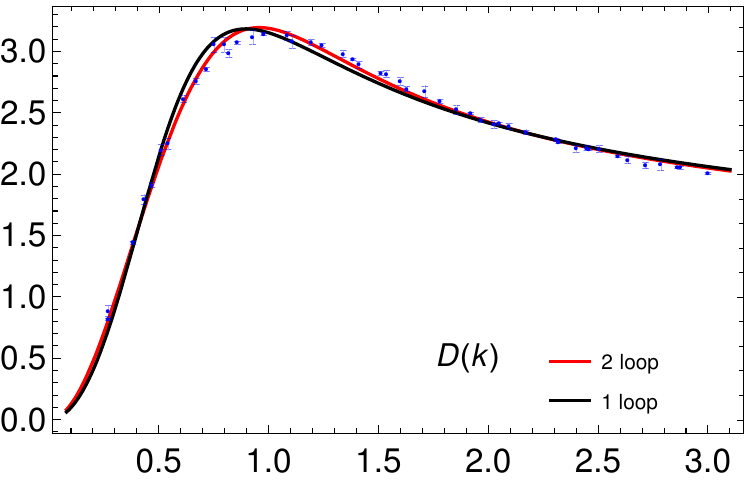}
\includegraphics[width=0.3\textwidth]{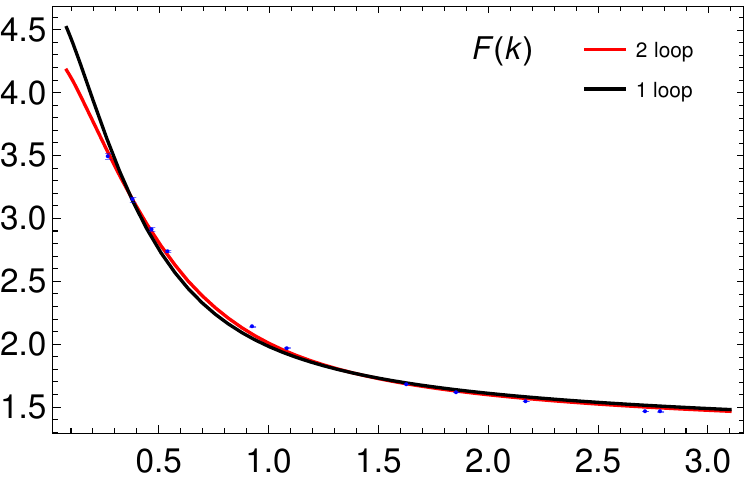}
\includegraphics[width=0.3\textwidth]{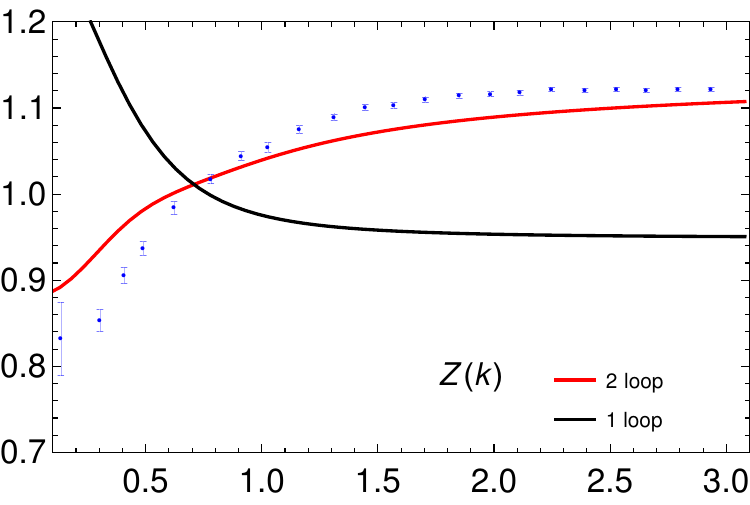}\\
\includegraphics[width=0.3\textwidth]{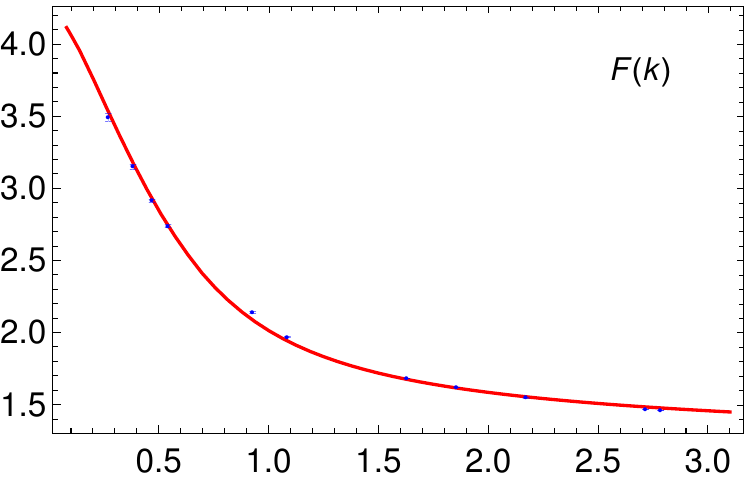}
\includegraphics[width=0.3\textwidth]{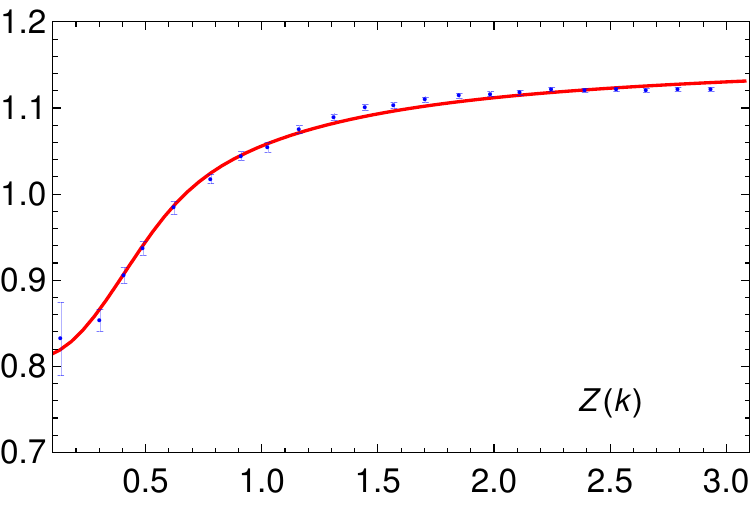}
\includegraphics[width=0.3\textwidth]{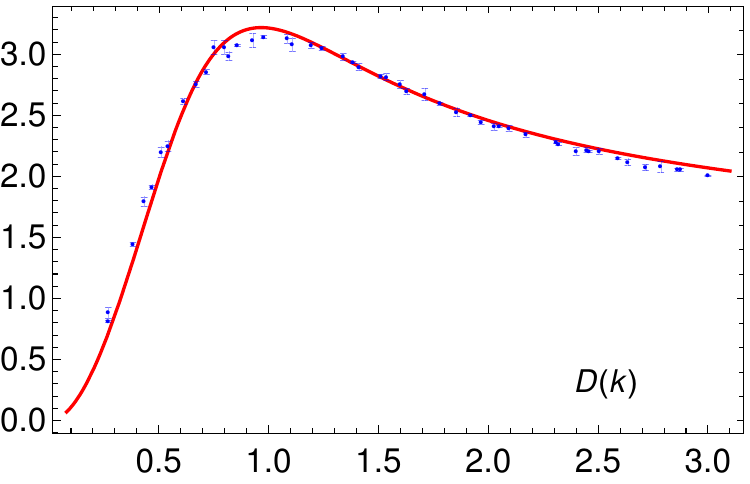}\\
\includegraphics[width=0.3\textwidth]{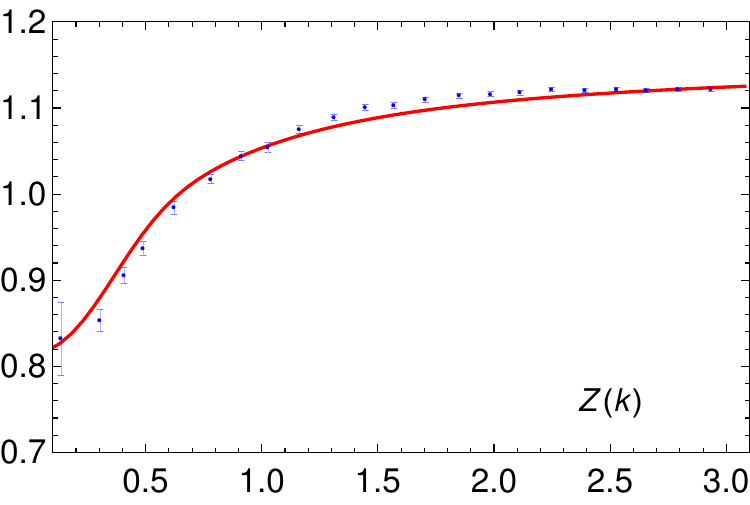}
\includegraphics[width=0.3\textwidth]{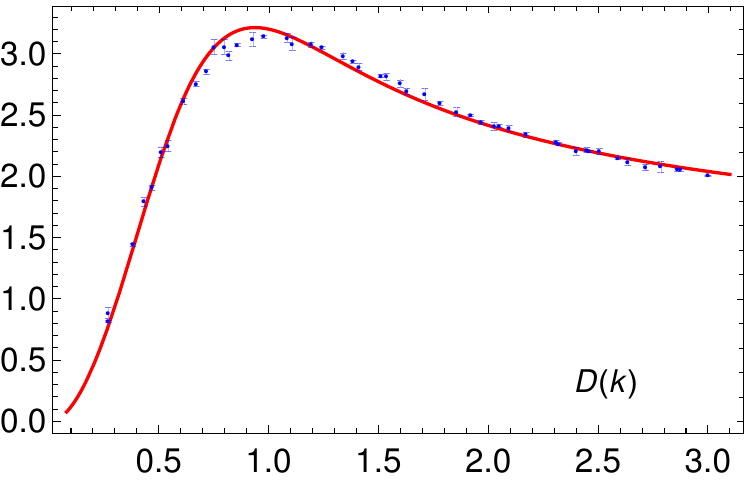}
\includegraphics[width=0.3\textwidth]{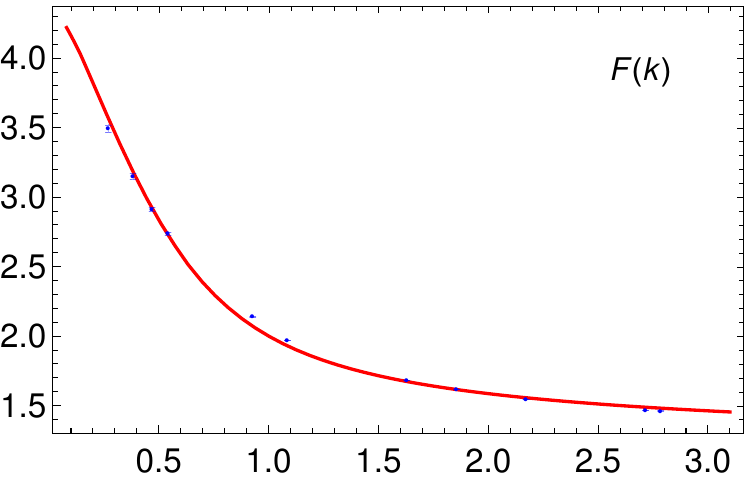}
 \caption{\label{fig:M150_2gg_ghostgluon} Fit of the two-loop CF results for the dressing functions $X$ (left) and $Y$ (middle) to the lattice data of Ref.~\cite{Sternbeck:2012qs,Oliveira:2018lln} using $\smash{M_\pi=150}$ MeV and the corresponding prediction of the third dressing function compared to data from the same references. The parameters (determined from a fit minimizing the reduced joint error $\chi_{XY}$) are found to be $\lambda_0=0.31$, $m_0=360$~MeV, $M_0=90$~MeV in the case $XY=DF$, $\lambda_0=0.38$, $m_0=400$~MeV, $M_0=250$~MeV in the case $XY=FZ$, and $\lambda_0=0.32$, $m_0=370$~MeV, $M_0=160$~MeV in the case $XY=ZD$.}
\end{figure*}

There are, therefore, three different possible fits that come from the minimization of $\chi_{DF}$, $\chi_{FZ}$ or $\chi_{ZD}$. The resulting errors are gathered in Tab.~\ref{table1FD} while the corresponding plots are shown in Figs.~\ref{fig:M426_2gg_ghostgluon} and \ref{fig:M150_2gg_ghostgluon}. Of course, one expects the error on the predicted function to increase as compared to the case where it was included in the global fit. The errors remain quite reasonable, however. The largest error is the one associated to the prediction of the gluon dressing function in the case $\smash{M_\pi=422}$\,MeV. This is understandable from the fact that the CF model rests on one phenomenological parameter related to the gluon field and fitting this parameter using correlation functions which do not directly involve gluons is probably not the best idea. In addition, for this value of the pion mass, the lattice data for the ghost dressing function contain only six points and none under 1 GeV which is not accurate enough to provide a ghost dressing fit of good quality in this range.

\begin{table}[h]
\begin{tabular}{|| c|c|c|c|c ||}
\hline
XY & $\chi_{XY} (\%)$ & $\chi_X(\%)$ & $\chi_Y(\%)$ & $\chi_{\rm pred.}(\%)$ \\
\hline \hline
DF\; & \;\;3.1\;\; & \;\;3.0\;\; & \;\;3.2\;\; & \;\;1.9\;\;\\
\hline
FZ\; & \;\;0.7\;\; & \;\;1.0\;\; & \;\;0.4\;\; & \;\;12.3\;\; \\
\hline
ZD\; & \;\;2.3\;\; & \;\;1.2\;\; & \;\;3.1\;\; & \;\;3.3\;\; \\
\hline
\end{tabular}

\vglue4mm

\begin{tabular}{|| c|c|c|c|c ||}
\hline
XY & $\chi_{XY} (\%)$ & $\chi_X(\%)$ & $\chi_Y(\%)$ & $\chi_{\rm pred.}(\%)$ \\
\hline \hline
\;DF\; & \;\;1.8\;\; & \;\;2.0\;\; & \;\;1.5\;\; & \;\;3.6\;\;\\
\hline
\;FZ\; & \;\;0.9\;\; & \;\;1.1\;\; & \;\;0.9\;\; & \;\;4.2\;\; \\
\hline
\;ZD\; & \;\;2.0\;\; & \;\;1.1\;\; & \;\;2.6\;\; & \;\;1.5\;\; \\
\hline
\end{tabular}
\caption{\label{table1FD} Global and individual errors as obtained from the partial fit of $X$ and $Y$ using the two-loop expressions and the corresponding error on the predicted dressing function in the case $M_\pi=422$~MeV (top) and in the case $M_\pi=150$~MeV (bottom).}
\end{table}

\subsection{The quark mass function}\label{sec:VB}

For completeness, let us here discuss the case of the quark mass function. This will allow us to illustrate the limitations of the perturbative approach within the CF model, which calls for the use of a more sophisticated, yet controlled approach, see Ref.~\cite{Pelaez:2020ups}. 

It is also to be stressed that the perturbative analysis of the quark mass function within the CF model is not totally academic. Indeed, the argument ruling out a priori the use of perturbation theory relies on the inability of the latter to describe the spontaneous breaking of chiral symmetry strictly in the limit of a vanishing bare quark mass.  Although it is most probable that no perturbative approach can describe the quark mass function for small enough bare quark masses, it is also reasonable to expect that perturbation theory becomes again valid for large enough bare quark masses. An intriguing question is then which value of the bare quark mass sets the frontier between a perturbative and a non-perturbative description of the quark mass function. The answer depends a priori on the implementation details of the perturbative approach, including the choice of gauge, the renormalization scheme or the particular modelling of the gauge fixing in the infrared. Consecuently, it is interesting to quantify more precisely the failure of the perturbative CF approach (within the renormalization scheme considered in this work) with regard to the quark mass function. 

We can try to address this question in various possible ways depending on how the parameters are fixed or the number of functions that are fitted to the lattice data. Since the overall picture that we will obtain eventually is similar in all cases, we shall refrain from including too many plots in this section and describe our results in the main text instead. In all the subsequent analyses, we use the following error function
\beq
\chi_M^2 & = & \frac{1}{2 N}\!\sum_i \!\left(\frac{1}{\bar{M}_\text{lt.}^2}\!+\!\frac{1}{M_\text{lt.}(k_i)^2} \right)\!(M_\text{lt.}(k_i)\!-\!M_{\rm th.}(k_i))^2\nonumber\\
\eeq
as an estimator of the quality of the quark mass function obtained within our approach. This formula corresponds to an average between the relative and absolute error (the latter is normalized by the maximal value $\bar{M}_\text{lt.}$ reached by the lattice quark mass function). The reason for this choice is that the quark mass function decreases rapidly in the UV, in such a way that the use of a pure relative error gives too much weight to the UV tail, while a pure absolute error gives too much weight to the IR tail. Since both regimes of momenta contain relevant information with regard to the spontaneous breaking of chiral symmetry (dynamically generated quark mass in the IR and quark condensate from the UV tail), we have chosen a compromise between these two definitions of the error. We mention that, if one aims at computing observables that are mostly sensitive to the IR region of the quark mass function, a different error function giving more weight to this range of momenta might be preferable. We shall comment on these other choices below.

\begin{figure}[t]
\includegraphics[width=0.45\textwidth]{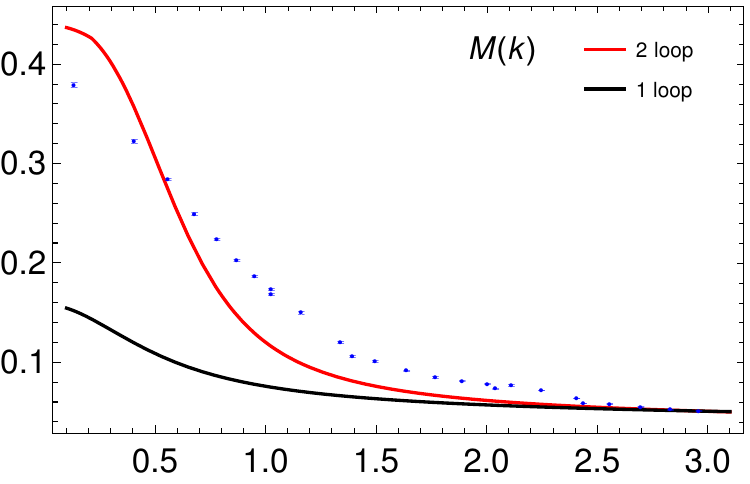}
\includegraphics[width=0.45\textwidth]{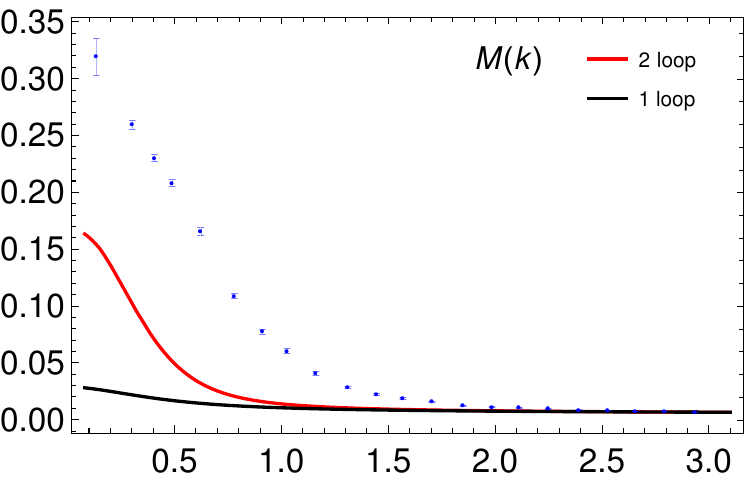}
\caption{\label{fig:M_prediction_UV} Prediction for the quark mass function from the two-loop CF expressions compared to the lattice data of Ref.~\cite{Sternbeck:2012qs,Oliveira:2018lln} in the cases $\smash{M_\pi=422}$ MeV (top) and $\smash{M_\pi=150}$ MeV (bottom). The parameters $m_0$ and $\lambda_0$ are determined from a global fit of the three functions $D$, $F$, $Z$ and the constraint that the quark mass parameter at the scale $\mu_0\simeq 3$\,GeV coincides with the lattice quark mass function at this value of the momentum. For the case $\smash{M_\pi=422}$ MeV, the parameters are found to be $\lambda_0=0.24$, $m_0=364$ MeV at one-loop order and $\lambda_0=0.34$ and $m_0=377$ MeV at two-loop order, while for $\smash{M_\pi=150}$ MeV, they are found to be $\lambda_0=0.28$ and $m_0=392$ MeV at one-loop order and  $\lambda_0=0.30$ and $m_0=291$ MeV at two-loop order. }
\end{figure}

We have first investigated how much quark mass is generated perturbatively within the CF model. That is, embracing our hypothesis that the dressing functions are essentially perturbative objects, we have used the parameters determined from the perturbative fits of these functions (see the previous section) to see how much quark mass is predicted in the two cases of study (far-from-chiral and close-to-physical). We should here mention that, when proceeding this way, part of the error on the quark mass function originates from a too na\"\i ve fixing of the quark mass parameter. Indeed, fixing the latter by fitting only the dressing functions is certainly not the best idea for none of these functions involves the quark mass parameter at tree-level. This is similar to fixing the gluon mass parameter of the CF model by fitting only the ghost and quark dressing functions, see above. For this reason, we proceed instead by fixing the coupling and the gluon mass parameter from fits of the dressing functions while the quark mass parameter is adjusted to agree with the lattice data for the quark mass function at some scale. When choosing this scale in the UV, the corresponding prediction for the quark mass function is shown in Fig.~\ref{fig:M_prediction_UV} and the corresponding errors are collected in Tab.~\ref{tableDFZ422_M_UV}.

\begin{table}[h]
\begin{tabular}{|| c|c|c|c|c|c ||}
\hline
order & $\chi_{DFZ}(\%)$ & $\chi_D(\%)$ & $\chi_F(\%)$ & $\chi_Z(\%)$ & $\chi_M(\%)$ \\
\hline \hline
\;1-loop\; & \;\;8.9\;\; & \;\;5.6\;\; &\;\;5.3\;\;  & \;\;12.1\;\; & \;\;34.6\;\;  \\
\hline
\;2-loop\; & \;\;3.0\;\;  & \;\;4.5\;\; &\;\;2.7\;\; & \;\;0.9\;\; & \;\;16.0\;\;\\
\hline
\end{tabular}

\vglue4mm

\begin{tabular}{|| c|c|c|c|c|c ||}
\hline
order & $\chi_{DFZ}(\%)$ & $\chi_D(\%)$ & $\chi_F(\%)$ & $\chi_Z(\%)$ & $\chi_M(\%)$ \\
\hline \hline
\;1-loop\; & \;\;11.0\;\; & \;\;3.2\;\; &\;\;7.0\;\;  & \;\;16.9\;\; & \;\;50.9\;\;  \\
\hline
\;2-loop\; & \;\;5.5\;\;  & \;\;6.6\;\; &\;\;3.0\;\; & \;\;6.1\;\; & \;\;41.8\;\; \\
\hline
\end{tabular}
\caption{\label{tableDFZ422_M_UV} Global and individual errors as obtained from the global fit of $D$, $F$, $Z$ enforcing the quark mass to be equal to a lattice value in the UV, $M(\text{3.0 GeV})=5.1$ MeV in the case $\smash{M_\pi=422}$ MeV and $M(\text{2.9 GeV})=6.6$ MeV in the case $\smash{M_\pi=150}$ MeV.}
\end{table}

In the case $\smash{M_\pi=422}$ MeV, the two-loop corrections greatly improve the one-loop result and, even though the two-loop error on the quark mass function is still  a few times larger than the one on the dressing functions, the trend from one- to two-loop order leaves room from improvement from higher order corrections. In contrast, in the case $\smash{M_\pi=150}$ MeV, although the two-loop corrections produce more quark mass in the IR than the one-loop expressions and the error on the quark mass function is reduced, the change is marginal and we are still far from reproducing the quark mass function. This is in line with the expectation that perturbation theory within the CF model cannot describe the quark mass function close to the physical case.  We also mention that the quality of the fit of the dressing functions deteriorates as compared to the case where the quark mass parameter was fixed by fitting the dressing functions. This can clearly be seen by comparing the second table in Tab.~\ref{tableDFZ422_M_UV} with Tab.~\ref{table2} and is yet an indication of the tension between the perturbative dressing functions and the perturbative quark mass function in the close-to-physical case.\footnote{The deterioration can also be seen in the plots (not shown), which do not look as neat as those in Figs.~\ref{fig:M422all} and \ref{fig:M150all}.}

\begin{figure}[t]
\includegraphics[width=0.4\textwidth]{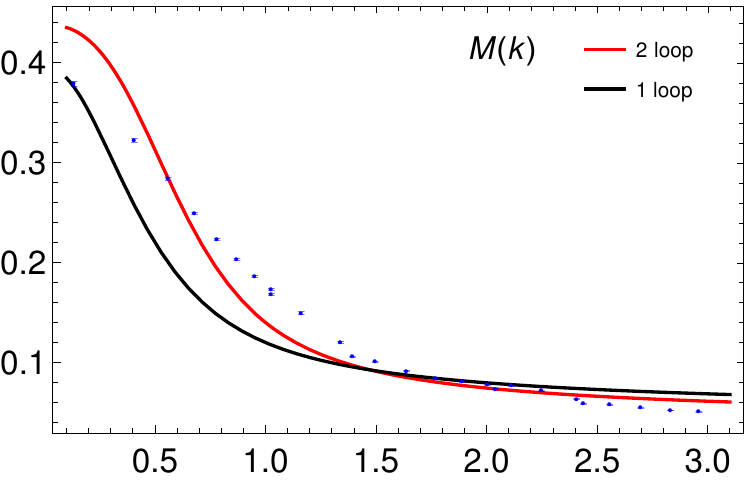}
\includegraphics[width=0.4\textwidth]{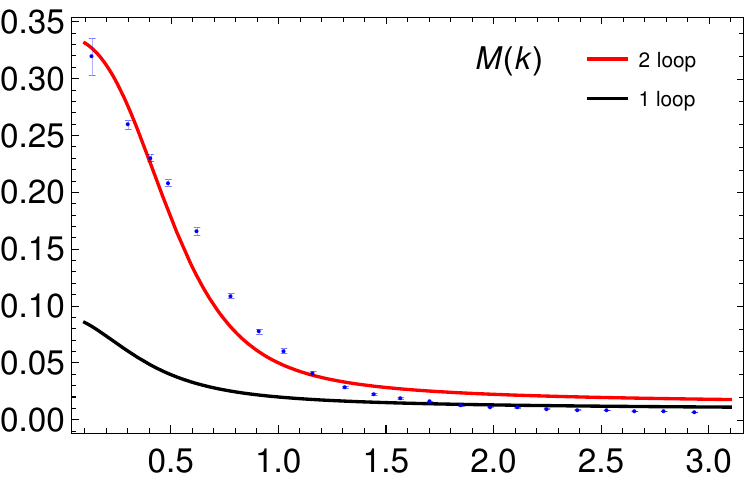}
\caption{\label{fig:DFMZ422} Fit of the two-loop CF results for the gluon, ghost and quark (bottom) dressing functions and quark mass (from top to bottom) Ref.~\cite{Sternbeck:2012qs,Oliveira:2018lln} using $\smash{M_\pi=422}$ MeV (top) and $\smash{M_\pi=150}$ MeV (bottom). In the case $\smash{M_\pi=422}$ MeV, the parameters are found to be $\lambda_0=0.42$, $m_0=420$ MeV, $M_0=120$ MeV at one-loop order, while in the case $\smash{M_\pi=150}$ MeV, the parameters are found to be $\lambda_0=0.43$, $m_0=430$ MeV, $M_0=20$ MeV at one-loop order and $\lambda_0=0.39$, $m_0=400$ MeV, $M_0=50$ MeV at two-loop order.} 
\end{figure}

In a second type of analysis, rather than trying to predict the quark mass function, we have investigated how the CF perturbative approach allows to globally describe the data for the dressing and quark mass functions and whether this perturbative description improves or worsens as the loop order is increased. To this purpose, we have performed a global fit of both the dressing functions and the quark mass function using the error function
\begin{equation}
\chi_{DFMZ}^2=\frac{1}{4}\left[\chi_D^2+\chi_F^2+\chi_M^2+\chi_Z^2\right].
\end{equation}
This is clearly less ambitious than trying to predict the quark mass function. The results for the quark mass function are shown in Figs. \ref{fig:DFMZ422} while the joint and individual errors are displayed in Tab.~\ref{tableDFMZ422}.

\begin{table}[h]
\begin{tabular}{|| c|c|c|c|c|c ||}
\hline
order & $\chi_{DFMZ}(\%)$ & $\chi_D(\%)$ & $\chi_F(\%)$ & $\chi_Z(\%)$ & $\chi_M(\%)$ \\
\hline \hline
\;1-loop\; & \;\;13.2\;\; & \;\;5.6\;\; &\;\;3.1\;\;  & \;\;15.9\;\; & \;\;18.1 \;\;\\
\hline
\;2-loop\; & \;\;5.9\;\;  & \;\;4.7\;\; &\;\;2.8\;\; & \;\;1.6\;\;& \;\;10.3\;\;\\
\hline
\end{tabular}

\vglue4mm

\begin{tabular}{|| c|c|c|c|c|c ||}
\hline
order & $\chi_{DFMZ}(\%)$ & $\chi_D(\%)$ & $\chi_F(\%)$ & $\chi_Z(\%)$ & $\chi_M(\%)$ \\
\hline \hline
\;1-loop\; & \;\;25.3\;\; & \;\;7.2\;\; &\;\;4.9\;\;  & \;\;21.5\;\; & \;\;45.0\;\;\\
\hline
\;2-loop\; & \;\;31.9\;\;  & \;\;9.1\;\; &\;\;4.7\;\; & \;\;3.4\;\;& \;\;62.8\;\;\\
\hline
\end{tabular}
\caption{\label{tableDFMZ422} Global and individual errors as obtained from the global fit of $D$, $F$, $M$ and $Z$ in the case $\smash{M_\pi=422}$~MeV (top) and  in the case $\smash{M_\pi=150}$ MeV (bottom).} 
\end{table}

Aside from a global deterioration in the quality of the dressing functions, we observe similar results as before for the quark mass function. Although we obtain a reasonable description of the quark mass function far from the chiral limit, the same function is poorly described in the close-to-physical case, with an error which is even larger at two-loop order than at one-loop order. The difference with the previous plots is that here the error comes dominantly from the UV tails.\\

 In conclusion, no matter what strategy is used, not all the features associated to chiral symmetry breaking can be reproduced: one has either a too low quark mass in the infrared or a not so accurate tail (and thus, probably, a not so accurate quark condensate) in the UV. We note, nonetheless, that, if one aims at computing observables that are mostly sensitive to the IR part of the quark mass function, our second strategy provides a rather reasonable description of the quark mass function in this range. In fact, we have checked that error functions that put more weight on the IR region typically give errors of the order of $15\%$ at two-loop order both far from the chiral limit and close to the physical case.

\subsection{Impact of the quark mass function on the\\ perturbative description of the dressing functions}\label{sec:VA3}
In the previous section we have illustrated the tension that exists between the two-loop perturbative evaluation of the dressing functions and the two-loop perturbative evaluation of the quark mass function within the CF model with regard to the QCD data.

\begin{figure*}[t]
\includegraphics[width=0.3\textwidth]{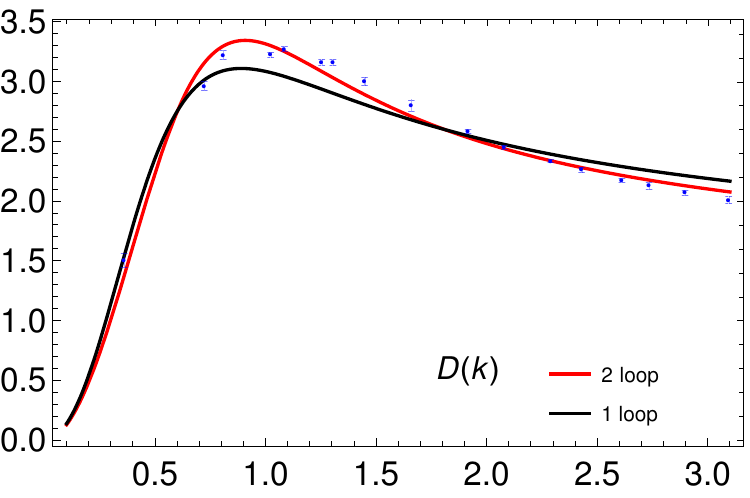}
\includegraphics[width=0.3\textwidth]{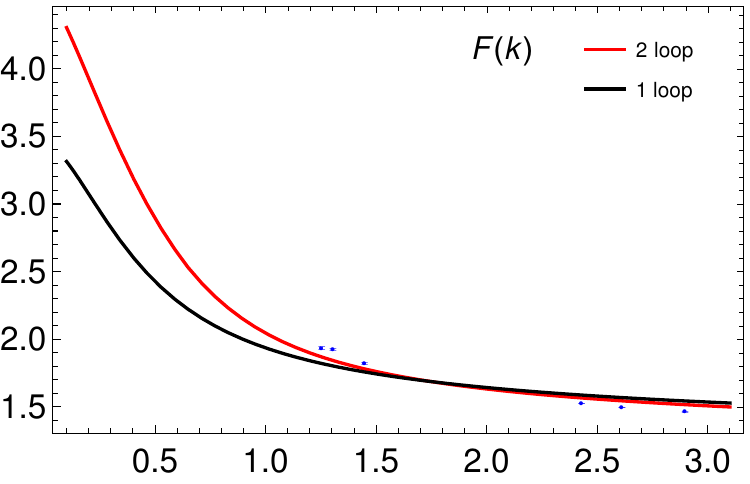}
\includegraphics[width=0.3\textwidth]{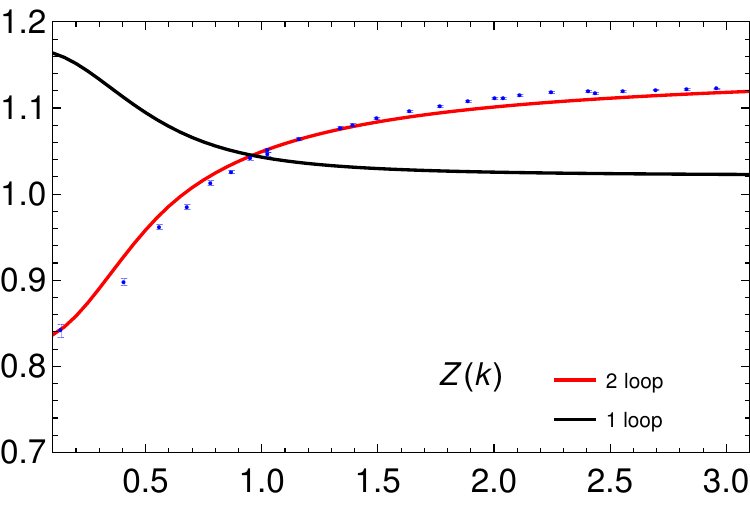}\\
\includegraphics[width=0.3\textwidth]{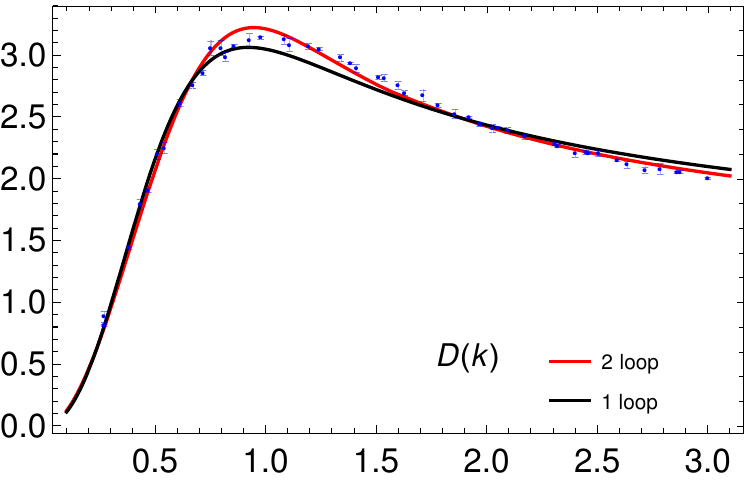}
\includegraphics[width=0.3\textwidth]{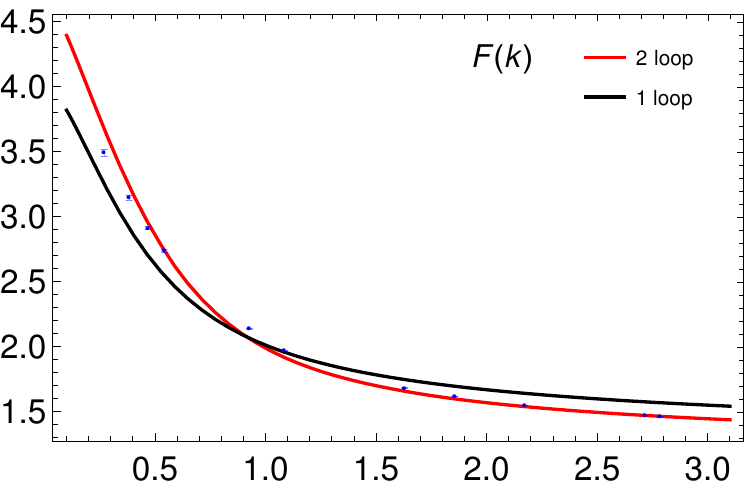}
\includegraphics[width=0.3\textwidth]{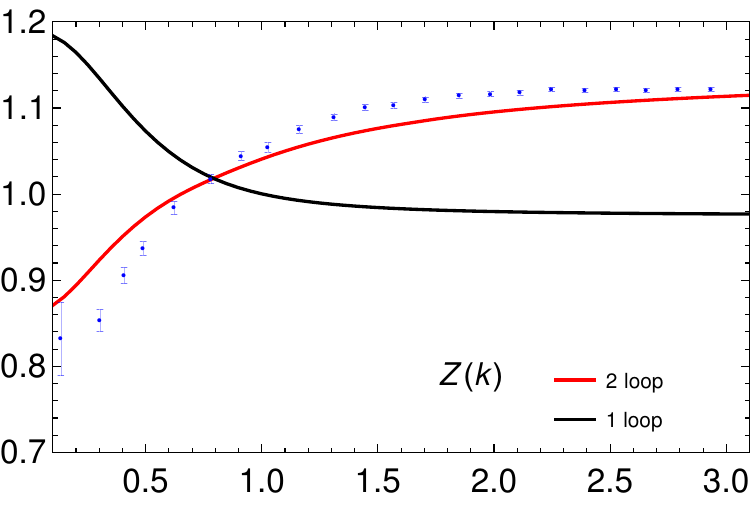}
\caption{\label{fig:DFZ422_m_int} Fit of the two-loop CF results for the gluon (left), ghost (middle) and quark (right) dressing functions Ref.~\cite{Sternbeck:2012qs,Oliveira:2018lln} in the case $\smash{M_\pi=422}$ MeV (top) and $\smash{M_\pi=150}$ MeV (bottom). The parameters are found to be $\lambda_0=0.31$, $m_0=360$ MeV and $\lambda_0=0.32$, $m_0=350$ at two-loop order, $\lambda_0=0.31$, $m_0=400$ MeV and $\lambda_0=0.26$, $m_0=380$ MeV at one-loop order for $\smash{M_\pi=422}$ MeV and $\smash{M_\pi=150}$ MeV respectively, and are determined from an interpolation of the lattice data for the quark mass and a global fit using the three functions $D$, $F$, $Z$.} 
\end{figure*}

One may argue that the appearance of this tension could jeopardize the perturbative picture for the various dressing functions which we have advertised above. In this subsection, we would like to demonstrate that this is not so. To this purpose, we reconsider the two-loop perturbative expressions for the dressing function, but rather than coupling them via the two-loop flow of the quark mass, we couple them using the actual non-perturbative flow, which we obtain from a simple interpolation of the lattice data for the quark mass function.\footnote{Here, we exploit the fact that, in the considered renormalization scheme, the flow of the quark mass and the quark mass function coincide.} We then minimize the joint error (\ref{joint_error}), leaving $\lambda_0$ and $m_0$ as free parameters. Of course, this procedure propagates the lattice data errors (on the quark mass function) to our results, but it still remains useful as a first approximation to study up to which extent the ghost, gluon and quark dressing functions are perturbative quantities once the actual quark mass is included in the game. In Fig.~\ref{fig:DFZ422_m_int}, we show the plots for $D$, $F$ and $Z$ for $\smash{M_\pi=422}$ MeV and $\smash{M_\pi=150}$ MeV and in Tab.~\ref{tableDFZ422_M_int} the corresponding joint and individual errors.

\begin{table}[h]
\begin{tabular}{|| c|c|c|c|c ||}
\hline
order & $\chi_{DFZ}(\%)$ & $\chi_D(\%)$ & $\chi_F(\%)$ & $\chi_Z(\%)$ \\
\hline \hline
\;1-loop\; & \;\;7.7\;\; & \;\;5.3\;\; &\;\;5.1\;\;  & \;\;11.1\;\;  \\
\hline
\;2-loop\; & \;\;2.8\;\;  & \;\;3.4\;\; &\;\;3.3\;\; & \;\;1.0\;\;\\
\hline
\end{tabular}

\vglue4mm

\begin{tabular}{|| c|c|c|c|c ||}
\hline
order & $\chi_{DFZ}(\%)$ & $\chi_D(\%)$ & $\chi_F(\%)$ & $\chi_Z(\%)$ \\
\hline \hline
\;1-loop\; & \;\;9.9\;\; & \;\;3.1\;\; &\;\;6.1\;\;  & \;\;15.4\;\;  \\
\hline
\;2-loop\; & \;\;2.6\;\;  & \;\;2.4\;\; &\;\;2.3\;\; & \;\;3.2\;\;\\
\hline
\end{tabular}
\caption{\label{tableDFZ422_M_int} Global and individual errors as obtained from the global fit of $D$, $F$, $Z$ and the interpolation of the quark mass lattice data in the case $M_\pi=422$~MeV (top) and $M_\pi=150$~MeV (bottom).} 
\end{table}

The quality of the fit for the dressing functions remains very good, supporting the claim that the functions $D$, $F$ and $Z$ remain perturbative, even when the actual quark mass is included in the analysis.

\section{Conclusions}
In this work, we have determined all two-point correlation functions of the Curci-Ferrari model in the presence of mass-degenerate fundamental quark flavors, at two-loop accuracy within the IR-safe renormalization scheme that was put forward in Ref.~\cite{Tissier:2011ey}. We have also compared them to QCD lattice data in the two flavour case, corresponding to various values of the pion mass; one that is relatively far from the chiral limit and another one that is closer to the physical value.

We find that those correlation functions that are not directly impacted by the spontaneous breaking of chiral symmetry are pretty well reproduced by the two-loop calculation within the CF model and, this, irrespectively of type of data that we try to reproduce. This includes the gluon and ghost dressing functions but also the quark dressing function. For the former two, the adequacy of the perturbative CF model to reproduce the lattice data was already seen at one-loop order \cite{Pelaez:2014mxa}. The two-loop contributions in this case represent tiny corrections that further improve the comparison to the data. In contrast, in the case of the quark dressing function, the two-loop corrections are pivotal as they drastically correct for the qualitatively inconsistent results obtained at one-loop order. As we have argued, they represent, in a sense, the true leading order contribution to the quark dressing function within the CF model, which provides as accurate results than for the other dressing functions.

As for the quark mass function, no strict perturbative approach can describe the spontaneous breaking of chiral symmetry in the limit of vanishing bare quark mass. From this fact, it is very reasonable to expect that no perturbative approach can describe the quark mass function close to the physical QCD case. The details, however, depend on the practical implementation of the perturbative approach. For completeness, we have then illustrated how the perturbative CF approach fails in reproducing the quark mass function at two-loop order.  Related to this question, we have studied the impact of the use of a non-perturbative running for the quark mass parameter extracted from the data (versus its two-loop CF version) on the perturbative determination of the quark dressing function. We find that the quality of the two-loop perturbative predictions for the dressing functions depends marginally on this consideration, confirming the perturbative nature of the dressing functions within the CF model.

These results are also of relevance for studies within the CF model beyond perturbation theory. As already mentioned, the RI-expansion of \cite{Pelaez:2017bhh,Pelaez:2020ups} captures the spontaneous breaking of chiral symmetry while dynamically generating the correct quark mass function. However, the quark dressing function is again badly reproduced. The problem is similar to the one in perturbation theory: at the order of approximation considered in \cite{Pelaez:2017bhh,Pelaez:2020ups}, the correction to the quark dressing function is abnormally small and requires one to push the Rainbow-Improved expansion scheme to a two-loop compatible level. The results in the present paper strongly suggest that this would allow to have both the correct dynamically generated quark mass function and an accurate quark dressing function, in line with what is observed in two-loop compatible DSE approaches \cite{Gao:2021wun}. One could even envisage a simpler, hybrid approach, combining the two-loop perturbative estimate for the quark dressing function and the quark mass function obtained from the Rainbow-Improved expansion at leading order.

In a future work, we also plan to extend the analysis to the quark-gluon vertex in those particular configurations where one of the external momenta vanishes, similar to the analysis of the ghost-antighost-gluon vertex given in Ref.~\cite{Barrios:2020ubx}. The challenge is here again to reduce all the Feynman integrals that enter the various form factors. However, since one of the external momenta vanishes, this is of the same complexity as the evaluation of the two-point form factors. Moreover, no additional renormalization group analysis needs to be carried out since all the relevant beta functions and anomalous dimensions have been evaluated in the present work.

\begin{acknowledgements} 
We are grateful to O.~Oliveira and A.~Sternbeck for kindly sharing the data of Refs.~\cite{Oliveira:2018lln} and \cite{Sternbeck:2012qs}. We also would like to thank J. Serreau, M. Tissier and N. Wschebor for insightful comments on the manuscript and M. Tissier once more for collaboration on a related work where part of the Mathematica routines used and extended here were written. J.A.G. gratefully acknowledges CNRS for a Visiting Fellowship and the hospitality of LPTMC, Sorbonne University, Paris where part 
of the work was carried out as well as the support of the German Research 
Foundation (DFG) through a Mercator Fellowship and partial support from STFC
via the Consolidated ST/T000988/1. N.~B. acknowledges the  financial support from the PEDECIBA program, the ANII-FCE-1-126412  project, the CAP ``Comisi\'on Acad\'emica de Posgrado" as well as the Laboratoire International Associ\'e of the CNRS, Institut Franco-Uruguayen de Physique. Several Feynman graphs were drawn with the {\sc Axodraw} package \cite{glmq10} and others with {\sc Jaxodraw} \cite{Binosi:2008ig}. Computations were carried out in part using the symbolic 
manipulation language {\sc Form}, \cite{glmq6,glmq7}.
\end{acknowledgements}

\pagebreak

\appendix

\section{Diagrams}

\subsection{Gluon two-point function}

\begin{figure}[h]

\includegraphics[width=0.08\textwidth]{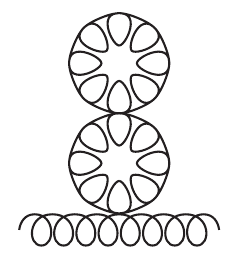}\quad\includegraphics[width=0.08\textwidth]{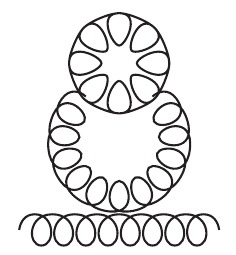}\quad\includegraphics[width=0.08\textwidth]{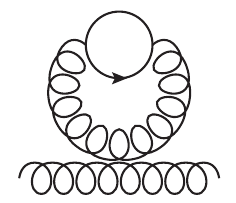}\quad\includegraphics[width=0.08\textwidth]{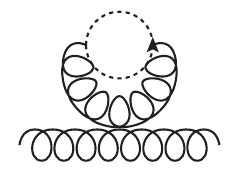}\\
\vglue7mm

\includegraphics[width=0.17\textwidth]{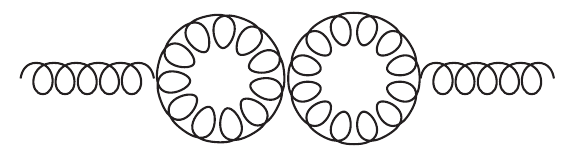}\\
\vglue6mm

\includegraphics[width=0.12\textwidth]{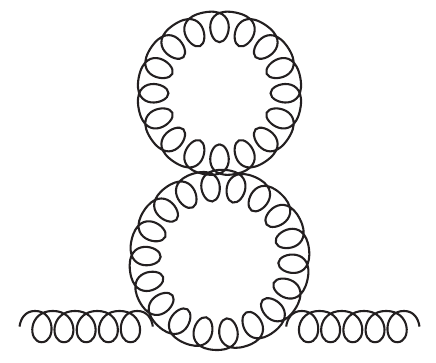}\quad\includegraphics[width=0.12\textwidth]{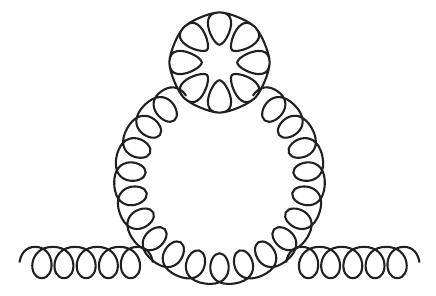}\\
\vglue6mm

\includegraphics[width=0.12\textwidth]{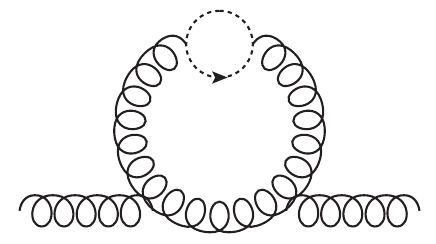}\quad\includegraphics[width=0.12\textwidth]{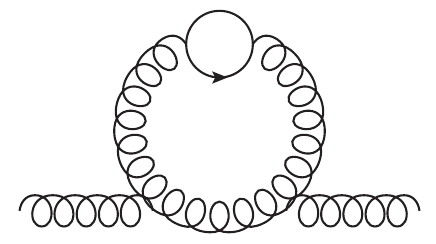}\\
\vglue6mm

\includegraphics[width=0.12\textwidth]{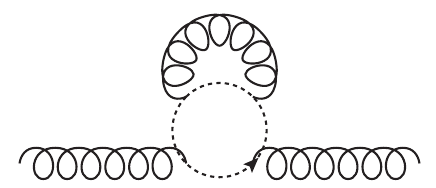}\quad\includegraphics[width=0.12\textwidth]{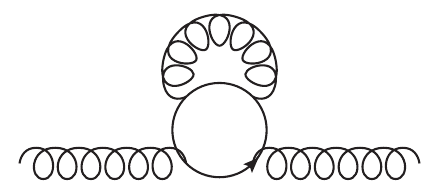}\\
\vglue6mm

\includegraphics[width=0.12\textwidth]{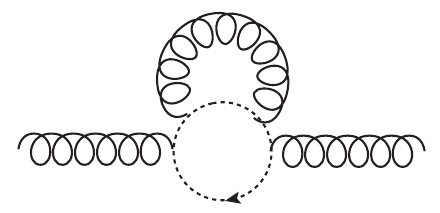}\quad\includegraphics[width=0.12\textwidth]{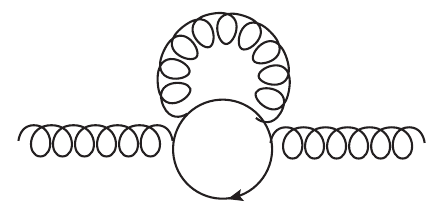}\\
\vglue6mm

\includegraphics[width=0.12\textwidth]{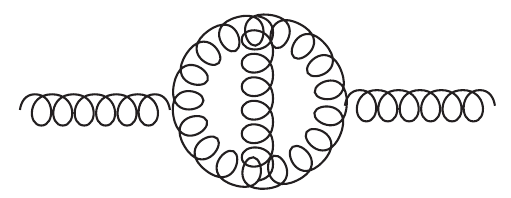}\quad\includegraphics[width=0.12\textwidth]{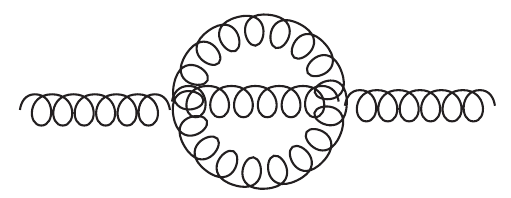}\\
\vglue6mm

\includegraphics[width=0.12\textwidth]{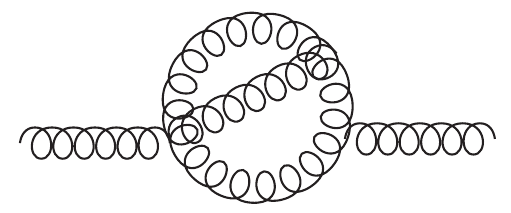}\quad\includegraphics[width=0.12\textwidth]{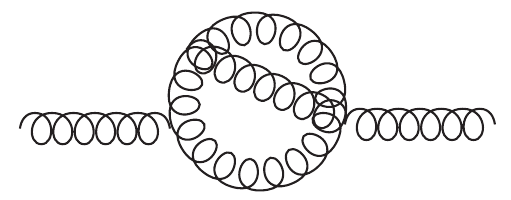}\\
\vglue6mm

\includegraphics[width=0.12\textwidth]{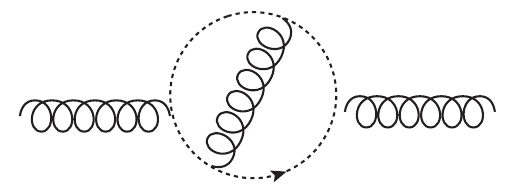}\quad\includegraphics[width=0.12\textwidth]{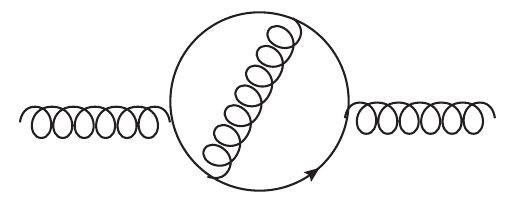}\\
\vglue6mm

\includegraphics[width=0.12\textwidth]{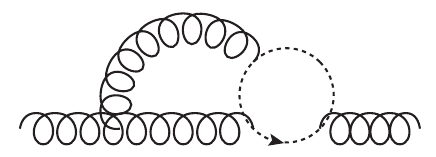}\quad\includegraphics[width=0.12\textwidth]{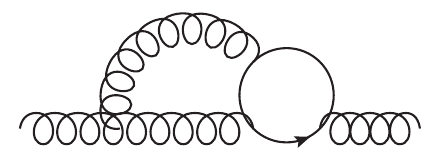}\\
\vglue6mm

\hglue-2mm\includegraphics[width=0.11\textwidth]{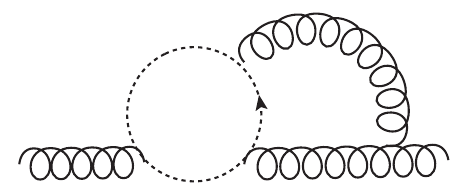}\quad\hglue3mm\includegraphics[width=0.11\textwidth]{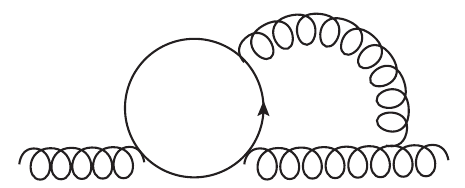}

\caption{Two-loop diagrams contributing to the gluon two-point function.}\label{fig:gluon_diag}

\end{figure}

The two-loop diagrams contributing to the gluon two-point function are displayed in Fig.~\ref{fig:gluon_diag}.

\pagebreak

\subsection{Ghost two-point function}
\begin{figure}[h]

\includegraphics[width=0.11\textwidth]{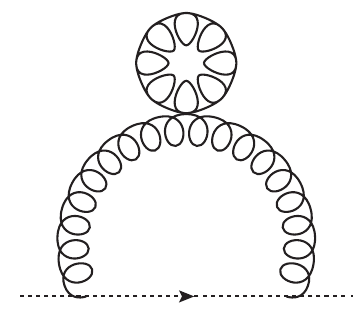}\quad\includegraphics[width=0.12\textwidth]{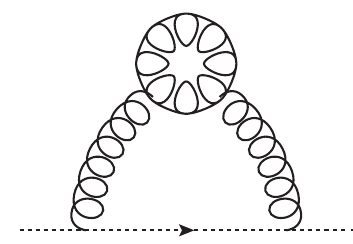}\\
\vglue7mm

\includegraphics[width=0.11\textwidth]{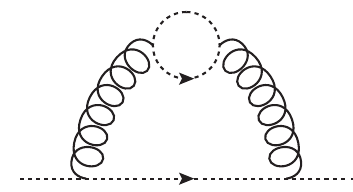}\quad\includegraphics[width=0.12\textwidth]{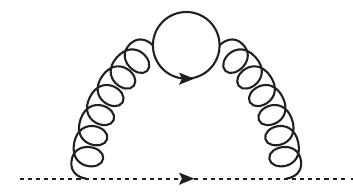}\\
\vglue7mm

\includegraphics[width=0.13\textwidth]{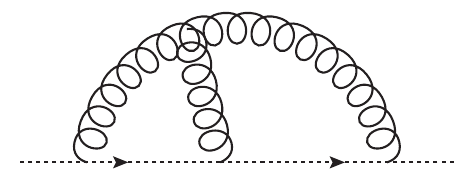}\quad\includegraphics[width=0.11\textwidth]{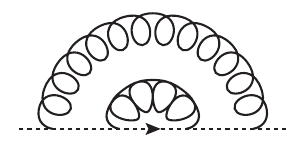}\\
\vglue9mm

\includegraphics[width=0.14\textwidth]{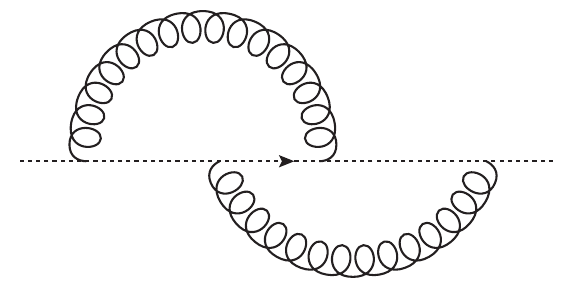}
\caption{Two-loop diagrams contributing to the ghost two-point function.}\label{fig:ghost_diag}
\end{figure}

The two-loop diagrams contributing to the ghost two-point function are displayed in Fig.~\ref{fig:ghost_diag}.

\subsection{Quark two-point function}

\begin{figure}[h]

\includegraphics[width=0.11\textwidth]{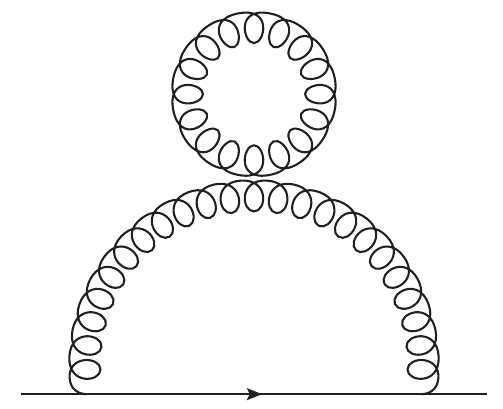}\quad\includegraphics[width=0.12\textwidth]{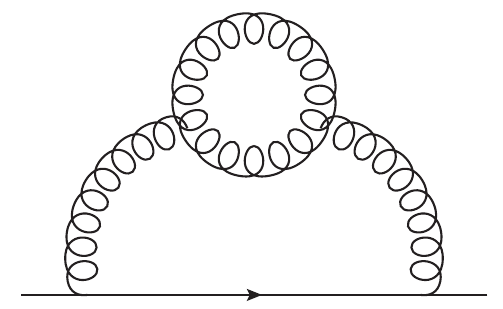}\\
\vglue7mm

\includegraphics[width=0.11\textwidth]{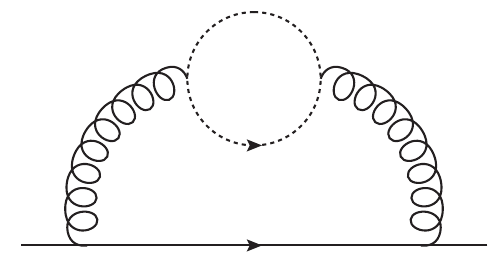}\quad\includegraphics[width=0.12\textwidth]{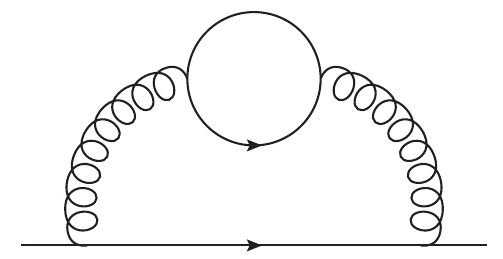}\\
\vglue7mm

\includegraphics[width=0.13\textwidth]{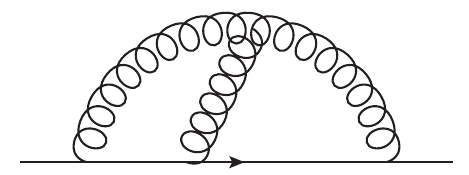}\quad\includegraphics[width=0.11\textwidth]{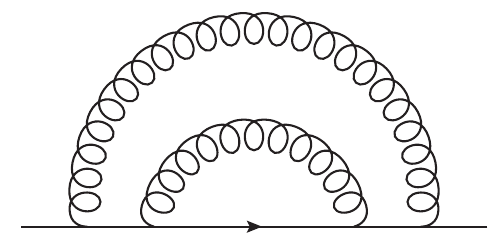}\\
\vglue9mm

\includegraphics[width=0.14\textwidth]{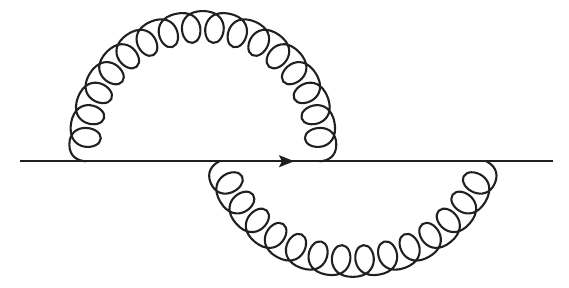}

\caption{Two-loop diagrams contributing to the quark two-point function.}\label{fig:quark_diag}
\end{figure}

The two-loop diagrams contributing to the quark two-point function are displayed in Fig.~\ref{fig:quark_diag}.

\section{Two-loop running}\label{app:2l_RG}
In this section, we derive general formulas for the two-loop anomalous dimensions and beta functions within a generic renormalization scheme defined from a given set of renormalization conditions, such as for instance the IR-safe conditions considered in this work. In Sec.~\ref{sec:MS}, for completeness, we shall also revisit the minimal subtraction scheme and see how it fits the general discussion (despite the absence of renormalization conditions in this case). 

We consider a field theory involving various bare fields $\varphi_{B,i}$ of bare square mass $m^2_{B,i}$. For simplicity, we assume that interactions are controlled by only one bare coupling, denoted by $\lambda_B$, but an extension to an arbitrary number of coupling constants is straightforward. We work in dimensional regularization, in which case the bare coupling has dimension $\smash{4-d=2\epsilon}$ and it is convenient to make this explicit by introducing a scale. We shall then operate the rescaling $\smash{\lambda_B\to\Lambda^{2\epsilon}\lambda_B}$ where the new $\lambda_B$ is dimensionless. As already mentioned in the main text, our notational choice $\Lambda$ (rather than $\mu$) is not innocent. It is meant to emphasize that this scale is  in general different from the renormalization scale $\mu$. The latter is introduced upon implementing a certain renormalization scheme via the renormalization conditions. On the other hand, the scale $\Lambda$ is a regulating scale that has nothing to do with the renormalization procedure.

To some extent, the scale $\Lambda$ should be put on the same footing as the cut-off scale in the cut-off regularization. This analogy needs to be taken with a pinch of salt of course because, in dimensional regularization, the regulating parameter $\epsilon$ is dissociated from the regulating scale $\Lambda$. In particular, the continuum limit is defined as the limit $\smash{\epsilon\to 0}$ and not as the limit $\smash{\Lambda\to\infty}$. However, as in any other regularization, we expect the continuum results obtained in the limit $\smash{\epsilon\to 0}$, to be independent of the regulating scale $\Lambda$, while they will in general depend on the renormalization scale $\mu$. This should apply in particular to the anomalous dimensions and the beta functions and we will check explicitly that this is indeed the case.

Let us mention that, in most approaches, the scale $\Lambda$ is identified with the scale $\mu$. This is a perfectly acceptable choice (and even a convenient one in some respects)\footnote{In particular, one does not need to introduce two scales in intermediate calculations. We stress however that continuum results do not depend on the scale $\Lambda$, so they depend only on one scale, $\mu$, even in the case where the choice $\Lambda\neq\mu$ is made.} since the anomalous dimensions and the beta functions do not depend on this choice and are in fact the same for any choice of dependence $\Lambda(\mu)$. However, the choice of a $\mu$-dependent $\Lambda$ tends to obscure the real source of $\mu$-dependence within the renormalization group, while unnecessarily complicating the evaluation of the anomalous dimensions and the beta functions (as we shall explicitly illustrate below). In what follows, we shall first derive the anomalous dimensions and the beta functions by taking $\Lambda$ independent from $\mu$ and then check that the so obtained functions do not depend on the choice of $\Lambda$, even when the latter is linked to $\mu$ in some way.

\subsection{RG basics}
Upon renormalization, the bare fields and the bare parameters are rescaled by renormalization factors as
\beq
\varphi_{B,i}=Z_{\varphi_i}^{1/2}\varphi_i\,, \,\,\, m^2_{B,i}=Z_{m^2_i}m^2_i\,,\,\,\, \lambda_B=Z_\lambda \lambda\,. ~~
\eeq
We shall denote the renormalization factors generically as $Z_X$ with $X\in\{\varphi_i,m^2_i,\lambda\}$. They depend a priori on the regulator $\epsilon$, the two scales $\Lambda$ and $\mu$, and the renormalized parameters $m^2_i$ and $g^2$.

The renormalized $n$-point functions are functions of the renormalization scale $\mu$. This $\mu$-dependence is controlled by the Callan-Szymanzik equation which, in its integrated form, is written as\footnote{We use the notation $\{k\}$ and $\{m^2\}$ to designate respectively the set of all external momenta of the considered $n$-point function and the set of all masses in the problem.}
\beq\label{eq:CZ}
& & \Gamma^{(n)}_{\varphi_{i_1}\dots\varphi_{i_n}}(\{k\};\{m^2_0\},\lambda_0,\mu_0)\\
& & \hspace{0.2cm}=\,\prod_{k=1}^n z_{i_k}^{-1/2}(\mu,\mu_0)\,\Gamma^{(n)}_{\varphi_{i_1}\dots\varphi_{i_n}}(\{k\};\{m^2(\mu)\},\lambda(\mu),\mu)\,,\nonumber
\eeq
and relates a given $n$-point function at a fixed scale $\mu_0$ to the same $n$-point function at the running scale $\mu$. We have already discussed the benefit of this type of equations in maintaining perturbative control when large logarithms of the form $\ln k/\mu_0$ are present. This is achieved by evaluating the right-hand side of Eq.~(\ref{eq:CZ}) with the choice $\mu=k$. This requires in turn the evaluation of the rescaling factor $z(\mu,\mu_0)$ as well as the running $m^2_i(\mu)$ and $\lambda(\mu)$ of the various parameters. 

The rescaling factor is given by
\beq
z_i(\mu,\mu_0)=\,\exp\left(\int_{\mu_0}^\mu d\nu\,\gamma_{\varphi_i}(\nu)\right),\label{eq:rescaling}
\eeq
where $\gamma_{\varphi_i}$ is the anomalous dimension of the field $\varphi_i$, related to the corresponding renormalization factor $Z_{\varphi_i}$ as
\beq
\gamma_{\varphi_i}\equiv \frac{d\ln Z_{\varphi_i}}{d\ln\mu}\,,\label{eq:gamma}
\eeq
where the $d/d\ln\mu$ derivatives are to be taken for fixed bare masses and dimensionful bare coupling $\Lambda^{2\epsilon}Z_\lambda\lambda$. On the other hand, the running of the parameters is given by the beta functions
\beq
\beta_\lambda\equiv \frac{d\lambda}{d\ln\mu}\,, \quad \beta_{m^2_i}\equiv \frac{d m^2_i}{d\ln\mu}\,.\label{eq:beta}
\eeq
By expressing that $\ln(Z_{m^2_i}m^2_i)$ and $\ln(\Lambda^{2\epsilon}Z_\lambda\lambda)$ do not depend on $\ln\mu$, one easily relates the beta functions to the anomalous dimensions associated with the parameters as\footnote{For the moment, we take $\Lambda$ as $\mu$-independent. We later discuss the case of a $\mu$-dependent $\Lambda$, including the conventional choice $\Lambda=\mu$.}
\beq\label{eq:beta_gamma}
0=\gamma_{m^2_i}+\frac{\beta_{m^2_i}}{m^2_i}\,, \quad 0=\gamma_\lambda+\frac{\beta_\lambda}{\lambda}\,,
\eeq
where
\beq
\gamma_{m^2_i}\equiv \frac{d\ln Z_{m^2_i}}{d\ln\mu} \quad \mbox{and} \quad \gamma_\lambda\equiv\frac{d\ln Z_\lambda}{d\ln\mu}
\eeq
It follows that the implementation of the renormalization group equation (\ref{eq:CZ}), requires the determination of the various anomalous dimensions 
\beq
\gamma_X\equiv \frac{d\ln Z_X}{d\ln\mu}\,,\label{eq:def}
\eeq
with $X\in\{\varphi_i,m^2_i,\lambda\}$. Because they correspond to infinitesimal variations of ratios of renormalization factors at different scales, which in turn can be written as ratios of renormalized correlation funtions, the anomalous dimensions are finite. Below, we evaluate these anomalous dimensions at one- and two-loop order.

We mention that, in deriving Eq.~(\ref{eq:beta_gamma}), we have made use of our assumption of a $\mu$-independent $\Lambda$. Were we to consider a $\mu$-dependent $\Lambda$, the right-hand side of the second equation of (\ref{eq:beta_gamma}) would involve an additional term $2\epsilon\,d\ln\Lambda/d\ln\mu$ which cannot be neglected because it can (and does) end up multiplying contributions proportional to $1/\epsilon$. By choosing a $\mu$-independent $\Lambda$, we do not need to worry about this subtlety. A related convenient feature of using a $\mu$-independent $\Lambda$ is that both $\beta_{m^2_i}/m^2_i$ and $\beta_\lambda/\lambda$ are of order $\lambda$, whereas with a $\mu$-dependent $\Lambda$, $\beta_\lambda/\lambda$ is of order $\lambda^0$ which leads to new contributions when evaluating the anomalous dimensions at a given order. We will show below that despite these implementation differences, the various additional contributions that one needs to consider in the case of a $\mu$-dependent $\Lambda$ cancel with each other, making the $\mu$-independent choice, the simpler one in practice.

\subsection{One-loop running}
To derive the anomalous dimension $\gamma_X$ at one-loop order, we start from the one-loop renormalization factor $Z_X$ expanded up to order $\epsilon^0$. We write it as
\beq
Z_X=1+\lambda\frac{z_{X,1}}{\epsilon}\,,\label{eq:1l}
\eeq
with
\beq
z_{X,1}=z_{X,11}+\epsilon z_{X,10}\,,
\eeq
and where $z_{X,11}$ and $z_{X,10}$ are a priori functions of $\Lambda$, $\mu$ and the masses $m^2_i$. We will see below that there are some constraints on the factors $z_{X,ab}$.

From (\ref{eq:def}) and (\ref{eq:1l}), the anomalous dimension becomes
\beq
\gamma_X & = & \frac{1}{Z_X}\left(\frac{\lambda}{\epsilon}\frac{\partial z_{X,1}}{\partial\ln\mu}+\frac{\beta_{\lambda}}{\lambda}\frac{\lambda}{\epsilon}z_{X,1}\right.\nonumber\\
& & \hspace{1.5cm}\left.+\,\sum_i\frac{\beta_{m^2_i}}{m^2_i}\frac{\lambda}{\epsilon}\frac{\partial z_{X,1}}{\partial\ln m^2_i}\right).\label{eq:g1}
\eeq
The term with the partial derivative $\partial/\partial\mu$ takes into account the explicit $\mu$-dependence of $z_{X,1}$, while the terms involving the beta functions, see Eq.~(\ref{eq:beta}), take into account the implicit $\mu$-dependence of $z_{X,1}$ via its dependence on $\lambda$ and $m^2_i$. Since $\beta_{m^2_i}/m^2_i$ and $\beta_{\lambda}/\lambda$ are of order $\lambda$, see the discussion above, we can neglect the terms proportional to the beta functions to the present order of accuracy. Moreover, we can replace $Z_X$ by $1$ in the denominator of (\ref{eq:g1}). We find eventually
\beq
\gamma_X=\frac{\lambda}{\epsilon}\frac{\partial z_{X,1}}{\partial\ln\mu}\,.\label{eq:gamma_X}
\eeq
Expanding to order $\epsilon^0$, this gives
\beq
\gamma_X=\frac{\lambda}{\epsilon}\frac{\partial z_{X,11}}{\partial\ln\mu}+\lambda\frac{\partial z_{X,10}}{\partial\ln\mu}\,.
\eeq
The finiteness of the anomalous dimensions imposes $z_{X,11}$ not to depend explicitly on $\mu$. This is not really a surprise since $z_{X,11}/\epsilon$ corresponds to the divergence of a one-loop Feynman integral and, as such, is a pure constant that does not depend on the considered renormalization scheme. We eventually arrive at
\beq
\gamma_X=\lambda\frac{\partial z_{X,10}}{\partial\ln\mu}\,.\label{eq:g1_final}
\eeq
We notice that the anomalous dimension could a priori still depend on $\Lambda$ (via the factor $z_{X,10}$). We will show below that this is not the case and also that the same expression could be obtained using a $\mu$-dependent $\Lambda$.

\subsection{Two-loop running}
In order to extend the anomalous dimension $\gamma_X$ to two-loop order, we need the renormalization factors to order $\lambda^2$ and $\epsilon^1$, which we write as
\beq
Z_X & = & 1+\lambda\frac{z_{X,1}}{\epsilon}+\lambda^2\frac{z_{X,2}}{\epsilon^2}\,,\label{eq:ZX} 
\eeq
with
\beq
z_{X,1} & = & z_{X,11}+z_{X,10}\epsilon+z_{X,1(-1)}\epsilon^2\,,\\
z_{X,2} & = & z_{X,22}+z_{X,21}\epsilon+z_{X,20}\epsilon^2\,.
\eeq
\vglue0.5mm
\noindent{We need to include $z_{X,1(-1)}$ because, although it is a contribution of order $\epsilon^1$ to $Z_X$, it contributes at order $\epsilon^0$ to the two-loop two-point functions; see the discussion in the main text. We will see below that it also contributes to the anomalous dimensions at this order.}

\begin{widetext}
From (\ref{eq:def}) and (\ref{eq:ZX}), the anomalous dimension becomes
\beq
\gamma_X=\frac{1}{Z_X}\left(\frac{\lambda}{\epsilon}\frac{\partial z_{X,1}}{\partial\ln\mu}+\frac{\lambda^2}{\epsilon^2}\frac{\partial z_{X,2}}{\partial\ln\mu}+\frac{\beta_{\lambda}}{\lambda}\left(\lambda\frac{z_{X,1}}{\epsilon}+2\lambda^2\frac{z_{X,2}}{\epsilon^2}\right)+\sum_i\frac{\beta_{m^2_i}}{m^2_i}\left(\frac{\lambda}{\epsilon}\frac{\partial z_{X,1}}{\partial\ln m^2_i}+\frac{\lambda^2}{\epsilon^2}\frac{\partial z_{X,2}}{\partial\ln m^2_i}\right)\right).
\eeq
Using the fact that $\beta_{m^2_i}/m^2_i$ and $\beta_{\lambda}/\lambda$ are both of order $\lambda$ and expanding $Z_X$ up to order $\lambda$, we find
\beq
\gamma_X=\frac{\lambda}{\epsilon}\frac{\partial z_{X,1}}{\partial\ln\mu}+\frac{\lambda^2}{\epsilon^2}\frac{\partial z_{X,2}}{\partial\ln\mu}-\frac{\lambda^2}{\epsilon^2}\frac{\partial z_{X,1}}{\partial\ln\mu}z_{X,1}-\gamma_{\lambda}\frac{\lambda}{\epsilon}z_{X,1}-\sum_i\gamma_{m^2_i}\frac{\lambda}{\epsilon}\frac{\partial z_{X,1}}{\partial\ln m^2_i}\,,\label{eq:gt}
\eeq
where $\gamma_{\lambda}$ and $\gamma_{m^2}$ are the gamma functions determined at one-loop order, prior to an expansion in $\epsilon$, see Eq.~(\ref{eq:gamma_X}). Using this latter equation, we find
\beq\label{eq:inter}
\gamma_X=\frac{\lambda}{\epsilon}\frac{\partial z_{X,1}}{\partial\ln\mu}+\frac{\lambda^2}{\epsilon^2}\left(\frac{\partial z_{X,2}}{\partial\ln\mu}-\left(\frac{\partial z_{X,1}}{\partial\ln\mu}+\frac{\partial z_{\lambda,1}}{\partial\ln\mu}\right)z_{X,1}-\sum_i\frac{\partial z_{m^2_i,1}}{\partial\ln \mu}\frac{\partial z_{X,1}}{\partial\ln m^2_i}\right),
\eeq
and expanding to order $\epsilon^0$, this gives
\beq
\gamma_X & = & \frac{\lambda^2}{\epsilon^2}\frac{\partial z_{X,22}}{\partial\ln\mu}+\frac{\lambda^2}{\epsilon}\left(\frac{\partial z_{X,21}}{\partial\ln\mu}-\left(\frac{\partial z_{X,10}}{\partial\ln\mu}+\frac{\partial z_{\lambda,10}}{\partial\ln\mu}\right)z_{X,11}\right)+\lambda\frac{\partial z_{X,10}}{\partial\ln\mu}\nonumber\\
& + & \lambda^2\!\left(\frac{\partial z_{X,20}}{\partial\ln\mu}\!-\!\left(\frac{\partial z_{X,10}}{\partial\ln\mu}\!+\!\frac{\partial z_{\lambda,10}}{\partial\ln\mu}\right)z_{X,10}\!-\!\left(\frac{\partial z_{X,1(-1)}}{\partial\ln\mu}\!+\!\frac{\partial z_{\lambda,1(-1)}}{\partial\ln\mu}\right)z_{X,11}\!-\!\sum_i\frac{\partial z_{m^2_i,10}}{\partial\ln \mu}\frac{\partial z_{X,10}}{\partial\ln m^2_i}\right),\label{eq:form}
\eeq
where we used that $z_{X,11}$ is a pure constant. The finiteness of the gamma function imposes that
\beq
\frac{\partial z_{X,22}}{\partial\ln\mu}=0 \quad \mbox{and} \quad \frac{\partial z_{X,21}}{\partial\ln\mu}-\left(\frac{\partial z_{X,10}}{\partial\ln\mu}+\frac{\partial z_{\lambda,10}}{\partial\ln\mu}\right)z_{X,11}=0\,.\label{eq:id2}
\eeq
The first constraint is again not a real surprise since $z_{X,22}/\epsilon^2$ has to do with the overall divergence of a two-loop Feynman integral and, as such, should be a pure constant that does not depend on the considered renormalization scheme. The second constraint is more subtle and relates to a similar well known identity in minimal subtraction which constrains $z_{X,22}$, $z_{X,11}$, $z_{\lambda,11}$, see below. Since $z_{X,11}$ is a pure constant, this second constraint can also be reformulated as stating that the combination $z_{X,21}-(z_{X,10}+z_{\lambda,10})z_{X,11}$ should not depend on $\mu$. In turn, this provides a cross-check for any two-loop determination of the $n$-point functions in a given scheme, which we have used in our particular application to the CF model. See the main text. 

We eventually arrive at the following finite expression for the two-loop anomalous dimension
\beq
\gamma_X=\lambda\frac{\partial z_{X,10}}{\partial\ln\mu}+\lambda^2\!\left(\frac{\partial z_{X,20}}{\partial\ln\mu}\!-\!\left(\frac{\partial z_{X,10}}{\partial\ln\mu}+\frac{\partial z_{\lambda,10}}{\partial\ln\mu}\right)\!z_{X,10}\!-\!\left(\frac{\partial z_{X,1(-1)}}{\partial\ln\mu}+\frac{\partial z_{\lambda,1(-1)}}{\partial\ln\mu}\right)\!z_{X,11}-\!\sum_i\frac{\partial z_{m^2_i,10}}{\partial\ln \mu}\frac{\partial z_{X,10}}{\partial\ln m^2_i}\right),\label{eq:gX_final}\nonumber\\
\eeq
in terms of the various factors $z_{X,ab}$. In the case $\smash{z_{X,11}\neq 0}$, this expression can be simplified using the second constraint in (\ref{eq:id2}). One finds
\beq
\gamma_X=\lambda\frac{\partial z_{X,10}}{\partial\ln\mu}+\lambda^2\left(\frac{\partial z_{X,20}}{\partial\ln\mu}-\frac{z_{X,10}}{z_{X,11}}\frac{\partial z_{X,21}}{\partial\ln\mu}-\left(\frac{\partial z_{X,1(-1)}}{\partial\ln\mu}+\frac{\partial z_{\lambda,1(-1)}}{\partial\ln\mu}\right)z_{X,11}-\sum_i\frac{\partial z_{m^2_i,10}}{\partial\ln \mu}\frac{\partial z_{X,10}}{\partial\ln m^2_i}\right).\label{eq:gX_final1}
\eeq
In the case $\smash{z_{X,11}=0}$, one cannot use the second constraint but the formula also gets simpler:
\beq
\gamma_X=\lambda\frac{\partial z_{X,10}}{\partial\ln\mu}+\lambda^2\left(\frac{\partial z_{X,20}}{\partial\ln\mu}-\left(\frac{\partial z_{X,10}}{\partial\ln\mu}+\frac{\partial z_{\lambda,10}}{\partial\ln\mu}\right)z_{X,10}-\sum_i\frac{\partial z_{m^2_i,10}}{\partial\ln \mu}\frac{\partial z_{X,10}}{\partial\ln m^2_i}\right).\label{eq:gX_final2}
\eeq
In the case $\smash{z_{X,11}\neq 0}$, we note that the anomalous dimensions involve $z_{X,1(-1)}$ and $z_{\lambda,1(-1)}$, that is order $\epsilon^1$ contributions to the renormalization factors. This, in turn, can be traced back to the fact that the one-loop anomalous dimensions that appear in Eq.~(\ref{eq:gt}) are multiplied by $1/\epsilon$ and, therefore, need to be expanded up to order $\epsilon^1$, contrary to the previous section where they were expanded up to order $\epsilon^0$ only. That the terms with $z_{X,1(-1)}$ and $z_{\lambda,1(-1)}$ are not present in the case $z_{X,11}=0$ is also visible in Eq.~(\ref{eq:gt}) since the just mentioned $1/\epsilon$ terms are not present.\\ 
\end{widetext}

\subsection{$\Lambda$-independence}
The formula (\ref{eq:gX_final}) and its simplified versions (\ref{eq:gX_final1}) and (\ref{eq:gX_final2}) are the ones we use in our implementation of the RG in Sec.~\ref{sec:RG}. We still need to clarify two questions however. 

First, the expression (\ref{eq:gX_final}) was derived assuming a $\mu$-independent $\Lambda$ and one is left wondering what would happen with a $\mu$-dependent $\Lambda$ (such as the standard choice $\smash{\Lambda=\mu}$). We will show that one obtains exactly the same expressions for the anomalous dimension $\gamma_X$, via a lengthier procedure however. Second, even though the expression (\ref{eq:gX_final}) does not depend explicitly on $\Lambda$, it could still depend implicitly on $\Lambda$ via the dependence of the factors $z_{X,ab}$. We will show that this is not so: the $\Lambda$-dependence cancels identically when the various $z_{X,ab}$ are combined into Eq.~(\ref{eq:gX_final}). 

A key remark in demystifying these two questions is that the only source of $\Lambda$-dependence in the renormalization factors appears via the $\epsilon$-expansion of $\Lambda^{2\epsilon}\lambda$ (since the scale $\Lambda$ is introduced as a rescaling of the coupling in the first place). In practice this means that, if the renormalization factors are written as
\beq
Z_X=1+\sum_{a\geq 1} \left(\frac{\lambda}{\epsilon}\right)^a z_{X,a}\,,
\eeq
one should have $\partial(\Lambda^{-2a\epsilon}z_{X,a})/\partial\ln\Lambda=0$, that is
\beq
\frac{\partial z_{X,a}}{\partial\ln\Lambda}-2a\epsilon z_{X,a}=0\,.\label{eq:La0}
\eeq
Writing each $z_{X,a}$ as
\beq
z_{X,a}=\sum_{b\leq a} z_{X,ab}\,\epsilon^{a-b}\,,
\eeq
the constraint (\ref{eq:La0}) can be rewritten as
\beq\label{eq:great}
\frac{\partial z_{X,ab}}{\partial\ln\Lambda}=2a z_{X,a(b+1)}\,,\label{eq:La}
\eeq
for $b<a$, and
\beq
\frac{\partial z_{X,aa}}{\partial\ln\Lambda}=0\,,
\eeq
this later result being totally trivial since the $z_{X,aa}$ are expected to be pure constants, independent of the considered renormalization scheme.

\subsubsection{Explicit $\Lambda$-dependence}
Keeping these remarks in mind, let us now re-derive the one-loop anomalous dimensions using a $\mu$-dependent $\Lambda$. There are two main differences with respect to the calculation that used a $\mu$-independent $\Lambda$. First, there is a new source of $\mu$-dependence in the renormalization factors, via $\Lambda$. This leads to the expression
\beq
\gamma_X & = & \frac{1}{Z_X}\left(\frac{\lambda}{\epsilon}\frac{\partial z_{X,1}}{\partial\ln\mu}+\frac{\lambda}{\epsilon}\frac{\partial z_{X,1}}{\partial\ln\Lambda}\frac{d\ln\Lambda}{d\ln\mu}\right.\nonumber\\
& & \left.+\,\frac{\beta_{\lambda}}{\lambda}\frac{\lambda}{\epsilon}z_{X,1}+\sum_i\frac{\beta_{m^2_i}}{m^2_i}\frac{\lambda}{\epsilon}\frac{\partial z_{X,1}}{\partial\ln m^2_i}\right),\label{eq:gL}
\eeq
where we note the presence of a new term proportional to $d\ln\Lambda/d\ln\mu$ as compared to (\ref{eq:g1}). Second, there is an additional term in the relation between the beta function and the anomalous dimension for $\lambda$, see (\ref{eq:beta_gamma}):
\beq
0=2\epsilon\frac{d\ln\Lambda}{d\ln\mu}+\gamma_{\lambda}+\frac{\beta_{\lambda}}{\lambda}\,.\label{eq:beta_gamma_p}
\eeq
When expanding the anomalous dimension (\ref{eq:gL}) up to order $\lambda$, this term cannot be neglected unlike $\gamma_{\lambda}$ because 1) it is of one order less in $\lambda$ as compared to $\gamma_{\lambda}$ and therefore produces a new order $\lambda$ contribution, and 2) this new contribution survives the continuum limit since it has the form $\epsilon\times 1/\epsilon$. One eventually arrives at
\beq
\gamma_X=\frac{\lambda}{\epsilon}\frac{\partial z_{X,1}}{\partial\ln\mu}+\lambda\left(\frac{1}{\epsilon}\frac{\partial z_{X,1}}{\partial\ln\Lambda}-2z_{X,1}\right)\frac{d\ln\Lambda}{d\ln\mu}\,.\label{eq:gL1} ~~~
\eeq
A similar but lengthier calculation at two-loop order leads to (\ref{eq:gt}) supplemented with the term
\beq
& &\left[\left(\lambda-\frac{\lambda^2}{\epsilon}z_{X,1}\right)\left(\frac{1}{\epsilon}\frac{\partial z_{X,1}}{\partial\ln\Lambda}-2z_{X,1}\right)\right.\nonumber\\
& & \hspace{1.8cm}\left.+\,\frac{\lambda^2}{\epsilon}\left(\frac{1}{\epsilon}\frac{\partial z_{X,2}}{\partial\ln\Lambda}-4z_{X,2}\right)\right]\frac{d\ln\Lambda}{d\ln\mu}\,.
\eeq
Owing to Eq.~(\ref{eq:La0}), it is easy to see that all these extra terms that one generates when evaluating the anomalous dimension with a $\mu$-dependent $\Lambda$ eventually cancel. As announced above, the final expression for the anomalous dimension in terms of the factors $z_{X,a}$ does not depend on the particular choice of $\Lambda$, and the fastest way to arrive at the result (avoiding unnecessary cancellations) is to use a $\mu$-independent $\Lambda$.

\subsubsection{Implicit $\Lambda$-dependence}
So far we have shown that the expressions (\ref{eq:gamma_X}) and (\ref{eq:gt}) have no explicit dependence on $\Lambda$. Obviously, this conclusion extends to (\ref{eq:g1_final}) and (\ref{eq:gX_final}) which are nothing but the order $\epsilon^0$ truncated versions of these expressions. However, there could still be an implicit dependence with respect to $\Lambda$ via the factors $z_{X,ab}$. We now show that this is not the case. 

Consider for instance (\ref{eq:g1_final}) and take a $\partial/\partial\ln\Lambda$ derivative. Owing to Eq.~(\ref{eq:great}), we have
\beq
\frac{\partial\gamma_X}{\partial\ln\Lambda}=\lambda\frac{\partial^2z_{X,10}}{\partial\ln\mu\partial\ln\Lambda}=2\lambda\frac{\partial z_{X,11}}{\partial\ln\mu}\,,
\eeq
which vanishes since $z_{X,11}$ is a pure constant.

A similar conclusion can be reached starting from the two-loop expression (\ref{eq:gX_final}) and exploiting (\ref{eq:great}). Focusing on the terms inside the bracket multiplying $\lambda^2$, we find
\begin{widetext}
\beq
\frac{\partial}{\partial\ln\Lambda}\Bigg(\dots\Bigg)_{\lambda^2} & = & 4\frac{\partial z_{X,21}}{\partial\ln\mu}-2\left(\frac{\partial z_{X,11}}{\partial\ln\mu}+\frac{\partial z_{\lambda,11}}{\partial\ln\mu}\right)z_{X,10}-2\left(\frac{\partial z_{X,10}}{\partial\ln\mu}+\frac{\partial z_{\lambda,10}}{\partial\ln\mu}\right)z_{X,11}\nonumber\\
& & -\,2\left(\frac{\partial z_{X,10}}{\partial\ln\mu}+\frac{\partial z_{\lambda,10}}{\partial\ln\mu}\right)z_{X,11}-\,\left(\frac{\partial z_{X,1(-1)}}{\partial\ln\mu}+\frac{\partial z_{\lambda,1(-1)}}{\partial\ln\mu}\right)\frac{\partial z_{X,11}}{\partial\ln\Lambda}\nonumber\\
& & -\,2\sum_i\frac{\partial z_{m^2_i,11}}{\partial\ln \mu}\frac{\partial z_{X,10}}{\partial\ln m^2_i}-2\sum_i\frac{\partial z_{m^2_i,10}}{\partial\ln \mu}\frac{\partial z_{X,11}}{\partial\ln m^2_i}=4\left(\frac{\partial z_{X,21}}{\partial\ln\mu}-\left(\frac{\partial z_{X,10}}{\partial\ln\mu}+\frac{\partial z_{\lambda,10}}{\partial\ln\mu}\right)z_{X,11}\right)=0\,,\nonumber\\
\eeq
where we have again used that $z_{X,11}$ is a pure constant, as well as the second constraint in (\ref{eq:id2}).\\
\end{widetext}

We mention that, contrary to the absence of explicit $\Lambda$-dependence, the absence of implicit $\Lambda$-dependence applies only to the anomalous dimensions in the continuum limit $\epsilon\to 0$. For instance, the one-loop anomalous dimension to order $\epsilon^1$
\beq
\gamma_X=\lambda\frac{\partial z_{X,10}}{\partial\ln\mu}+\lambda\frac{\partial z_{X,1(-1)}}{\partial\ln\mu}\,,
\eeq
depends implicitly on $\Lambda$ since one has
\beq
\frac{\partial\gamma_X}{\partial\ln\Lambda}=\lambda\frac{\partial^2z_{X,1(-1)}}{\partial\ln\mu\partial\ln\Lambda}=2\lambda\frac{\partial z_{X,10}}{\partial\ln\mu}\,,\label{eq:Ldep}
\eeq
which is usually not zero. As we already discussed above, this $\Lambda$-dependent, order $\epsilon^1$ one-loop anomalous dimension is crucial for the correct evaluation of the order $\epsilon^0$ two-loop anomalous dimension since it is multiplied by a $1/\epsilon$ factor in Eq.~(\ref{eq:gt}). In this case, however, the $\Lambda$-dependence (\ref{eq:Ldep}) gets cancelled by other $\Lambda$-dependent terms in the $\epsilon^0$ order two-loop anomalous dimension, thus ensuring the $\Lambda$-independence of the latter.

\subsection{Non-renormalization theorems}
The formula for the two-loop anomalous dimensions that we have derived above is general. It may happen, as in the model considered in this work, that some of the renormalization factors obey a non-renormalization theorem stating that their product $\prod_i Z_{X_i}$ is finite and allowing one to consider a scheme where this product is set equal to $1$. This, in turn, implies the relation $\sum_i \gamma_{X_i}=0$ between the corresponding anomalous dimensions.

Let us here check that the general formula (\ref{eq:inter}) is compatible with this expectation. The non-renormalization theorem implies
\beq
0 & = & \sum_i z_{X_i,1}\,,\label{eq:7}\\
0 & = & 2\sum_i z_{X_i,2}+\sum_{i\neq j} z_{X_i,1}z_{X_j,1}\,.\label{eq:8}
\eeq
Owing to (\ref{eq:7}),  it is trivial to check that the terms of (\ref{eq:inter}) that are linear in $z_{X,1}$ cancel in the sum $\sum_i \gamma_{X_i}$. To check that the remaining terms cancel as well, we use (\ref{eq:7}) in order to rewrite (\ref{eq:8}) as
\beq
0=2\sum_i z_{X_i,2}-\sum_i z_{X_i,1}^2\,.
\eeq
Then, the remaining terms in (\ref{eq:inter}) are proportional to
\beq
& &\sum_i \left(\frac{\partial z_{X_i,2}}{\partial\ln\mu}-\frac{\partial z_{X_i,1}}{\partial\ln\mu}z_{X_i,1}\right)\nonumber\\
& & \hspace{0.5cm}=\,\frac{1}{2}\frac{\partial}{\partial\ln\mu}\sum_i(2z_{X_i,2}- z_{X_i,1}^2)=0\,.
\eeq

\section{Minimal subtraction}\label{sec:MS}
Up until now, we have restricted our attention to renormalization schemes associated with renormalization conditions. In this section, for completeness, we revisit the minimal subtraction scheme which relies, not on renormalization conditions, but rather on the strict absorption of $1/\epsilon$ poles in the renormalization factors. We will show that, despite appearances, this scheme fits the general discussion of the previous section. 

In the minimal subtraction scheme, renormalization factors contain purely divergent terms in the limit $\epsilon\to 0$ (whose pre-factors are pure constants), and do not involve any finite part. At two-loop order for instance, we have
\beq
Z_X=1+\lambda\frac{z^{\mbox{\tiny $\overline{MS}$}}_{X,1}}{\epsilon}+\lambda^2\frac{z^{\mbox{\tiny $\overline{MS}$}}_{X,2}}{\epsilon^2}\,,\label{eq:zms}
\eeq
with $\smash{z^{\mbox{\tiny $\overline{MS}$}}_{X,1}=z_{X,11}}$ and $\smash{z^{\mbox{\tiny $\overline{MS}$}}_{X,2}=z_{X,22}+\epsilon z^{\mbox{\tiny $\overline{MS}$}}_{21}}$,\footnote{The values of $z_{X,11}$ and $z_{X,22}$ are the same as in the previous section since they are scheme independent. On the other hand, $z^{\mbox{\tiny $\overline{MS}$}}_{X,10}$, $z^{\mbox{\tiny $\overline{MS}$}}_{X,1(-1)}$, $z^{\mbox{\tiny $\overline{MS}$}}_{X,21}$ and $z^{\mbox{\tiny $\overline{MS}$}}_{X,20}$ have no reason to be the same as those in the previous section.} and thus $\smash{z^{\mbox{\tiny $\overline{MS}$}}_{X,10}=z^{\mbox{\tiny $\overline{MS}$}}_{X,1(-1)}=z^{\mbox{\tiny $\overline{MS}$}}_{X,20}=0}$.  Na\"\i vely, it seems that one cannot use the formulas (\ref{eq:g1_final}) and (\ref{eq:gX_final}) for these would give simply $0$. Moreover, it seems that there is no point in distinguishing between a renormalization scale $\mu$ and a regulating scale $\Lambda$ as we did above since there are no renormalization conditions to introduce the renormalization scale $\mu$ in the first place. 

On the other hand, the only scale $\mu$ that is introduced in minimal subtraction is the scale that makes the bare coupling dimensionless. As we have already mentioned, this is a regulating scale a priori, which has nothing to do with renormalization (denoting it as $\mu$ is not enough to qualify it as a renormalizaiton scale) and it is not clear how such a scale could control the renormalization group flow. 

In this section, we first derive the minimal subtraction anomalous dimensions in the standard way, without paying much attention to these considerations. We then revisit the same calculations using a point of view more in line with the general discussion of the previous section. While making the minimal subtraction scheme fit the general picture, this point of view clarifies the true source of $\mu$-dependence in this scheme and makes the determination of anomalous dimensions simpler and compatible with the formulas (\ref{eq:g1_final}) and (\ref{eq:gX_final}).

\subsection{Standard derivation}
From Eq.~(\ref{eq:zms}) at one-loop order, because the factors $z^{\mbox{\tiny $\overline{MS}$}}_{X,ab}$ are constants and because the only source for $\mu$-dependence is $\lambda$, we find
\beq
\gamma_X=\frac{\beta_{\lambda}/\lambda}{Z_X}\lambda\frac{z_{X,11}}{\epsilon}\,.\label{eq:gms}
\eeq
According to Eq.~(\ref{eq:beta_gamma_p}) with $\Lambda=\mu$, $\beta_{\lambda}/\lambda$ starts at order $\lambda^0$ with the contribution $-2\epsilon$. To obtain the anomalous dimension at order $\lambda$, we just need to keep this leading contribution to $\beta_{\lambda}/\lambda$ and replace $Z_X$ by $1$ in the denominator of Eq.~(\ref{eq:gms}). One finds eventually
\beq
\gamma_X=-2z_{X,11}\lambda\,,\label{eq:gmsf}
\eeq
where we note that the $\epsilon$ coming from $\beta_{\lambda}/\lambda$ has combined with the $1/\epsilon$ in (\ref{eq:gms}) to produce an order $\epsilon^0$ anomalous dimension.

One can proceed similarly at two-loop order. Starting from Eq.~(\ref{eq:zms}), one finds
\beq
\gamma_X=\frac{\beta_{\lambda}/\lambda}{Z_X}\left(\lambda\frac{z_{X,11}}{\epsilon}+2\lambda^2\frac{z_{X,22}+\epsilon z_{X,21}}{\epsilon^2}\right).
\eeq
This time, $\beta_{\lambda}/\lambda$ (as well as $Z_X$) needs to be expanded to order $\lambda$. This includes the contribution $-2\epsilon$ but also $\gamma_{\lambda}$ as given by Eq.~(\ref{eq:gmsf}) with $X=\lambda$. One finds
\beq
\gamma_X & = & -4\lambda^2\frac{z_{X,22}-z_{X,11}(z_{X,11}+z_{\lambda,11})/2}{\epsilon}\nonumber\\
& & -(2z_{X,11}\lambda+4z^{\mbox{\tiny $\overline{MS}$}}_{X,21}\lambda^2)\,.
\eeq
The finiteness of the beta function imposes that
\beq
z_{X,22}=\frac{1}{2}z_{X,11}(z_{X,11}+z_{\lambda,11})\,,\label{eq:idms}
\eeq
and we finally arrive at
\beq
\gamma_X=-\big(2z_{X,11}\lambda+4z^{\mbox{\tiny $\overline{MS}$}}_{X,21}\lambda^2\big)\,.\label{eq:gmsf2}
\eeq

\subsection{Connecting to the general discussion}
Let us now re-derive these results with a slightly different perspective that makes the minimal subtraction fit the general discussion. In particular, Eqs.~(\ref{eq:gmsf}) and (\ref{eq:gmsf2}) will appear as particular cases of Eqs.~(\ref{eq:g1_final}) and (\ref{eq:gX_final}).

Consider a slight generalization of the minimal subtraction scheme, which we refer to as $\Lambda\overline{MS}$, defined by the renormalization factors
\beq
Z^{\mbox{\tiny $\Lambda\overline{MS}$}}_X=1+\lambda\frac{z^{\mbox{\tiny $\Lambda\overline{MS}$}}_{X,1}}{\epsilon}+\lambda^2\frac{z^{\mbox{\tiny $\Lambda\overline{MS}$}}_{X,2}}{\epsilon^2}\,,\label{eq:lzms}
\eeq
with
\beq
z_{X,a}^{\mbox{\tiny $\Lambda\overline{MS}$}}=\left(\frac{\Lambda}{\mu}\right)^{2a\epsilon}z_{X,a}^{\mbox{\tiny $\overline{MS}$}}\,,\label{eq:rel}
\eeq
or, equivalently
\beq
z_{X,aa}^{\Lambda\overline{MS}} & = & z_{X,aa}^{\overline{MS}}\,,\\
z_{X,a(a-1)}^{\Lambda\overline{MS}} & = & z_{X,a(a-1)}^{\overline{MS}}+2az_{X,aa}^{\overline{MS}}\ln\frac{\Lambda}{\mu}\,,\\
& \dots & \nonumber
\eeq
where the dots represent $z_{X,ab}$ for $\smash{b<a-1}$. In fact, this defines a family of schemes parametrized by $\Lambda$, of which the standard minimal subtraction corresponds to the choice $\Lambda=\mu$. The scale $\Lambda$ plays the role of the regulating scale, while the scale $\mu$ is the renormalization scale and the flow with respect to this latter scale needs to be determined for $m^2_B=Z_{m^2}m^2$ and $\lambda_B=\Lambda^{2\epsilon}Z_{\lambda}\lambda$ fixed. In particular the anomalous dimensions should again be independent of the choice of $\Lambda$ in the continuum limit (we will check this explicitly below), thus providing an alternative way to obtain the anomalous dimensions in minimal subtraction. The benefit of this approach is that, upon the appropriate introduction of $Z_{\lambda}$ factors, the only way the coupling appears is via the combination $\Lambda^{2\epsilon}Z_{\lambda}\lambda$. Therefore one never needs to consider $\beta_\lambda/\lambda$ and cancellations of the type $\epsilon\times 1/\epsilon$.

At one-loop order for instance, up to higher order corrections, one writes
\beq
Z_X=1 & + & Z_{\lambda}\lambda\left(\frac{\Lambda}{\mu}\right)^{2\epsilon}\frac{z_{X,11}}{\epsilon}\,.
\eeq
The only dependence on $\mu$ is via the factor $\mu^{-2\epsilon}$ and one obtains immediately
\beq
\gamma_X=-2z_{X,11} \frac{Z_{\lambda}}{Z_X}\lambda\left(\frac{\Lambda}{\mu}\right)^{2\epsilon}=-2z_{X,11}\lambda\left(\frac{\Lambda}{\mu}\right)^{2\epsilon}\!\!, ~~~
\eeq
which boils down to (\ref{eq:gmsf}) in the continuum limit. Similarly, at two-loop order, one would write
\beq\label{eq:derivation}
Z_X & = & 1+\lambda\left(\frac{\Lambda}{\mu}\right)^{2\epsilon}\frac{z_{X,11}}{\epsilon}\nonumber\\
& + & \lambda^2\left(\frac{\Lambda}{\mu}\right)^{4\epsilon}\left(\frac{z_{X,22}}{\epsilon^2}+\frac{z_{X,21}}{\epsilon}\right)\nonumber\\
& = & 1+Z_{\lambda}\lambda\left(\frac{\Lambda}{\mu}\right)^{2\epsilon}\frac{z_{X,11}}{\epsilon}\nonumber\\
& + & Z_{\lambda}^2\lambda^2\left(\frac{\Lambda}{\mu}\right)^{4\epsilon}\left(\frac{z_{X,22}-z_{X,11}z_{\lambda,11}}{\epsilon^2}+\frac{z_{X,21}}{\epsilon}\right).\nonumber\\
\eeq
Again, the only $\mu$-dependence is via the factors $\mu^{-2a\epsilon}$ and one recovers (\ref{eq:gmsf2}) together with the constraint (\ref{eq:idms}). In fact, with this approach, it is not difficult to see that the minimal subtraction anomalous dimension is given at any order by
\beq
\gamma_X=-\sum_{a\geq 1} 2^a z_{X,a1}^{\mbox{\tiny $\overline{MS}$}} g^{2a}\,.
\eeq

We mention also that the $\Lambda\overline{MS}$-scheme is no different from the generic schemes considered in the previous section and, as such, the expressions (\ref{eq:gmsf}) and (\ref{eq:gmsf2}) should be compatible with (\ref{eq:g1_final}) and (\ref{eq:gX_final}). This is easily seen after noting that, from (\ref{eq:rel}), the $\mu$-dependence of the factors $z^{\mbox{\tiny $\Lambda\overline{MS}$}}_{X,a}$ is controlled by the same equation that controls the $\Lambda$-dependence, see Eq.~(\ref{eq:La0}), up to a sign:
\beq
\frac{\partial z^{\mbox{\tiny $\Lambda\overline{MS}$}}_{X,a}}{\partial\ln\mu}+2a\epsilon z^{\mbox{\tiny $\Lambda\overline{MS}$}}_{X,a}=0\,.
\eeq
This in turn implies
\beq
\frac{\partial z^{\mbox{\tiny $\Lambda\overline{MS}$}}_{X,ab}}{\partial\ln\mu}=-2a z^{\mbox{\tiny $\Lambda\overline{MS}$}}_{X,a(b+1)}\,.
\eeq
Using these identities, it is easily seen that, in the minimal subtraction scheme, (\ref{eq:gmsf}) and (\ref{eq:gmsf2}) are compatible with (\ref{eq:g1_final}) and (\ref{eq:gX_final}). Moreover, the identity (\ref{eq:id2}) is nothing but a rewriting of (\ref{eq:idms}).

The present discussion also clarifies the true source of $\mu$-dependence within the standard minimal subtraction scheme. By revisiting the derivation (\ref{eq:derivation}) with $\Lambda=\mu$, we see that the scale $\mu$ that appears in the numerator of the factor $(\mu/\mu)^\epsilon$, and that stems from the rescaling of the coupling, has nothing to do with the RG running. The running originates instead from the scale $\mu$ that appears in the denominator of the factor $(\mu/\mu)^\epsilon$. Contrary to the former which is nothing but a regulating scale needed to make the coupling dimensionless in dimensional regularization, this second occurrence of $\mu$ is the renormalization scale. It is introduced here not by renormalization conditions, but rather by the minimal subtraction requirement that the renormalization factors do not depend explicitly on any scale and in particular on the regulating scale.

\subsection{Integrating the one-loop flow}
For completeness, let us recall here how the minimal subtraction beta functions and anomalous dimensions are integrated out at one- and two-loop orders. At one-loop order, we have
\beq
\frac{\beta_{\lambda}}{\lambda}=-\gamma_{\lambda}=2z_{\lambda,11}\lambda\,.
\eeq
This is rewritten as
\beq
\frac{d\lambda}{\lambda^2} & = & z_{\lambda,11}\,d\ln\mu^2\,,
\eeq
which integrates to
\beq
\lambda(\mu)=\frac{\lambda_0}{1-z_{\lambda,11}\lambda_0\ln\frac{\mu^2}{\mu_0^2}}=\frac{1}{-z_{\lambda,11}\ln\frac{\mu^2}{\Lambda^2_{\rm LP}}}
\eeq
with
\beq
\Lambda^2_{\rm LP}=\mu_0^2\,\exp\left(\frac{1}{z_{\lambda,11}\lambda_0}\right).
\eeq
The scale $\Lambda_{\rm LP}$ is the Landau pole and, if $z_{\lambda,11}$ is negative, the flow makes sense only for $\mu>\Lambda_{\rm LP}$. We restrict to this case from now on.

Next, we write
\beq
\frac{\beta_m^2/m^2}{\beta_\lambda/\lambda}=\frac{\gamma_{m^2}}{\gamma_{\lambda}}=\frac{z_{m^2,11}}{z_{\lambda,11}}\,,
\eeq
which is nothing but
\beq
d\ln m^2=\frac{z_{m^2,11}}{z_{\lambda,11}}\,d\ln \lambda=d\ln (\lambda)^{\frac{z_{m^2,11}}{z_{\lambda,11}}}\,.
\eeq
It follows that
\beq
\frac{m^2}{m_0^2}=\left(\frac{\lambda}{\lambda_0}\right)^{\frac{z_{m^2,11}}{z_{\lambda,11}}}\,.\label{eq:C34}
\eeq
Finally, we write
\beq
\frac{\gamma_\varphi}{\beta_{\lambda}/\lambda}=-\frac{\gamma_\varphi}{\gamma_{\lambda}}=-\frac{z_{\varphi,11}}{z_{\lambda,11}}\,,
\eeq
which is nothing but
\beq
d\ln z_i=-\frac{z_{\varphi_i,11}}{z_{\lambda,11}}d\ln \lambda=d\ln (\lambda)^{-\frac{z_{\varphi_i,11}}{z_{\lambda,11}}}\,,
\eeq
where $z_i$ is the rescaling factor (\ref{eq:rescaling}). It follows that
\beq
z_i(\mu,\mu_0)=\left(\frac{\lambda}{\lambda_0}\right)^{-\frac{z_{\varphi,11}}{z_{\lambda,11}}}\,.\label{eq:C37}
\eeq
\vglue4mm

\subsection{Integrating the two-loop flow}
We have
\beq
\frac{\beta_{\lambda}}{\lambda}=-\gamma_{\lambda}=2z_{\lambda,11}\lambda+4z_{\lambda,21}\lambda^2\,,\label{eq:bginf2}
\eeq
which is nothing but
\beq
\frac{d\lambda}{\lambda^2\left(1+2\frac{z_{\lambda,21}}{z_{\lambda,11}}\lambda\right)}=z_{\lambda,11}d\ln\mu^2\,.
\eeq
We can rewrite this conveniently as
\beq
z_{\lambda,11}d\ln\mu^2 & = & -\frac{\frac{1}{\lambda}}{\left(\frac{1}{\lambda}+2\frac{z_{\lambda,21}}{z_{\lambda,11}}\right)}d\left(\frac{1}{\lambda}\right)\nonumber\\
& = & \left[-1+\frac{2\frac{z_{\lambda,21}}{z_{\lambda,11}}}{\frac{1}{\lambda}+2\frac{z_{\lambda,21}}{z_{\lambda,11}}}\right]d\left(\frac{1}{\lambda}\right),\label{eq:C40}
\eeq
which integrates to
\beq
z_{\lambda,11}\ln\frac{\mu^2}{\mu^2_0}=\frac{1}{\lambda_0}-\frac{1}{\lambda(\mu)}+2\frac{z_{\lambda,21}}{z_{\lambda,11}}\ln\frac{\frac{1}{\lambda(\mu)}+2\frac{z_{\lambda,21}}{z_{\lambda,11}}}{\frac{1}{\lambda_0}+2\frac{z_{\lambda,21}}{z_{\lambda,11}}}\,,\label{eq:C41}\nonumber\\
\eeq
and gives $\mu$ as a function of $\lambda$. We notice that, because the running of $g$ is logarithmic, the second term is sub-leading in the UV and we recover the one-loop running. We can estimate the correction to the one-loop behavior by replacing $\lambda(\mu)$ by $\lambda_{\rm 1loop}(\mu)$ in the logarithm. We obtain
\beq
\frac{1}{\lambda(\mu)}=\frac{1}{\lambda_0}-z_{\lambda,11}\ln\frac{\mu^2}{\mu^2_0}+2\frac{z_{\lambda,21}}{z_{\lambda,11}}\ln\frac{-z_{\lambda,11}\ln\frac{\mu^2}{\mu^2_0}}{\frac{1}{\lambda_0}+2\frac{z_{\lambda,21}}{z_{\lambda,11}}}\,.\nonumber\\
\eeq
We note that
\beq
\lambda_{\rm 2loop}-\lambda_{\rm 1loop} & = & -2\lambda_{\rm 1loop}\lambda_{\rm 2loop}\frac{z_{\lambda,21}}{z_{\lambda,11}}\ln\frac{-z_{\lambda,11}\ln\frac{\mu^2}{\mu^2_0}}{\frac{1}{\lambda_0}+2\frac{z_{\lambda,21}}{z_{\lambda,11}}}\,,\nonumber\\
\eeq
and thus the corrections are not that small. We mention also that (\ref{eq:C41}) provides corrections to the Landau pole defined by the scale at which $1/\lambda(\mu)$ vanishes. One finds
\beq
\Lambda^2_{\rm LP}=\mu_0^2\left(\frac{1}{1+\frac{z_{\lambda,11}}{2z_{\lambda,21}\lambda_0}}\right)^{2\frac{z_{\lambda,21}}{z^2_{\lambda,11}}}\!\!\!\exp\left(\frac{1}{z_{\lambda,11} \lambda_0}\right)\,,
\eeq
in terms of which we have
\beq
z_{\lambda,11}\ln\frac{\mu^2}{\Lambda^2_{\rm LP}}=-\frac{1}{\lambda(\mu)}+2\frac{z_{\lambda,21}}{z_{\lambda,11}}\ln\left(1+\frac{z_{\lambda,11}}{2z_{\lambda,21}}\frac{1}{\lambda(\mu)}\right)\,.\nonumber\\
\eeq

\vglue4mm

Next, we write
\beq
\frac{\beta_{m^2}}{m^2}=-\gamma_{m^2}=2z_{m^2,11}\lambda+4z_{m^2,21}\lambda^2\,,\label{eq:bminf2}
\eeq
which is nothing but
\beq
d\ln m^2=\Big(z_{m^2,11}\lambda+2z_{m^2,21}\lambda^2\Big)d\ln\mu^2\,.
\eeq
\begin{widetext}

Upon using (\ref{eq:C40}), this is rewritten as
\beq
z_{\lambda,11}d\ln m^2 & = & -\frac{z_{m^2,11}+2z_{m^2,21}\lambda}{\frac{1}{\lambda}+2\frac{z_{\lambda,21}}{z_{\lambda,11}}}d\left(\frac{1}{\lambda}\right)=-\left[\frac{z_{m^2,11}}{\frac{1}{\lambda}+2\frac{z_{\lambda,21}}{z_{\lambda,11}}}+\frac{2z_{m^2,21}}{\frac{1}{\lambda}\left(\frac{1}{\lambda}+2\frac{z_{\lambda,21}}{z_{\lambda,11}}\right)}\right]d\left(\frac{1}{\lambda}\right)\nonumber\\
& = & -\left[\frac{z_{m^2,11}}{\frac{1}{\lambda}+2\frac{z_{\lambda,21}}{z_{\lambda,11}}}+\frac{z_{\lambda,11}}{z_{\lambda,21}}z_{m^2,21}\left(\frac{1}{\frac{1}{\lambda}}-\frac{1}{\frac{1}{\lambda}+2\frac{z_{\lambda,21}}{z_{\lambda,11}}}\right)\right]d\left(\frac{1}{\lambda}\right)\,,
\eeq
which is easily integrated to
\beq
\ln\frac{m^2}{m_0^2} & = & \frac{z_{m^2,21}}{z_{\lambda,21}}\ln\frac{\lambda}{\lambda_0}+\left(\frac{z_{m^2,21}}{z_{\lambda,21}}-\frac{z_{m^2,11}}{z_{\lambda,11}}\right)\ln\frac{\frac{1}{\lambda(\mu)}+2\frac{z_{\lambda,21}}{z_{\lambda,11}}}{\frac{1}{\lambda_0}+2\frac{z_{\lambda,21}}{z_{\lambda,11}}}\nonumber\\ 
& = & \frac{z_{m^2,11}}{z_{\lambda,11}}\ln\frac{\lambda}{\lambda_0}+\left(\frac{z_{m^2,21}}{z_{\lambda,21}}-\frac{z_{m^2,11}}{z_{\lambda,11}}\right)\ln\frac{1+2\frac{z_{\lambda,21}}{z_{\lambda,11}}\lambda}{1+2\frac{z_{\lambda,21}}{z_{\lambda,11}}\lambda_0}\,,
\eeq
or
\beq
\frac{m^2}{m_0^2} & = & \left(\frac{\lambda}{\lambda_0}\right)^{\frac{z_{m^2,11}}{z_{\lambda,11}}}\left(\frac{1+2\frac{z_{\lambda,21}}{z_{\lambda,11}}\lambda}{1+2\frac{z_{\lambda,21}}{z_{\lambda,11}}\lambda_0}\right)^{\frac{z_{\lambda,11}z_{m^2,21}-z_{m^2,11}z_{\lambda,21}}{z_{\lambda,21}z_{\lambda,11}}}\,.\label{eq:48}
\eeq
The second factor approaches $1$ in the deep UV and we recover the one-loop running.\\

Finally, if we solve formally (\ref{eq:bginf2}) and (\ref{eq:bminf2}) for $\lambda$ and $\lambda^2$, we find
\beq
\lambda=\frac{\frac{\beta_{\lambda}}{\lambda}z_{m^2,21}-\frac{\beta_{m^2}}{m^2}z_{\lambda,21}}{2(z_{\lambda,11}z_{m^2,21}-z_{m^2,11}z_{\lambda,21})} \quad {\rm and} \quad \lambda^2=\frac{\frac{\beta_{m^2}}{m^2}z_{\lambda,11}-\frac{\beta_{\lambda}}{\lambda}z_{m^2,11}}{4(z_{\lambda,11}z_{m^2,21}-z_{m^2,11}z_{\lambda,21})}\,.
\eeq
\end{widetext}
Plugging this back into
\beq
\gamma_{\varphi_i}=2z_{\varphi_{i,11}}\lambda+4z_{\varphi_{i,21}}\lambda^2\,,
\eeq 
gives
\beq
\gamma_\varphi & = & -\frac{z_{\varphi,11}z_{m^2,21}-z_{m^2,11}z_{\varphi, 21}}{z_{\lambda,11}z_{m^2,21}-z_{m^2,11}z_{\lambda,21}}\frac{\beta_{\lambda}}{\lambda}\nonumber\\
& & -\,\frac{z_{\varphi,11}z_{\lambda,21}-z_{\lambda,11}z_{\varphi, 21}}{z_{m^2,11}z_{\lambda,21}-z_{\lambda,11}z_{m^2,21}}\frac{\beta_{m^2}}{m^2}
\eeq
from which it follows that
\beq
z_i(\mu,\mu_0) & = & \left(\frac{\lambda}{\lambda_0}\right)^{-\frac{z_{\varphi,11}z_{m^2,21}-z_{m^2,11}z_{\varphi, 21}}{z_{\lambda,11}z_{m^2,21}-z_{m^2,11}z_{\lambda,21}}}\nonumber\\
& & \times\,\left(\frac{m^2}{m_0^2}\right)^{-\frac{z_{\varphi,11}z_{\lambda,21}-z_{\lambda,11}z_{\varphi, 21}}{z_{m^2,11}z_{\lambda,21}-z_{\lambda,11}z_{m^2,21}}}\,.~~~\label{eq:C53}
\eeq
To recover the one-loop behavior, we notice that deep in the UV, Eq.~(\ref{eq:C34}) holds, and this leads to
\beq
z(\mu,\mu_0) & = & \left(\frac{\lambda}{\lambda_0}\right)^{-\frac{z_{\varphi,11}z_{m^2,21}-z_{m^2,11}z_{\varphi, 21}}{z_{\lambda,11}z_{m^2,21}-z_{m^2,11}z_{\lambda,21}}}\nonumber\\
& & \times\,\left(\frac{\lambda}{\lambda_0}\right)^{-\frac{z_{m^2,11}}{z_{\lambda,11}}\frac{z_{\varphi,11}z_{\lambda,21}-z_{\lambda,11}z_{\varphi, 21}}{z_{m^2,11}z_{\lambda,21}-z_{\lambda,11}z_{m^2,21}}}\,.\nonumber\\
\eeq
We notice that the terms proportional to $z_{\varphi,21}$ in the numerator of the exponent cancel and we are left with
\beq
z_i(\mu,\mu_0) & = & \left(\frac{\lambda}{\lambda_0}\right)^{-\frac{z_{\varphi,11}\left(z_{m^2,21}-\frac{z_{\lambda,21}}{z_{\lambda,11}}z_{m^2,11}\right)}{z_{\lambda,11}z_{m^2,21}-z_{m^2,11}z_{\lambda,21}}}\nonumber\\
& = & \left(\frac{\lambda}{\lambda_0}\right)^{-\frac{z_{\varphi,11}}{z_{\lambda,11}}}\,,
\eeq
in agreement with Eq.~(\ref{eq:C37}).

We mention that the previous derivation is not valid for a massless field and the formula (\ref{eq:C53}) is plagued by singularities (since $z_{m^2,ab}=0$). However, plugging (\ref{eq:48}) into (\ref{eq:C53}), we can combine the various powers of $\lambda/\lambda_0$ just as before and we arrive at
\beq
z_i(\mu,\mu_0)=\left(\frac{\lambda}{\lambda_0}\right)^{-\frac{z_{\varphi,11}}{z_{\lambda,11}}}\left(\frac{1+2\frac{z_{\lambda,21}}{z_{\lambda,11}}\lambda}{1+2\frac{z_{\lambda,21}}{z_{\lambda,11}}\lambda_0}\right)^{\frac{z_{\varphi, 11}}{z_{\lambda, 11}}-\frac{z_{\varphi, 21}}{z_{\lambda, 21}}}.\nonumber\\
\eeq
These formulas do not make any reference to the mass of the fields and apply, therefore, to a massless field as well.

\section{Asymptotic expansion in the UV}\label{app:UV_asymptotic}
In this section, we collect the next-to-leading order UV and IR asymptotic expansions of the various two anomalous dimensions as computed in the IR-safe scheme. The corresponding expansions for the beta functions for $\lambda$ and $m^2$ can be deduced from the non-renormalization theorems, whereas that for $M$ can be deduced directly from its relation to $\gamma_M$.  

In the UV, at next-to-leading order of the asymptotic expansion, we find for the gluon and ghost anomalous dimensions
\begin{widetext}
\begin{eqnarray}
\gamma_A & = & \lambda\left\{\left[-\frac{13}{3}+\left(\frac{65}{4}+\frac{3}{2}\ln\frac{\mu^2}{m^2}\right)\frac{m^2}{\mu^2}\right]+\frac{N_f}{N}\left[\frac{4}{3}-8\frac{M^2}{\mu^2}\right]\right\}\nonumber\\
& + & \lambda^2\left\{-\frac{85}{6}+\left(\frac{18343}{96}+\frac{\pi^2}{48}+\frac{171}{4}\zeta(3)-\frac{891}{16}S_2+\frac{205}{16}\ln\frac{\mu^2}{m^2}+\frac{35}{8}\ln^2\frac{\mu^2}{m^2}\right)\frac{m^2}{\mu^2}\right.\nonumber\\
& & \hspace{0.5cm} +\,\frac{N_f}{N}\left[\frac{17}{3}-\left(\frac{8}{3}+48\zeta(3)\right)\frac{m^2}{\mu^2}-\left(\frac{281}{3}+16\zeta(3)\right)\frac{M^2}{\mu^2}+2\left(\frac{m^2}{\mu^2}-2\left(1+\frac{M^2}{m^2}\right)\frac{M^2}{\mu^2}\right)\tilde I_{mMM}\right.\nonumber\\
& & \hspace{2.0cm} \left.+\,\left(2\ln\frac{\mu^2}{m^2}-2\ln\frac{\mu^2}{m^2}\ln\frac{\mu^2}{M^2}+\ln^2\frac{\mu^2}{M^2}\right)\frac{m^2}{\mu^2}-2\left(\ln\frac{\mu^2}{m^2}+2\ln\frac{\mu^2}{M^2}\right)\frac{M^2}{\mu^2}\right]\nonumber\\
& & \hspace{0.5cm} \left.+\,\frac{N_f}{N}\frac{C_F}{N}\left[4-\left(\frac{128}{3}-32\zeta(3)\right)\frac{m^2}{\mu^2}-48\frac{M^2}{\mu^2}\right]\right\}\,,\\
\gamma_c & = & \lambda\left\{-\frac{3}{2}-\left(\frac{3}{4}-\frac{3}{2}\ln\frac{\mu^2}{m^2}\right)\frac{m^2}{\mu^2}\right\}\nonumber\\
& + & \lambda^2\left\{-\frac{17}{4}+\left(-\frac{211}{8}+\frac{\pi^2}{48}+\frac{3}{4}\zeta(3)-\frac{891}{16}S_2+\frac{103}{8}\ln\frac{\mu^2}{m^2}+\frac{35}{8}\ln^2\frac{\mu^2}{m^2}\right)\frac{m^2}{\mu^2}\right.\nonumber\\
& & \hspace{0.5cm} +\,\frac{N_f}{N}\left[\frac{1}{2}+\frac{3}{2}\frac{m^2}{\mu^2}+11\frac{M^2}{\mu^2}+2\left(\frac{m^2}{\mu^2}-2\left(1+\frac{M^2}{m^2}\right)\frac{M^2}{\mu^2}\right)\tilde I_{mMM}\right.\nonumber\\
& & \hspace{2.0cm} \left.\left.+\,\left(\ln\frac{\mu^2}{m^2}-2\ln\frac{\mu^2}{m^2}\ln\frac{\mu^2}{M^2}+\ln^2\frac{\mu^2}{M^2}\right)\frac{m^2}{\mu^2}-2\left(\ln\frac{\mu^2}{m^2}+2\ln\frac{\mu^2}{M^2}\right)\frac{M^2}{\mu^2}\right]\right\}\,,
\eeq
with 
\beq
S_2\equiv\frac{4}{9\sqrt{3}}\,{\rm Im}\,{\rm Li}_2(e^{i\pi/3})\,,
\eeq 
and
\beq
\tilde I_{mMM}=-m\,{\rm Re}\left\{\frac{\sqrt{m^2-4M^2}}{m^2-4M^2}\left[\frac{\pi^2}{6}-\frac{1}{2}\ln^2\frac{M^2}{m^2}+\ln^2\left(\frac{1}{2}-\frac{\sqrt{m^2-4M^2}}{2m}\right)-2{\rm Li}_2\left(\frac{1}{2}-\frac{\sqrt{m^2-4M^2}}{2m}\right)\right]\right\}\!\,,\nonumber\\
\eeq
where ${\rm Li}_2$ denotes the dilogarithm function. It is easily checked that the term between square brackets in $\tilde I_{mMM}$ vanishes linearly as $m\to 2M$ and thus the above expressions for $\gamma_A$ and $\gamma_c$ are regular in this limit. 

In mass-independent schemes, the coupling beta function is two-loop universal,  whereas in mass-dependent schemes, such as the IR-safe scheme considered here, it is two-loop universal in the UV. Using $\beta_\lambda/\lambda=\gamma_A+2\gamma_c$, we have checked that we recover indeed the two-loop univeral behavior in the UV \cite{vanRitbergen:1997va}.\\

Similarly, for the quark anomalous dimensions, we find
\beq
\gamma_\psi & = & \lambda\frac{C_F}{N}\left(\frac{9}{2}-3\ln\frac{\mu^2}{m^2}\right)\frac{m^2}{\mu^2}\nonumber\\
& + & \lambda^2\frac{C_F}{N}\left\{\frac{25}{2}+\left(\frac{695}{8}-\frac{\pi^2}{24}-45\zeta(3)+\frac{891}{8}S_2-\frac{47}{2}\ln\frac{\mu^2}{m^2}-\frac{35}{4}\ln^2\frac{\mu^2}{m^2}\right)\frac{m^2}{\mu^2}-\Big(31-18\zeta(3)\Big)\frac{M^2}{\mu^2}\right.\nonumber\\
& & \hspace{1.0cm}+\,\frac{C_F}{N}\left[-3-8\Big(5-6\zeta(3)\Big)\frac{m^2}{\mu^2}+24\frac{M^2}{\mu^2}\right]\nonumber\\
& & \hspace{1.0cm}+\,\frac{N_f}{N}\left[-2-5\frac{m^2}{\mu^2}-18\frac{M^2}{\mu^2}-4\left(\frac{m^2}{\mu^2}-2\left(1+\frac{M^2}{m^2}\right)\frac{M^2}{\mu^2}\right)\tilde I_{mMM}\right.\nonumber\\
& & \hspace{2.0cm} \left.\left.-\,2\left(\ln\frac{\mu^2}{m^2}-2\ln\frac{\mu^2}{m^2}\ln\frac{\mu^2}{M^2}+\ln^2\frac{\mu^2}{M^2}\right)\frac{m^2}{\mu^2}+4\left(\ln\frac{\mu^2}{m^2}+2\ln\frac{\mu^2}{M^2}\right)\frac{M^2}{\mu^2}\right]\right\}\,,
\eeq
and
\beq
\gamma_M & = & \lambda\frac{C_F}{N}\left\{6-\left(\frac{9}{2}+3\ln\frac{\mu^2}{m^2}\right)\frac{m^2}{\mu^2}-6\frac{M^2}{\mu^2}\ln\frac{\mu^2}{M^2}\right\}\nonumber\\
& + & \lambda^2\frac{C_F}{N}\left\{\frac{67}{2}-\left(\frac{41}{2}+\frac{\pi^2}{24}+69\zeta(3)-\frac{891}{8}S_2+\frac{49}{4}\ln\frac{\mu^2}{m^2}+\frac{35}{4}\ln^2\frac{\mu^2}{m^2}\right)\frac{m^2}{\mu^2}+\left(3-36\zeta(3)-46\ln\frac{\mu^2}{M^2}\right)\frac{M^2}{\mu^2}\right.\nonumber\\
& & \hspace{1.0cm}+\,\frac{C_F}{N}\left[3-\left(\frac{35}{2}-48\zeta(3)\right)\frac{m^2}{\mu^2}+2\Big(7-12\zeta(3)\Big)\frac{M^2}{\mu^2}-72 \tilde{I}_{mMM}\frac{M^2}{\mu^2}\right.\nonumber\\
& & \hspace{5.0cm}\left.  -45\ln\frac{\mu^2}{m^2}\frac{m^2}{\mu^2}+18\left(3-2\ln\frac{\mu^2}{M^2}\right)\ln\frac{\mu^2}{M^2}\frac{M^2}{\mu^2}\right]\nonumber\\
& & \hspace{1.0cm}+\,\frac{N_f}{N}\left[-2+5\frac{m^2}{\mu^2}-10\frac{M^2}{\mu^2}-4\left(\frac{m^2}{\mu^2}-2\left(1+\frac{M^2}{m^2}\right)\frac{M^2}{\mu^2}\right)\tilde I_{mMM}\right.\nonumber\\
& & \hspace{2.0cm} \left.\left.-\,2\left(\ln\frac{\mu^2}{m^2}-2\ln\frac{\mu^2}{m^2}\ln\frac{\mu^2}{M^2}+\ln^2\frac{\mu^2}{M^2}\right)\frac{m^2}{\mu^2}+4\left(\ln\frac{\mu^2}{m^2}+3\ln\frac{\mu^2}{M^2}\right)\frac{M^2}{\mu^2}\right]\right\}\,.
\end{eqnarray}

\section{Asymptotic expansion in the IR}\label{app:IR_asymptotic}
In order to obtain the IR asymptotic expansion of the two-loop anomalous dimensions at next-to-leading order, we first checked that all the master integrals required to obtain the anomalous dimensions to order $\mu^4/m^4$ and $\mu^4/M^4$ are either known analytically or such that one can always root the external momentum through massive propagators. For this second type of master integrals, one can employ the strategy of Ref.~\cite{Davydychev:1992mt} that we briefly reviewed in Sec.~\ref{sec:IR}. For completeness, we here provide the resulting expansions.\\

In the case of $S_{abc}$, assuming $a\neq 0$, it is convenient to choose the loop momenta as follows
\beq\label{eq:S}
S_{abc}(k)=\int_p \int_q \frac{1}{(p+k)^2+a}\frac{1}{(q+p)^2+b}\frac{1}{q^2+c}\,.
\eeq
One can then expand the massive propagator carrying the external momentum $k$:
\beq
\frac{1}{(p+k)^2+a}=\sum_{n=0}^\infty (-1)^n\frac{(2(p\cdot k)+k^2)^n}{(p^2+a)^{n+1}}=\sum_{n=0}^\infty (-1)^n\sum_{\ell=0}^n\frac{n!}{\ell!(n-\ell)!}\frac{(2(p\cdot k))^\ell(k^2)^{n-\ell}}{(p^2+a)^{n+1}}\,.
\eeq
This yields
\beq
S_{abc}(k)=\sum_{n=0}^\infty (-1)^n\sum_{\ell=0}^n\frac{n!}{\ell!(n-\ell)!}\int_p \int_q \frac{(2p\cdot k)^\ell (k^2)^{n-\ell}}{(p^2+a)^{n+1}}\frac{1}{(q+p)^2+b}\frac{1}{q^2+c}\,.
\eeq
The $p$-integral vanishes for $\ell$ odd, whereas for $\ell$ even, one can use the formula 
\beq
\int\frac{d^dp}{(2\pi)^d}\,f(p^2)\,(2\,p\cdot k)^\ell=\frac{\ell!}{(\ell/2)!}\frac{(k^2)^{\ell/2}}{(d/2)_{\ell/2}}\int\frac{d^dp}{(2\pi)^d}\,f(p^2)\,(p^2)^{\ell/2}\,,\label{eq:Davy}
\eeq 
given in Ref.~\cite{Davydychev:1993pg}, where $(a)_n\equiv a(a+1)\cdots (a+n-1)$ is the Pochhammer symbol. One then arrives at
\beq
S_{abc}(k) & = & \sum_{n=0}^\infty (-1)^n\sum_{\ell=0}^{[n/2]}\frac{n!}{\ell!(n-2\ell)!}\frac{(k^2)^{n-\ell}}{(d/2)_\ell}\int_p \int_q \frac{(p^2)^\ell}{(p^2+a)^{n+1}}\frac{1}{(q+p)^2+b}\frac{1}{q^2+c}\nonumber\\
& = & \sum_{n=0}^\infty \sum_{\ell=0}^{[n/2]}\sum_{h=0}^\ell (-1)^{n+\ell-h}\frac{n!}{(n-2\ell)!h!(\ell-h)!}\frac{(k^2)^{n-\ell}}{(d/2)_\ell}a^{\ell-h}\,I_{(n+1-h)11}(a,b,c),\label{eq:Sexp}
\eeq
where we redefined $\ell\to 2\ell$ (since it is even) and $[n/2]$ denotes the integer part of $n/2$. 

We have thus expressed the IR expansion of $S_{abc}$ in terms of the integrals $I_{(n+1-h)11}$ which are essentially nothing but multiple derivative of $I_{111}(a,b,c)$ with respect to $a$. These multiple derivatives can be conveniently obtained by repeated use of Eq.~(\ref{eq:derI}).\footnote{One could wonder why it is not possible to simply take multiple derivative of the explicit expression for $I_{111}(a,b,c)$. Although possible this leads to cumbersome combinations of hypergeometric functions and their derivatives. It is much more convenient to first express the multiple derivatives algebraically in terms of $I_{111}(a,b,c)$ using Eq.~(\ref{eq:derI}) and only then do the substitution of $I_{111}(a,b,c)$ by its explicit expression.} We note that $\ell\leq n/2$ and thus the exponent of $k^2$ in (\ref{eq:Sexp}) is such that $n-\ell\geq n/2$. This implies that terms with $n>2p$ contribute to powers of $k^2$ with an exponent strictly larger than $p$. In other words, to obtain the expansion up to order $(k^2)^p$, it is enough to truncate the sum over $n$ up to and including $n=2p$.\\

In the case of $U_{abcd}(k)$, assuming $a\neq 0$, we write
\beq
U_{abcd}(k)=\int_p \int_q \frac{1}{(p+k)^2+a}\frac{1}{p^2+b}\frac{1}{q^2+c}\frac{1}{(q+p)^2+d}\,,
\eeq
which is similar to (\ref{eq:S}) with $b\to d$ and an additional propagator $1/(p^2+b)$. It is then clear that by using the same technique as above, we arrive at
\beq
U_{abcd}(k) & = & \sum_{n=0}^\infty (-1)^n\sum_{\ell=0}^{[n/2]}\frac{n!}{\ell!(n-2\ell)!}\frac{(k^2)^{n-\ell}}{(d/2)_\ell}\int_p \int_q \frac{(p^2)^\ell}{(p^2+a)^{n+1}}\frac{1}{p^2+b}\frac{1}{q^2+c}\frac{1}{(q+p)^2+d}\nonumber\\
& = & \sum_{n=0}^\infty \sum_{\ell=0}^{[n/2]}\sum_{h=0}^\ell (-1)^{n+\ell-h}\frac{n!}{(n-2\ell)!h!(\ell-h)!}\frac{(k^2)^{n-\ell}}{(d/2)_\ell}a^{\ell-h}\nonumber\\
& & \hspace{3.0cm}\times\int_p \int_q \frac{1}{(p^2+a)^{n+1-h}}\frac{1}{p^2+b}\frac{1}{q^2+c}\frac{1}{(q+p)^2+d}\,.
\eeq
In the case where $a=b$, we then obtain
\beq
U_{aacd}(k) & = & \sum_{n=0}^\infty \sum_{\ell=0}^{[n/2]}\sum_{h=0}^\ell (-1)^{n+\ell-h}\frac{n!}{(n-2\ell)!h!(\ell-h)!}\frac{(k^2)^{n-\ell}}{(d/2)_\ell}a^{\ell-h}\,I_{(n+2-h)11}(a,c,d)\,.
\eeq
In the case $a\neq b$, we write
\beq
\alpha_{n+1} & \equiv & \frac{1}{(p^2+a)^{n+1}}\frac{1}{p^2+b}=\frac{1}{(p^2+a)^{n}}\frac{1}{p^2+a}\frac{1}{p^2+b}= \frac{1}{b-a}\frac{1}{(p^2+a)^{n}}\left[\frac{1}{p^2+a}-\frac{1}{p^2+b}\right]\nonumber\\
& = & \frac{1}{b-a}\frac{1}{(p^2+a)^{n+1}}-\frac{1}{b-a}\alpha_n=\frac{1}{b-a}\frac{1}{(p^2+a)^{n+1}}-\frac{1}{(b-a)^2}\frac{1}{(p^2+a)^n}+\frac{1}{(b-a)^2}\alpha_{n-1}\nonumber\\
& = & \frac{1}{b-a}\frac{1}{(p^2+a)^{n+1}}-\frac{1}{(b-a)^2}\frac{1}{(p^2+a)^n}+\dots +\frac{(-1)^n}{(b-a)^{n+1}}\frac{1}{p^2+a}-\frac{(-1)^n}{(b-a)^{n+1}}\alpha_{0}\nonumber\\
& = & \sum_{j=0}^n\frac{(-1)^j}{(b-a)^{j+1}}\frac{1}{(p^2+a)^{n+1-j}}-\frac{(-1)^n}{(b-a)^{n+1}}\frac{1}{p^2+b}\,,\label{eq:recurrence}
\eeq
and then
\beq
& & U_{abcd}(k)=-\sum_{n=0}^\infty \sum_{\ell=0}^{[n/2]}\sum_{h=0}^\ell (-1)^\ell\frac{n!}{(n-2\ell)!h!(\ell-h)!}\frac{(k^2)^{n-\ell}}{(d/2)_\ell}\frac{a^{\ell-h}}{(b-a)^{n+1-h}}\,I_{111}(b,c,d)\nonumber\\
& & \hspace{0.5cm}+\,\sum_{n=0}^\infty \sum_{\ell=0}^{[n/2]}\sum_{h=0}^\ell\sum_{j=0}^{n-h} (-1)^{n+\ell-h+j}\frac{n!}{(n-2\ell)!h!(\ell-h)!}\frac{(k^2)^{n-\ell}}{(d/2)_\ell}\frac{a^{\ell-h}}{(b-a)^{j+1}}\,I_{(n+1-h-j)11}(a,c,d)\,.
\eeq
As before, to obtain the expansion up to order $(k^2)^p$, we need to consider the sum over $n$ up to $n=2p$.\\

Let us finally consider the case of $M_{abcde}$, assuming $a\neq 0$ and $b\neq 0$. We write
\beq
M_{abcde}(k)=\int_p \int_q \frac{1}{(p+k)^2+a}\frac{1}{(q+k)^2+b}\frac{1}{p^2+c}\frac{1}{q^2+d}\frac{1}{(p-q)^2+e}\,,
\eeq
where we assume $a\neq 0$ and $b\neq 0$. The expansion of the two propagators carrying $k$ leads to
\beq
M_{abcde}(k) & = & \sum_{n_1=0}^\infty\sum_{n_2=0}^\infty\sum_{\ell_1=0}^{n_1}\sum_{\ell_2=0}^{n_2}\frac{(-1)^{n_1+n_2}n_1!n_2!}{\ell_1!\ell_2!(n_1-\ell_1)!(n_2-\ell_2)!}(k^2)^{n_1+n_2-\ell_1-\ell_2}\nonumber\\
& & \times\,\int_p \int_q \frac{(2(p\cdot k))^{\ell_1}}{(p^2+a)^{n_1+1}}\frac{(2(q\cdot k))^{\ell_2}}{(q^2+b)^{n_2+1}}\frac{1}{p^2+c}\frac{1}{q^2+d}\frac{1}{(p-q)^2+e}\,.
\eeq
This can be simplified using the last formula in the Appendix of Ref.~\cite{Davydychev:1993pg}
\beq
M_{abcde}(k) & = & \sum_{n_1=0}^\infty\sum_{n_2=0}^\infty\sum_{\ell_1=0}^{n_1}\sum_{\ell_2=0}^{n_2}\sum_{2h_{1/2}+h_3=\ell_{1/2}}\frac{(-1)^{n_1+n_2}n_1!n_2!}{(n_1-\ell_1)!(n_2-\ell_2)!h_1!h_2!h_3!}\frac{(k^2)^{n_1+n_2-(\ell_1+\ell_2)/2}}{(d/2)_{(\ell_1+\ell_2)/2}}\nonumber\\
& & \times\,\int_p \int_q \frac{(p^2)^{h_1}}{(p^2+a)^{n_1+1}}\frac{(q^2)^{h_2}}{(q^2+b)^{n_2+1}}\frac{1}{p^2+c}\frac{1}{q^2+d}\frac{(2p\cdot q)^{h_3}}{(p-q)^2+e}\,.
\eeq
Using $2p\cdot q= p^2+q^2+e-(p-q)^2-e$, this rewrites
\beq
M_{abcde}(k) & = & \sum_{n_1=0}^\infty\sum_{n_2=0}^\infty\sum_{\ell_1=0}^{n_1}\sum_{\ell_2=0}^{n_2}\sum_{2h_{1/2}+h_3=\ell_{1/2}}\sum_{j_1+j_2+j_3+j_4=h_3}\frac{(-1)^{n_1+n_2+j_4}n_1!n_2!}{(n_1-\ell_1)!(n_2-\ell_2)!h_1!h_2!j_1!j_2!j_3!j_4!}\nonumber\\
& & \times\,\frac{(k^2)^{n_1+n_2-(\ell_1+\ell_2)/2}}{(d/2)_{(\ell_1+\ell_2)/2}}e^{j_3}\int_p \int_q \frac{(p^2)^{h_1+j_1}}{(p^2+a)^{n_1+1}}\frac{(q^2)^{h_2+j_2}}{(q^2+b)^{n_2+1}}\frac{1}{p^2+c}\frac{1}{q^2+d}\frac{1}{((p-q)^2+e)^{1-j_4}}\nonumber\\
& = & \sum_{n_1=0}^\infty\sum_{n_2=0}^\infty\sum_{\ell_1=0}^{n_1}\sum_{\ell_2=0}^{n_2}\sum_{2h_{1/2}+h_3=\ell_{1/2}}\sum_{j_1+j_2+j_3+j_4=h_3}\sum_{p_1=0}^{h_1+j_1}\sum_{p_2=0}^{h_2+j_2}\nonumber\\
& & \times\,\frac{(-1)^{n_1+n_2+j_4+h_1+j_1-p_1+h_2+j_2-p_2}n_1!n_2!(h_1+j_1)!(h_2+j_2)!}{(n_1-\ell_1)!(n_2-\ell_2)!h_1!h_2!j_1!j_2!j_3!j_4!p_1!p_2!(h_1+j_1-p_1)!(h_2+j_2-p_2)!}\frac{(k^2)^{n_1+n_2-(\ell_1+\ell_2)/2}}{(d/2)_{(\ell_1+\ell_2)/2}}\nonumber\\
& & \times\,a^{h_1+j_1-p_1}b^{h_2+j_2-p_2}e^{j_3}\int_p \int_q \frac{(p^2+c)^{-1}}{(p^2+a)^{n_1+1-p_1}}\frac{(q^2+d)^{-1}}{(q^2+b)^{n_2+1-p_2}}\frac{1}{((p-q)^2+e)^{1-j_4}}\,.\nonumber\\
\eeq
In the case $a=c$ and $b=d$, we arrive at
\beq
M_{ababe}(k) & = & \sum_{n_1=0}^\infty\sum_{n_2=0}^\infty\sum_{\ell_1=0}^{n_1}\sum_{\ell_2=0}^{n_2}\sum_{2h_{1/2}+h_3=\ell_{1/2}}\sum_{j_1+j_2+j_3+j_4=h_3}\sum_{p_1=0}^{h_1+j_1}\sum_{p_2=0}^{h_2+j_2}\nonumber\\
& & \times\,\frac{(-1)^{n_1+n_2+j_4+h_1+j_1-p_1+h_2+j_2-p_2}n_1!n_2!(h_1+j_1)!(h_2+j_2)!}{(n_1-\ell_1)!(n_2-\ell_2)!h_1!h_2!j_1!j_2!j_3!j_4!p_1!p_2!(h_1+j_1-p_1)!(h_2+j_2-p_2)!}\nonumber\\
& & \times\,\frac{(k^2)^{n_1+n_2-(\ell_1+\ell_2)/2}}{(d/2)_{(\ell_1+\ell_2)/2}}a^{h_1+j_1-p_1}b^{h_2+j_2-p_2}e^{j_3}I_{(n_1+2-p_1)(n_2+2-p_2)(1-j_4)}(a,b,e)\,.
\eeq
In the other cases, we need to make use of (\ref{eq:recurrence}). We note that $\ell_i\leq n_i$ and thus $n_i-\ell_i/2\geq n_i/2$, so terms with $n_1+n_2> 2p$ contribute to powers of $k^2$ with exponent $n_1+n_2-(\ell_1+\ell_2)/2>p$. In other words, to obtain the expansion up to order $(k^2)^p$, we need to truncate the double sum over $n_1$ and $n_2$ such that it includes all terms with $n_1+n_2\leq 2p$. For $j_4=0$, we need to relate $I_{(n_1+2-p_1)(n_2+2-p_2)1}(a,b,e)$ to $I_{111}(a,b,e)$ by repeated use of (\ref{eq:derI}). For $j_4\geq 1$, we can relate $I_{(n_1+2-p_1)(n_2+2-p_2)(1-j_4)}(a,b,e)$ to
\beq
J_{\alpha,\beta}(a)\equiv\int_p \frac{1}{(p^2+a)^\alpha (p^2)^\beta}=\frac{a^{2-\alpha-\beta-\epsilon}}{(4\pi\Lambda^2)^{-\epsilon}}\frac{\Gamma(2-\beta-\epsilon)\Gamma(\alpha+\beta-2+\epsilon))}{\Gamma(2-\epsilon)\Gamma(\alpha)}\,.
\eeq
instead. More precisely
\beq
& & I_{(n_1+2-p_1)(n_2+2-p_2)(1-j_4)}(a,b,e)\nonumber\\
& & \hspace{0.5cm}=\,\int_p \int_q \frac{1}{(p^2+a)^{n_1+2-p_1}}\frac{1}{(q^2+b)^{n_2+2-p_2}}((p-q)^2+e)^{j_4-1}\nonumber\\
& & \hspace{0.5cm}=\,\sum_{q_1+q_2+q_3+q_4=j_4-1}\frac{(j_4-1)!}{q_1!q_2!q_3!q_4!}e^{q_4}\int_p \int_q \frac{(p^2)^{q_1}}{(p^2+a)^{n_1+2-p_1}}\frac{(q^2)^{q_2}}{(q^2+b)^{n_2+2-p_2}}(-2p\cdot q)^{q_3}\nonumber\\
& & \hspace{0.5cm}=\,\sum_{q_1+q_2+2q_3+q_4=j_4-1}\frac{(j_4-1)!}{q_1!q_2!(2q_3)!q_4!}\frac{(2q_3)!}{q_3!}\frac{e^{q_4}}{(d/2)_{q_3}}\int_p \int_q \frac{(p^2)^{q_1+q_3}}{(p^2+a)^{n_1+2-p_1}}\frac{(q^2)^{q_2+q_3}}{(q^2+b)^{n_2+2-p_2}}\nonumber\\
& & \hspace{0.5cm}=\,\sum_{q_1+q_2+2q_3+q_4=j_4-1}\frac{(j_4-1)!}{q_1!q_2!q_3!q_4!}\frac{e^{q_4}}{(d/2)_{q_3}}J_{n_1+2-p_1,-q_1-q_3}(a)J_{n_2+2-p_2,-q_2-q_3}(b)\,,
\eeq
where in the last steps we have used Eq.~(\ref{eq:Davy}).\\

For instance up to order $\mu^2$, the gluon and ghost anomalous dimensions in the IR are found to be
\begin{align}
\gamma_A=&\lambda\left\{\frac{1}{3}-\frac{217}{180}\frac{\mu^2}{m^2}+\frac{4 N_f}{5 N}\frac{\mu^2}{M^2}  \right\}\nonumber\\
&+\lambda^2\frac{\mu^2}{m^2}\left\{\frac{38687}{25920}-\frac{37}{288}\pi^2+\frac{3647}{288}S_2-\frac{179}{360}\ln\frac{\mu^2}{m^2}+\frac{13}{144}\ln^2\frac{\mu^2}{m^2}\right.\nonumber\\
& \hspace{1.4cm}\left.\left.+\frac{N_f}{N}\left[\left(\frac{8}{9}-16x^2+\frac{994}{9}x^4-\frac{2756}{9}x^6+\frac{520}{9}x^8+\frac{7216}{9}x^{10}-\frac{1984}{3}x^{12} \right) \frac{\tilde{I}_{1xx}}{(1-4x^2)^4} \right. \right. \right.\nonumber \\
&\hspace{2.0cm}\left. \left. +\left(\frac{151}{90}-\frac{3334}{135}x^2+\frac{3280}{27}x^4-\frac{33112}{135}x^6+\frac{3112}{9}x^8-\frac{992}{3}x^{10}  \right) \frac{\ln x^2}{(1-4x^2)^4}\right. \right. \nonumber \\
&\hspace{7.0cm}\left. \left.-\frac{25+1122x^2-12128x^4+36760x^6-44640 x^8}{270 (1-4x^2)^3}\right] \right. \nonumber\\
& \hspace{1.4cm}\left. +\frac{C_F}{N}\frac{N_f}{N}\left[-\left(\frac{16}{9}-32x^2+\frac{1952}{9}x^4-\frac{5888}{9}x^6+\frac{1664}{3}x^8+\frac{1280}{9}x^{10} \right)\frac{\tilde{I}_{1xx}}{(1-4x^2)^4} \right. \right. \nonumber\\
&\hspace{2.7cm}\left. \left. -\left(\frac{4+504x^2-8056x^4+47792x^6-78432x^8+19840x^{10}+9600x^{12}}{135(1-x^2)^2} \right)\frac{\ln x^2}{(1-4x^2)^4}\right. \right.\nonumber\\
& \hspace{9.0cm}\left. \left. -\frac{4-416x^2+3904x^4-5376x^6+4800x^8}{135(1-4x^2)^3(1-x^2)}\right.\right\},
\end{align}
and
\begin{align}
\gamma_c=&\lambda \left( -\frac{5}{12}+\frac{1}{2}\ln\frac{\mu^2}{m^2} \right)\frac{\mu^2}{m^2}\nonumber\\
&+\lambda^2\frac{\mu^2}{m^2}\left\{ -\frac{4295}{576}+\frac{5}{72}\pi^2+\frac{459}{16}S_2 +\frac{1}{12}\ln\frac{\mu^2}{m^2}\right.\nonumber \\
&\hspace{1.5cm}\left.   +\frac{N_f}{N}\left[\frac{5}{9}+4x^2+4x^4\tilde{I}_{1xx}+\left(\frac{1}{3}+2x^2 \right)\ln x^2 \right]\right\},
\end{align}
where we have set $x\equiv M/m$. In deriving these expressions, we have used that $\psi_1(1/3)+\psi_1(1/6)=8\pi^2/3+81S_2$, where $\psi_1$ denotes the trigamma function. In the quenched limit $N_f\to 0$, we recover the results obtained in Ref.~\cite{glmq8}. As already noticed in this reference, the gluon anomalous dimension is entirely controlled by the one-loop result in the infrared because the two-loop contribution carries an extra factor $\mu^2/m^2$. In the case of the ghost anomalous dimension, both contributions have a factor $\mu^2/m^2$ and feature logarithms of the form $\ln\mu^2/m^2$. We note however that the logarithm in the two-loop contribution has the same power than the logarithm in the one-loop contribution and it is thus under perturbative control in the deep infrared where the coupling approaches $0$ \cite{Reinosa:2017qtf}. Similar (but lengthier) expressions can be obtained for the anomalous dimensions $\gamma_\psi$ and $\gamma_M$.\\

\end{widetext}

\end{document}